**Spontaneous Twist of Ferroelectric Smectic Blocks in Polar Fluids**

*Hiroya Nishikawa[1]\*, Yasushi Okumura[2], Dennis Kwaria[1], Atsuko Nihonyanagi[1], and Fumito Araoka[1]\**

H. Nishikawa, D. Kwaria, A. Nihonyanagi, F. Araoka
RIKEN Center for Emergent Matter Science, 2-1 Hirosawa, Wako, Saitama 351-0198, Japan
E-mail: hiroya.nishikawa@riken.jp; fumito.araoka@riken.jp

Y. Okumura
Institute for Materials Chemistry and Engineering Kyushu University, 6-1 Kasuga-Koen, Kasuga, Fukuoka 816–8580, Japan

Keywords: ferroelectric nematic, ferroelectric smectic, chiral ground state, liquid crystal, polar fluid

In soft matter, the polar orientational order of molecules can facilitate the coexistence of structural chirality and ferroelectricity. The ferroelectric nematic ($N_F$) state, exhibited by achiral calamitic molecules with large dipole moments, serves as an ideal model for the emergence of spontaneous structural chirality. This chiral ground state arises from a left- or right-handed twist of polarization due to depolarization effects. In contrast, the ferroelectric smectic state, characterized by a polar lamellar structure with lower symmetry, experiences significantly higher energy associated with layer-twisting deformations and the formation of domain walls, thus avoiding a continuously twisted layered structure. In this study, we report two types of achiral molecules (**BOE-NO$_2$** and **DIOLT**) that possess different molecular structure but exhibit a $N_F$–ferroelectric smectic phase sequence. We demonstrate that the chiral ground state of $N_F$ is inherited in the ferroelectric smectic phases of **BOE-NO$_2$**, which features larger dipole moments and a steric hindrance moiety, thereby triggering the formation of the twisted polar smectic blocks.

## 1. Introduction

A strong correlation between structural chirality and ferroelectricity is a rare phenomenon. A notable example is achiral triglycine sulfate, discovered by Matthias et al. in 1956.[1] This crystalline material is a proper ferroelectric that exhibits spontaneous polarization ($P_s$), which is linked to the paraelectric-ferroelectric phase transition via a change in molecular orientation. Remarkably, this system demonstrates chirality inversion, where the direction of optical rotation reverses in response to polarization switching under electric ($E$-) field reversal.[2] The fact that chirality and ferroelectricity can interrelate is particularly intriguing, as it opens the possibility of functionalization, such as a polarization-chirality nonvolatile dual binary memory in materials. The correlation between chirality and ferroelectricity is also observed in soft matter systems. A well-known example is the chiral smectic C phase, the first ferroelectric liquid crystal (LC) reported by Meyer et al. in 1975. In this case, ferroelectricity arises from the introduction of a chiral carbon into a calamitic molecule, reducing symmetry.[3] However, it was later discovered that chirality and ferroelectricity can also correlate in achiral LC systems. For example, in achiral bent-core molecules, spontaneous symmetry breaking occurs due to the tilted lamellar structure, leading to layer chirality in the liquid crystalline smectic fluid.[4] The electrically switchable phases among these include the so-called B2 (SmCP) phases, in which the chiral ($\pm$)-$SmC_sP_F$ structure exhibits ferroelectricity. More recently, achiral rod-like molecules with large permanent dipoles (> 10 Debye) were found to spontaneously exhibit chiral ferroelectric (helielectric) phases[5,6] with polar nematic/smectic ordering. This new type of chirality-ferroelectricity correlation has generated significant interest among scientists, a discovery made possible by the development of ferroelectric nematic ($N_F$) LCs. [7]

The $N_F$ phase is the ferroelectric counterpart of the nematic (N) phase in three-dimensional fluids, and in some instances, the hierarchical structure transitions sequentially to the ferroelectric smectic A ($SmA_F$)[8] and ferroelectric smectic C ($SmC_F$[9a,9b], $SmC_P$[9c]) phases, which are axially polar and exhibit periodic molecular positional distribution. The unique characteristics of these polar phases are reflected in their substantial polarization, which has garnered significant attention due to the distinct physical phenomena[10] and functionalities[11] they present. Interestingly, in the $N_F$ phase, emergent chirality-ferroelectricity correlations have been observed in bulk systems through mechanisms distinct from those found in known systems. For instance, in confined $N_F$ pillars, left-/right-handed polar topologies are spontaneously generated as a result of the balance between polarization strength and elasticity.[10j] Conversely, in confined $N_F$ slabs where torque is not influenced by the surface, periodic domains of alternating left-/right-handed twists are formed, bounded by domain walls to mitigate the energy cost of electrostatic interactions.[10m] This behavior indicates that the $N_F$ system naturally favors a chiral ground state (**Figure 1**a). It raises the question of whether such a depolarization-driven chiral ground state can also be observed in ferroelectric smectic phases (i.e., SmA$_F$/SmC$_F$/SmC$_P$). However, since the increase in elastic energy due to twisting in the smectic phase is much higher than that in the N$_F$ phase, the twisting of smectic blocks (Figure 1b) is rare in achiral systems.

In this study, we discover a twisted ferroelectric smectic structure in novel polar tolan-based molecules, **nBOE-NO$_2$** and **nDIOLT** series, which potentially exhibit the N$_F$–ferroelectric smectic phase sequence. We experimentally confirm that **nBOE-NO$_2$** (n = 3–5) and **nDIOLT** (n = 2,3) follow the N$_F$–SmA$_F$–(SmX$_F$) sequence. In an antiparallel rubbed cell, twisted ferroelectric smectic (T-SmA$_F$, T-SmX$_F$) phases are generated, inheriting the twisted N$_F$ structure. Notably, we demonstrate that the **nBOE-NO$_2$** homologue preserves the twisted-ferroelectric-nematic/smectic structure even under a polar anchoring–degenerated planar anchoring system.

## 2. Results

### 2.1. Molecular design

Recently, we reported on the unique properties of cyano-tolan polar molecules bearing a bicycloorthoester unit (**BOE**).[5b] It is well-established that strong head-tail dipole–dipole interactions are crucial for stabilizing the N$_F$ phase. However, longer alkyl chains tend to destabilize this phase. Through our systematic study of a series of **BOE** compounds, we concluded that additional interactions—such as π···π, C–F···π, acetylene···π, and H···F interactions—also play a significant role in stabilizing the N$_F$ phase. Our previous work found that the cyano (CN) group connected to the **BOE** is a key substituent for the N$_F$ to heliconical ferroelectric nematic (N$_{TBF}$[5a], $^{HC}$N$_F$[5b]) phase transition. To substantiate this, we further synthesized a nitro-version of **BOE**, i.e., **4BOE-NO$_2$**, as a control material which exhibited N$_F$–SmA$_F$ phase transition (Supporting Figure 13 in reference [5b]). Therefore, the **nBOE-NO$_2$** models (with **5BOE-NO$_2$** shown in **Figure 2**a) are promising candidates for investigating the direct phase transition between N$_F$ and SmA$_F$ phases. Similar to **BOE**, we confirmed the high thermodynamic stability of the N$_F$ phase for **nBOE-NO$_2$**. To explore the effect of the rigid tolan unit on the phase sequence, we designed another model by incorporating the tolan unit into polar molecules previously developed by Karcz et al.[5a] and Gibb & Hobbs et al.[6a] In this study, this model is referred to as **nDIOLT** (refer to Figure 2b for **3DIOLT**). This mesogenic structure based on 1,3-dioxane (DIO) with the long tolan (LT) unit is referred to as **DIOLT** hereafter. We calculated the optimized structures, permanent dipoles ($\mu$), and dipolar deviation angles ($\beta$) for **nBOE-NO$_2$** (n = 3–5) and **nDIOLT** (n = 1–3). As shown in Figure S7 and Table S1, Supporting Information, **nBOE-NO$_2$** and **nDIOLT** exhibited high $\mu$ values of approximately 15.0 and 13.0 Debye, respectively. Very small dipole angles (0.2° < $\beta$ < 1.5°) were observed for **nBOE-NO₂**, owing to its straight and rigid structure. In contrast, **nDIOLT** showed $\beta$ values between 11° and 12°, which were smaller than those of conventional models, as the tolan unit contributes to the molecular linearity and rigidity. As anticipated, **nBOE-NO₂** and **nDIOLT** exhibited a direct phase transition from $N_F$ to $SmA_F$ phase, with **5BOE-NO₂** and **nDIOLT** (n = 2,3) displaying the ferroelectric smectic ($SmX_F$) phase.

## 2.2 Phase transition behavior

Figure 2c, 2d and Figure S8, S9 (Supporting Information) present the DSC curves for **5BOE-NO₂** and **3DIOLT** upon 1st cooling and 2nd heating (additional details can be found in Tables S2 and S3, Supporting Information). Both compounds exhibit nematic (N), antiferroelectric mesomorphic ($M_{AF}$)[12], ferroelectric nematic ($N_F$), ferroelectric smectic A ($SmA_F$), and unspecified ferroelectric smectic ($SmX_F$) phases. Corresponding polarized optical microscopic (POM) images are shown in Figure 2e–j and Figure S10 (Supporting Information). The $SmX_F$ phase is observed only during cooling, whereas the $SmA_F$, $M_{AF}$, and $N_F$ phases are thermodynamically stable, indicating enantiotropic LC. For **5BOE-NO₂**, the peaks at 220.4 °C ($\Delta H$ = 0.02 kJ mol⁻¹), 212.8 °C ($\Delta H$ = 0.59 kJ mol⁻¹), and 170.1 °C ($\Delta H$ = 0.11 kJ mol⁻¹) correspond to the N–$M_{AF}$, $M_{AF}$–$N_F$, and $N_F$–$SmA_F$ phase transitions, respectively. For **3DIOLT**, the peaks at 127.1 °C ($\Delta H$ = 0.13 kJ mol⁻¹) and 123.4 °C ($\Delta H$ = 0.05 kJ mol⁻¹) correspond to the N–$N_F$ and $N_F$–$SmA_F$ transitions, respectively. The transition enthalpy ($\Delta H$) for the $SmA_F$–$SmX_F$ phase transition is not determined due to a weak and uncertain peak, suggesting a weak first- or second-order phase transition.

In both molecules, in the parallel rubbed cell (thickness: 10 μm), uniform $N_F$ (U-$N_F$) and uniform $SmA_F$ (U-$SmA_F$) textures are observed, where the molecular director (**n**) aligns along the rubbing direction (Figure S11, Supporting Information). Under crossed polarizers, the dark field is seen in these states when the directions of the polarizer (**P**) and rubbing (**R**) coincide (i.e., at extinction position). The $SmX_F$ phase, in contrast, displays numerous stripe domains with faint birefringence. In the antiparallel rubbed cell (thickness: 10 μm), π-twist walls are generated in the $N_F$ phase (Figure 2e). These walls separate neighboring chiral domains, where **P** and **n** are twisted by 180° either clockwise or counterclockwise.[13] Such domain walls persist in the $SmA_F$ phase and even in the $SmX_F$ phase, indicating that the twisted structure is inherited from the twisted $N_F$ (T-$N_F$) phase, henceforth referred to as T-$SmA_F$ and T-$SmX_F$ phases. Interestingly, at the T-$N_F$–T-$SmA_F$ phase transition for **5BOE-**

**NO₂**, multiple triangular domains form across the entire cell (Figure 2f, Movie S1, Note S1, Supporting Information), showing color changes upon rotation but without extinction positions. These domains then merge to form a uniform domain (T-SmA$_F$ state) (Figure 2g). A characteristic grid pattern emerges as the system cools further into the T-SmX$_F$ phase (Figure 2h).

Although the characteristic textures of T-SmA$_F$ and T-SmX$_F$ states are observed in both compounds, U-SmA$_F$ and U-SmX$_F$ states coexist in **3DIOLT** (Figure 2i and 2j). Additionally, the transient triangular pattern does not appear in **3DIOLT**. It is also notable that, in the case of **3DIOLT**, domain walls between T-/U-SmA$_F$ (T-/U-SmX$_F$) states move at the N$_F$–SmA$_F$ transition to eliminate the twisted domains, forming a planar structure in most regions (Figure 2i and Figure S12, Supporting Information). In contrast, for **5BOE-NO₂**, the twisted structure remains highly stable, and the planar structure is hardly formed.

## 2.3. Polarization behavior

The results of broadband dielectric spectroscopy (BDS), reversal polarization current (PRC), and second harmonic generation (SHG) studies for **5BOE-NO₂** are presented in **Figure 3**, Figure S13 and S14 (Supporting Information). For BDS, PRC, SHG studies, we used a ITO glass cell (thickness: 9 μm), a IPS cell (thickness: 5 μm) and a IPS cell (thickness: 10 μm), respectively. For BDS, the apparent dielectric permittivity ($\varepsilon'_{app}$) at 100 Hz reaches ≈10k in the N$_F$ phase, then decreases to ≈6k in the SmA$_F$ phase (Figure 3a). It subsequently increases again to 7.4k in the SmX$_F$ phase. The corresponding dielectric loss ($\varepsilon''$) is shown in Figure 3b. During the phase transitions from N$_F$ to SmA$_F$ to SmX$_F$, the relaxation frequency in the N$_F$ phase ($f_r \approx 10^4$ Hz) gradually shifts to a lower frequency in the SmA$_F$ phase, and then suddenly jumps back to $f_r \approx 10^4$ Hz in the SmX$_F$ phase. The dielectric relaxation in the SmA$_F$ phase ($f_r \approx 10^2$ Hz) is likely due to the suppression of polarization fluctuations within the lamellar structure.[9b] In contrast, the dielectric behavior in the SmX$_F$ phase ($f_r \approx 10^4$ Hz) is akin to the Goldstone mode behavior due to tilt direction fluctuations observed in the SmC$_F$ phase[14] Recently, the large dielectric permittivity in the N$_F$ phase is interpreted as apparent values,[15] while observing such a signature behavior is still useful for phase identification. In addition, the recent BDR research suggests that the major dielectric relaxation process of the NF phase is not due to non-collective modes associated with molecular rotation but is mainly due to the Goldstone mode (collective mode) related to director orientation fluctuation. Figure 3c–g display the PRC data recorded in an in-plane switching cell (antiparallel rubbing ∥ **E**) for various phases. At 216 °C, the switching current curve exhibits four peaks within one period, indicating an antiferroelectric response. Therefore, we assign this as the $M_{AF}$ phase. In the $N_F$, $SmA_F$, and $SmX_F$ phases, two distinct reversal peaks are observed, indicating that these three phases have two distinct polar ground states, characterizing them as ferroelectric. Moreover, in the $SmA_F$ and $SmX_F$ phases, the field reversal brings about the polarization reversal through the reorientation of the layers, further confirming the proper ferroelectric nature of these phases. At the $N_F$–$SmA_F$ phase transition, multiple current peaks within each half-period of the applied voltage are observed, suggesting the presence of multiple polar domains.

The spontaneous polarization ($P_s$) and coercive electric field ($E_c$) as a function of temperature are shown in Figure 3h and 3i, respectively. The value of $P_s$ gradually increases to 3.7 μC cm$^{-2}$ in the $N_F$ phase, then rises sharply to ≈6.0 μC cm$^{-2}$ in the $SmA_F$ phase. In the $SmX_F$ phase, Ps shows only a slight increase. As shown in Figure 3j, strong SH activity is observed throughout the $N_F$, $SmA_F$, and $SmX_F$ phases. Results for the polarization behavior of **3DIOLT** are presented in Figure S15–S17 (Supporting Information). The temperature window of the $N_F$ phase is narrow, with a small $P_s$ of 1.7 μC cm$^{-2}$ observed. Over the entire temperature range of the $SmA_F$ and $SmX_F$ phases, $P_s$ gradually increases to 3.7 μC cm$^{-2}$. $E_c$ in the $N_F$ phase of **5BOE-NO₂** is approximately half (≈0.25 kV cm$^{-1}$) of that observed for **3DIOLT**. In the $SmA_F$ phase, $E_c$ ranges from 1.3 to 3.3 kV cm$^{-1}$, which is also about half that of **3DIOLT** ($2.3 < E_c < 6.0$ kV cm$^{-1}$). As shown in Figure 3j and Figure S15i (Supporting Information), the $N_F$, $SmA_F$, and $SmX_F$ phases for **5BOE-NO₂** and **3DIOLT** exhibit strong SHG. Intriguingly, for **5BOE-NO₂**, the distinct circular polarized SHG was detected in the chiral neighboring domains. We estimated g-values in the T-$N_F$ and T-$SmA_F$ states to be −1.51/1.08 and −0.81/0.73, respectively, using the following formula:

$$g = \frac{I_L - I_R}{\frac{1}{2}(I_L + I_R)} \qquad (1)$$

, where $g$, $I_L$ and $I_R$ represent g-value, SHG intensity in the domains with left/right chirality, respectively. Another chiroptical properties are discussed in section 3.

## 2.4. Structural characterization

**Figure 4**a–h display 2D small-angle X-ray scattering (SAXS) patterns for **5BOE-NO₂** (free-standing sample). In the $N_F$ phase, two diffuse diffraction spots are observed due to flow-induced alignment (Figure 4a). Interestingly, at the phase transition from $N_F$ to $SmA_F$, these two spots split into four distinct peaks (Figure 4c and 4b). As cooling progresses, the split angle changes, and the four peaks persist in both the $SmA_F$ and $SmX_F$ phases (Figure 4d). Figure 4i presents a 1D X-ray diffractogram derived from the box scan analysis of the 2D X-

ray profile under a magnetic field (*M*-field, ≈1 T) (Figure 4e, 4f and Figure S18, S19, Supporting Information). For the free-standing sample, a line scan through a pair of diffraction peaks passing through the beam center is shown in Figure S20 (Supporting Information). The variation of $2\pi/q$ and $\Delta q_{FWHM}$ as a function of temperature is displayed in Figure 4j. During the phase transition from $N_F$ to $SmA_F$, $2\pi/q$ increases, while the full width at half maximum ($\Delta q_{FWHM}$) rapidly decreases, indicating a significant increase in correlation length. This suggests that both the $SmA_F$ and $SmX_F$ phases exhibit long-range positional order. The characteristics of the $SmX_F$ phase, such as the larger $2\pi/q$ value and the higher relaxation frequency compared to the $SmA_F$ phase, display a trend similar to that previously reported for the $SmX_F$ phase in the **BOE** series.[5b] When the $SmA_F$ and $SmX_F$ phases are placed in a magnetic field (*M*-field, ≈1 T), only a single pair of diffraction peaks appears along the *M*-field (**n ∥ B**). For the *E*-field-aligned sample, the SAXS peak in the SmAF phase showed $2\pi/q$ = 2.78 nm, which was almost identical to the single-molecule length (2.68 nm), indicating the lamellar structure with a $d \approx 2.78$ nm period. The pair of SAXS peaks for the free-standing sample also showed same value ($2\pi/q$ = 2.79 nm). Besides, over the entire $SmA_F$ regime, the $\Delta q_{FWHM}$ of aligned and non-aligned samples are matched (Figure 4j). Thus, a uniform lamellar structure was formed even in the absence of the *M*-field. In this case, the orientation of the lamellar structure is likely influenced by the constraint at the air-LC interface. Assuming that the two pairs of diffraction peaks observed in the non-aligned sample are attributed to twisted smectic blocks, it is reasonable to conclude that the smectic blocks align unidirectionally into a uniform smectic (U-$SmA_F$) structure under the influence of the *M*-field. To investigate the dynamic coalescence of the smectic blocks, we examined the 2D XRD patterns before and after applying an electric field (*E*-field). As shown in Figure 4g, one pair of diffraction peaks appears on the equator, while another pair appears at a different angle. Upon applying the *E*-field in the equatorial direction, the diffraction peak on the equator remains unchanged, while the position of the other pair rotates toward the equator, resulting in the degenerated U-$SmA_F$ structure. This observation provides evidence that two or more smectic blocks, responsive to the *E*-field, twist along the cell normal in the $SmA_F$ phase. It suggests that these twisted smectic blocks inherit the twisted structure of the inherently twisted $N_F$ phase, where the molecules at the air/LC interface are pinned by anchoring. After removing the *E*-field, the degenerated U-$SmA_F$ structure does not revert to the T-$SmA_F$ configuration. Next, we investigated the twist angle of the polar smectic blocks. Assuming that two polar smectic blocks facing the air twist in the free-standing sample, Figure 4k, 4l display 1D XRD profiles as a function of chi ($\chi$) angles at various temperatures. Here, the twist angles of the smectic blocks are defined by the average values of $\chi_1$ and $\chi_2$. The temperature dependence of the twisting angles is shown in Figure 4m. The twist angle reaches its maximum (≈88°) during the $N_F$–$SmA_F$ phase transition, then immediately decreases to ≈55° in the deep $SmA_F$ phase, and slightly increases to ≈58° in the $SmX_F$ phase.

## 3. Discussion

DSC, POM, BDS, PRC, SHG, and XRD studies have demonstrated that **nBOE-NO$_2$** (n = 3, 4) and **nDIOLT** (n = 2, 3) exhibit ferroelectric phases, including the $N_F$ and $SmA_F$ ($SmX_F$) phases (Figure S13, S16, S18, and S21–S37, Supporting Information). As mentioned, in the parallel rubbed cell, the planar alignment is preferred for the $N_F$, $SmA_F$ and $SmX_F$ phases, where **P** and **n** orient parallel to the rubbing direction (Figure S11 and S25, Supporting Information). In contrast, in the antiparallel rubbed cell, a twisted structure is observed in the $N_F$ phase. In this twisted configuration, both **P** and **n** are twisted 180° from the upper to the lower substrate surfaces, with clockwise and counterclockwise twisting directions. This results in domain segregation into two regions, bounded by π-walls. Similarly, domain boundaries are also observed in the T-$SmA_F$ and T-$SmX_F$ structures. For all three phases, no extinction position is observed; instead, the contrast between neighboring domains is reversed by rotating the analyzer. Therefore, the two twisted domains observed in the T-$N_F$, T-$SmA_F$, and T-$SmX_F$ states are in a macroscopic chiral configuration (Figure 2e–2j and Figure S26, Supporting Information). To detect the macroscopic chirality, we performed UV-vis (Figure S38, Supporting Information) and micro-microscopic circular dichroism (micro-CD) spectroscopies. **Figure 5**a–c shows POM images, while Figure 5d displays the corresponding micro-CD spectra with positive and negative signals recorded in adjacent domains of the T-$SmA_F$ phase. As shown in Figure 5a–c, when the analyzer was rotated by ±20°, the image contrast reversed, providing evidence of spontaneous chirality breaking.

When comparing **5BOE-NO$_2$** and **3DIOLT** with the POM images of the twisted smectic phase, **5BOE-NO$_2$** shows a twisted structure induced throughout the entire cell. In contrast, **3DIOLT** exhibits a similar twisted structure formed only partially in the cell, with planar orientation predominantly dominant in the equilibrium state (Figure S12, Supporting Information). For **nBOE-NO$_2$**, the T-$SmA_F$ structure readily forms in all homologs (n = 3–5) by slow cooling from the T-$N_F$ state. Similarly, **2DIOLT** predominantly exhibits T-$SmA_F$ and T-$SmX_F$ regions throughout the cell, although planar orientation is also observed in some areas (Figure S26d, Supporting Information). In the SAXS results for free-standing samples, diffraction patterns reflecting the twisted smectic blocks are observed for all **nBOE-NO$_2$** series (Figure S35a, S35b, and S36, Supporting Information). However, no distinct split pattern is observed for **2DIOLT** and **3DIOLT**, with **3DIOLT** in particular showing a ring-shaped diffraction, suggesting a preference for random planar alignment of mixed T-/U-Smectic domains (Figure S35c and S35d, Supporting Information). **nDIOLT** likely experiences higher twisted elastic energy in the smectic phase than **BOE-NO₂**, making it energetically more favorable to reorient the structure into a uniform alignment to eliminate the twist deformation. Therefore, based on the POM and XRD studies of the **DIOLT** series, it is reasonable to conclude that uniform orientation is preferred.

Then, why does the **BOE-NO₂** series predominantly form a twisted structure? One possible explanation lies in the differences in elastic changes during the $N_F$-$SmA_F$ transition between **nBOE-NO₂** and **nDIOLT**, as well as the elasticity of the $SmA_F$ phase itself. Such an effect is likely due to the difference in molecular structure, dipole moment, dipole angle as well as flexibility. As observed in the POM and PRC studies, **nBOE-NO₂** exhibited multiple current peaks in the temperature range of ≈6–20 K during the $N_F$-$SmA_F$ transition, accompanied by the formation of a multiple triangular texture. In contrast, **nDIOLT** displayed a narrower temperature window (≈2–4 K), with multiple current peaks but no triangular texture within the $N_F$-$SmA_F$ transition. The twisted structure of $SmA_F$ ($SmX_F$) may form via a mechanism similar to that of the chiral ground state in the $N_F$ phase. In the $N_F$ phase, elastic energy increases due to the spontaneous twisted structure and the defect walls that arise from domain division. However, such an unfavorable twisted structure is likely eliminated by the local screening effect of ion doping, suggesting that spontaneous twisting is induced by electrostatic effects.

To further investigate this, we compared the texture of the T-$SmA_F$ state in the **BOE-NO₂** and **DIOLT** series doped with [BMIM][PF₆] (1 wt%) in the antiparallel rubbed cell. **nBOE-NO₂** (n = 3–5) exhibited a partially unwound ground state (Figure 5e, 5f, and Figure S39a, S39b, Supporting Information). As shown in Figure S39c (Supporting Information) and Figure 5g, 5h, doped **2DIOLT** and **3DIOLT** exhibited partially and completely unwound structures, respectively. These results indicate that the T-$N_F$/T-$SmA_F$ structures are influenced by local ionic screening.

Furthermore, we demonstrated the spontaneous twist of the $N_F$ and $SmA_F$ phases in a hybrid cell, which consisted of a rubbed-PI layer and a non-rubbed-polystyrene (PS) layer, corresponding to polar and degenerated planar anchoring[10m,16], respectively. Figure 5i–5k shows POM images of the $N_F$ phase for **5BOE-NO₂**. On one substrate, the polarization **P** is pinned at the interface, while on the counter substrate, azimuthally degenerate in-plane alignment of **P** is allowed. When depolarization within the bulk is energetically more penalized than elastic distortion, a left-handed/right-handed twist, i.e., a chiral ground state, is formed. Remarkably, this twisted state is maintained even after the phase transition to the SmA$_F$ phase, indicating that the polar smectic blocks spontaneously tend to twist each other in the SmA$_F$ phase (Figure 5l–5n). The twist of the smectic block was also confirmed by confocal laser scanning microscopy (Note S2, Supporting Information). The U-SmA$_F$ phase was also partially formed, leading us to suggest that the T-SmA$_F$ structure emerges through an extrinsic formation process that reflects the T-N$_F$ structure.

## 4. Conclusion

In summary, to investigate the twisted ferroelectric smectic structure, we developed two types of tolan-based polar molecules, **nBOE-NO$_2$** and **nDIOLT**, capable of exhibiting the N$_F$–SmA$_F$ phase transition. Among these, **nBOE-NO$_2$** (n = 3–5) and **nDIOLT** (n = 2–3) exhibited the N$_F$–SmA$_F$ phase transition, with **5BOE-NO$_2$**, **2DIOLT**, and **3DIOLT** showing an additional SmX$_F$ phase just below the SmA$_F$ phase. In all cases, a trend was observed where the N$_F$ phase was destabilized in molecules with longer terminal chains, while the SmA$_F$ (SmX$_F$) phases were stabilized. BDS, PRC, and SHG studies confirmed the presence of ferroelectricity in the N$_F$, SmA$_F$, and SmX$_F$ phases. In the antiparallel cell, the N$_F$, SmA$_F$, and SmX$_F$ phases displayed no extinction position and displayed defect lines. POM and micro-CD studies revealed that these defect lines clearly separated the chiral neighboring domains. Notably, only the **BOE-NO$_2$** series exhibited the transient multiple triangular texture during the N$_F$–SmA$_F$ phase transition. The formation of the twisted structure (T-SmA$_F$ and T-SmX$_F$) was also supported by XRD studies. Specifically, the XRD profile of the free-standing sample revealed two distinct pairs of diffraction peaks with finite angles (≈60°) in the SmA$_F$ and SmX$_F$ phases. Moreover, in the antiparallel cell, doping the SmA$_F$ and SmX$_F$ phases with the ionic liquid [BMIM][PF$_6$] (1 wt%) resulted in the observation of a partially unwound structure. When the SmA$_F$ and SmX$_F$ phases were confined between rubbed PI and PS plates, which provided azimuthally degenerated polar anchoring, both twisted and partially non-twisted structures appeared. This observation suggests that the spontaneous twist of the polar smectic blocks was inherited from the chiral ground state of the N$_F$ phase, driven by depolarization. Importantly, the T-SmA$_F$ and T-SmX$_F$ structures were overwhelmingly observed in **nBOE-NO$_2$**, while **nDIOLT** preferred a non-twisted structure. These differences are likely determined by factors such as molecular steric hindrance, depolarization, and elasticity. Understanding this kind of polarity-induced twist may deliver the possibility of manipulating macroscopic chiroptical properties in higher-order polar fluids and/or polar crystals, opening new prospects for chirality engineering.

## Acknowledgements

We are grateful to Dr. H. Koshino (RIKEN, CSRS) and Dr. H. Sato (RIKEN, CEMS) for allowing us to use JNM-ECZ500 (500 MHz, JEOL) and QTOF compact (BRUKER), respectively. We would like to acknowledge the Hokusai GreatWave Supercomputing Facility (project no. RB230008) at the RIKEN Advanced Center for assistance in computing and communication. This work was partially supported by JSPS KAKENHI (JP22K14594; H.N.,

**Contributions**

H.N. and F.A. conceived the project and designed the experiments. F.A. constructed the optical, SHG and a micro-CD measurement system. H.N. designed molecules. A.N. synthesized and characterized all compound. H.N. carried out DFT, DSC, POM, BDS, PRC and micro-CD studies. Y.O. constructed the confocal laser scanning microscopy system and observed confocal fluorescent microscopic images. D.K. performed SHG studies. H.N. D.K. and F.A. took place XRD measurements. H.N. and F.A. analyzed data and discussed the results. H.N. and F.A wrote the manuscript, and all authors approved the final manuscript.

**Corresponding authors**

Correspondence to Hiroya Nishikawa or Fumito Araoka

**Competing interests**

The authors declare no competing interests.

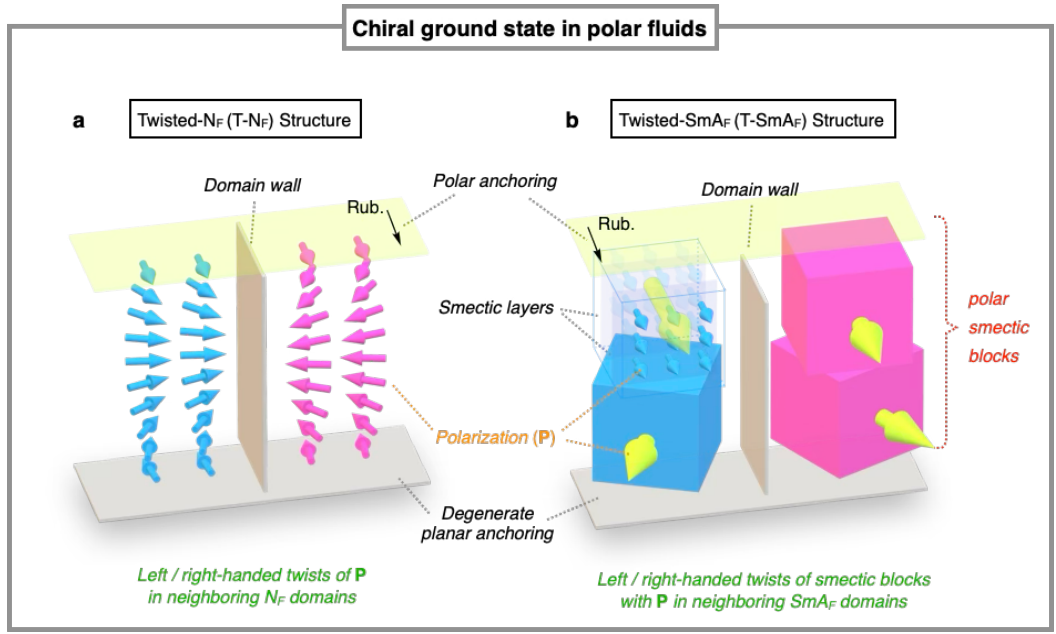

**Figure 1** Schematic illustration of chiral ground states: twisted-$N_F$ (T-$N_F$) (a) and twisted-$SmA_F$ (T-$SmA_F$) (b) structures. Upper and bottom plates are treated by rubbed alignment (polyimide, polar anchoring) and azimuthally degenerated in-plane alignment (polystyrene, degenerate planar anchoring), respectively.

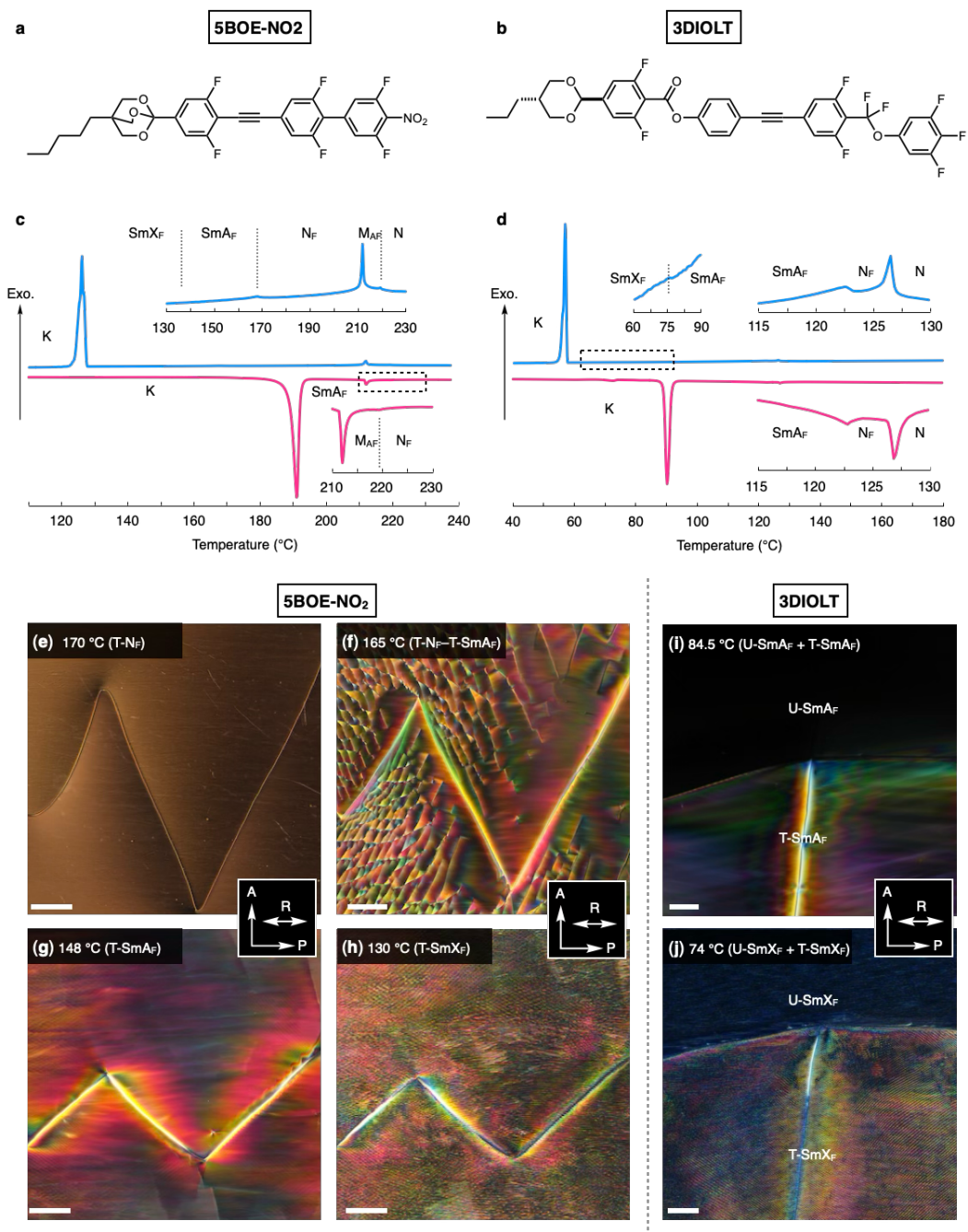

**Figure 2** Chemical structure of **5BOE-NO₂** (a) and **3DIOLT** (b). DSC curves for **5BOE-NO₂** (c) and **3DIOLT** (d). Rate: 10 K min$^{-1}$. Polarized optical microscopic images under cross polarizers for **5BOE-NO₂** (e–h) and **3DIOLT** (i,j). Conditions: antiparallel rubbed cell (thickness: 10 μm).

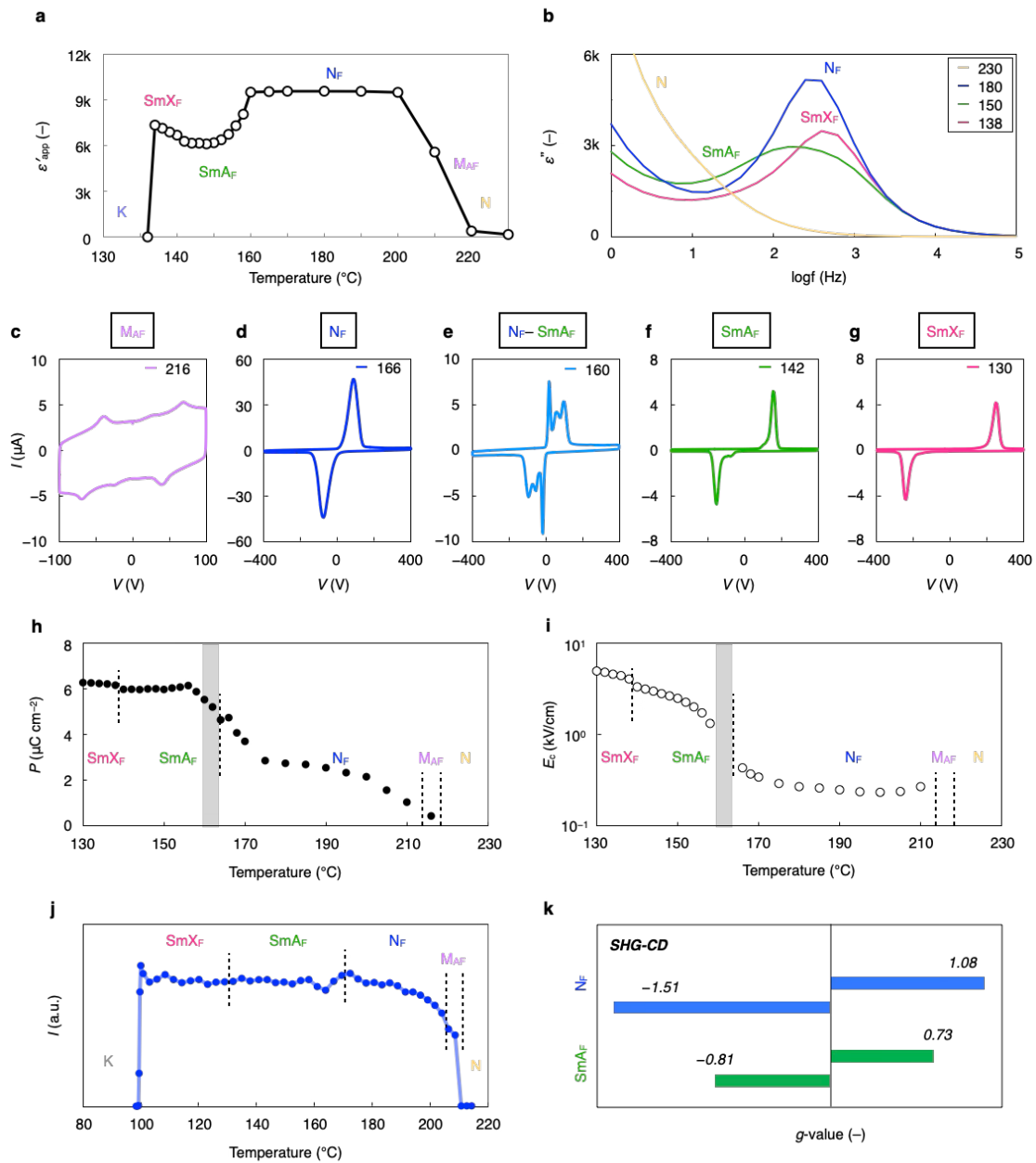

**Figure 3** Polarization behavior for **5BOE-NO$_2$**. DR properties: a) Apparent dielectric permittivity vs temperature ($f$ = 100 Hz), b) apparent dielectric loss vs frequency. PRC properties: c–g) current vs applied voltage, h) polarization density ($P$) vs temperature, i) coercive electric field ($E_c$) vs temperature. SHG properties: j) SH intensity vs temperature; k) $g$-values in two domains with left-/right-handed chirality for T-N$_F$ (blue bar) and T-SmA$_F$ (green bar) states. Note: to avoid thermal damage, no detail record was kept in the high-temperature range (> 200 °C). PRC and SHG studies were performed using an antiparallel rubbed IPS cell (PRC: 5 μm thickness, SHG: 10 μm thickness). h,i) Multiple triangular domains were observed within temperature region (gray-colored). In this region, $E_c$ could not be estimated due to multiple peaks.

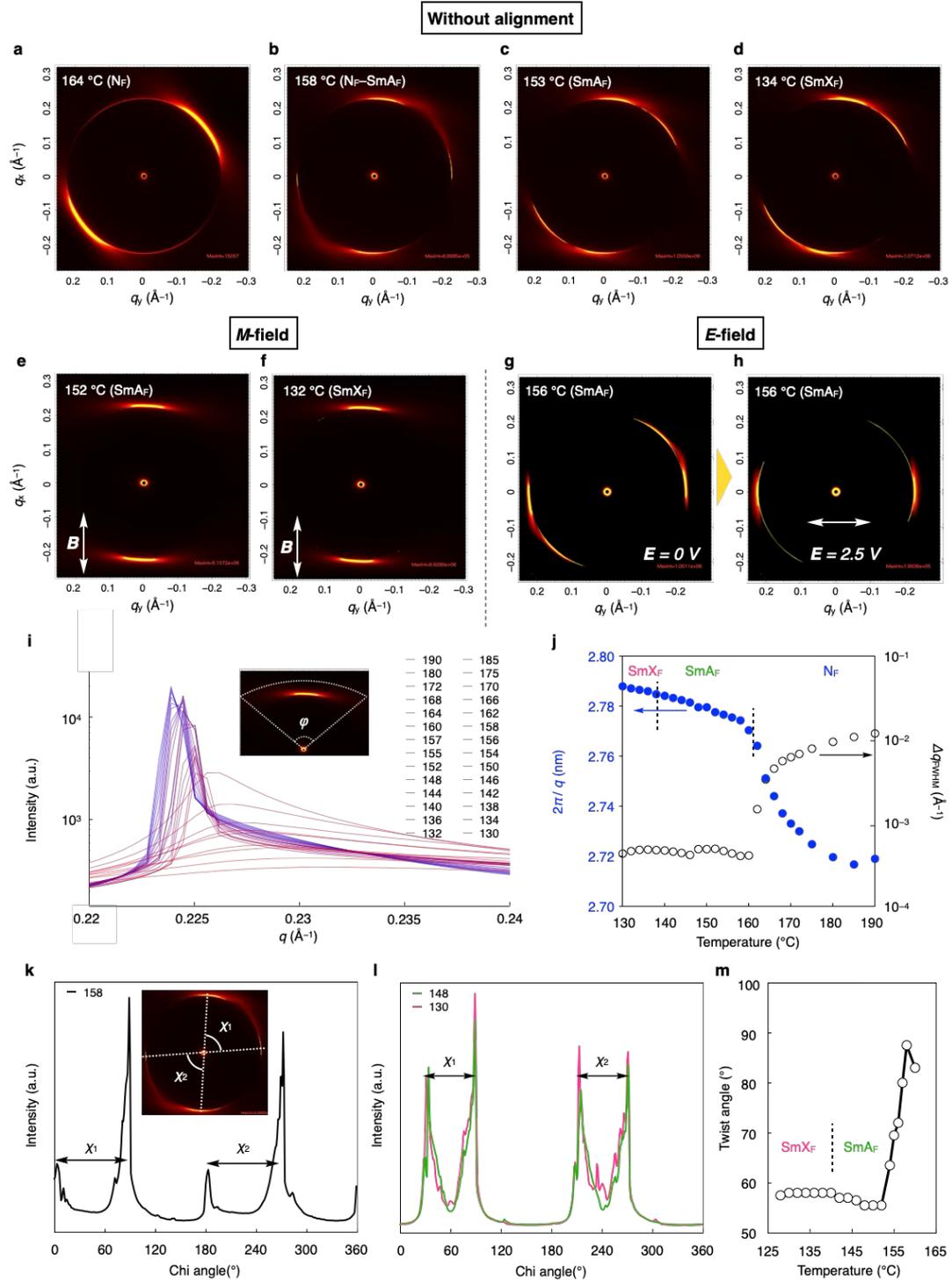

**Figure. 4** XRD profiles for **5BOE-NO$_2$**. 2D SAXS pattern of free standing sample (a–d); sample with *M*-field (≈1 T) (e,f); sample without/with *E*-field (0/2.5 V) (g,h). i) 1D X-ray diffractogram of *M*-field-aligned sample in various temperatures. The intensity profile was obtained by integration within range of $\varphi = 100°$. j) $2\pi/q$ and $\Delta q_{FWHM}$ as a function of temperature. k,l) Intensity vs chi angle. m) Twist angle between two pair of diffraction peaks as a function of temperature. k,l) A chi scan at the primary diffraction peak integrated over a range of $\delta q = \pm 0.003$ Å$^{-1}$.

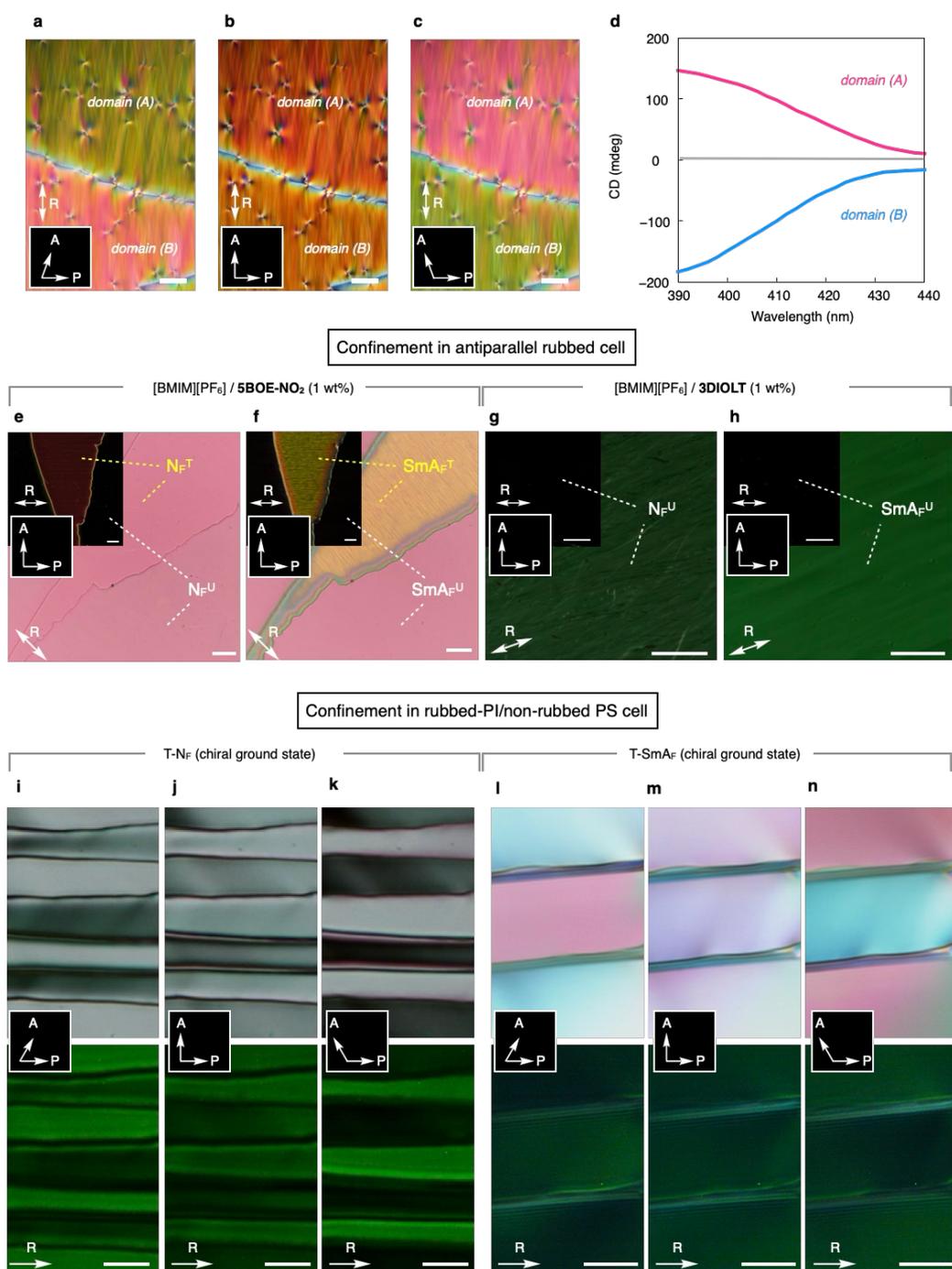

**Figure. 5** Chiroptical profiles for **5BOE-NO₂**: POM images under crossed (b) and slight de-crossed (a,c) polarizers in the antiparallel rubbed cell (thickness: 10 μm). Scale bar: 100 μm. d) Micro-CD spectra in the antiparallel rubbed cell (thickness: 5 μm). POM images in the antiparallel rubbed cell (thickness: 5 μm) for [BMIM][PF₆]/**5BOE-NO₂** (1 wt%) (e: $N_F$ phase, f: $SmA_F$ phase) and [BMIM][PF₆]/**3DIOLT** (1 wt%) (g: $N_F$ phase, h: $SmA_F$ phase). Scale bar: (e,f) 200 μm; (g,h) 100 μm. i–n) POM images in the rubbed-PI/non-rubbed-PS cell (thickness: 5 μm) for **5BOE-NO₂** in the $N_F$ (i–k) and $SmA_F$ (l–n) phases under de-crossed (i,k,l,n) and (j,m) crossed polarizers. Scale bar: 100 μm.

Supporting Information

## Spontaneous Twist of Ferroelectric Smectic Blocks in Polar Fluids

*Hiroya Nishikawa*[*], *Yasushi Okumura, Dennis Kwaria, Atsuko Nihonyanagi and Fumito Araoka*[*]

[*]To whom correspondence should be addressed.

E-mail: hiroya.nishikawa@riken.jp (H.N.), fumito.araoka@riken.jp (F.A.)

## Methods

### 1. General and materials

**Nuclear magnetic resonance (NMR) spectroscopy**: $^1$H, $^{13}$C, and $^{19}$F NMR spectra were recorded on JNM-ECZ500 (JEOL) operating at 500 MHz, 126 MHz, and 471 MHz for $^1$H [$^1$H{$^{19}$F}], $^{13}$C{$^1$H} [$^{13}$C{$^1$H,$^{19}$F}] and $^{19}$F [$^{19}$F{$^1$H}] NMR, respectively, using the TMS (trimethylsilane) as an internal standard for $^1$H NMR and the deuterated solvent for $^{13}$C NMR. The absolute values of the coupling constants are given in Hz, regardless of their signs. Signal multiplicities were abbreviated by s (singlet), d (doublet), t (triplet), q (quartet), quint (quintet), sext (sextet), and dd (double–doublet), respectively.

**High-resolution mass (HRMS) spectroscopy**: The quadrupole time-of-flight high-resolution mass spectrometry (QTOF-HRMS) was performed on COMPACT (BRUKER). The calibration was carried out using LC/MS tuning mix, for APCI/APPI (Agilent Technologies).

**Density Functional Theory (DFT) Calculation**: Calculations were performed using the Chem3D (pro, 22.2.0.3300) and Gaussian 16 (G16, C.01) softwares (installed at the RIKEN Hokusai GreatWave Supercomputing facility) for MM2 and DFT calculations, respectively. GaussView 6 (6.0.16) software was used to visually analyze the calculation results. Positions of hydrogens of molecules were optimized using the B3LYP/6-31G++ level Gaussian 16 program package. [S1] Dipole moments of molecules were calculated using b3lyp/6-311+g(d,p) and B3LYP-gd3bj which were added for empirical dispersion corrections to the standard B3LYP. The calculation method is as follows: opt=tight freq b3lyp/6-311+g(d,p) geom=connectivity empiricaldispersion=gd3bj int=ultrafine.

**Polarized optical microscopy**: Polarized optical microscopy was performed on a polarizing microscope (Eclipse LV100 POL, Nikon) with a hot stage (HSC402, INSTEC) on the rotation stage. Unless otherwise noted, the sample temperature was controlled using the INSTEC temperature controller and a liquid nitrogen cooling system pump (mk2000 and LN2-P/LN2-D2, INSTEC).

**Differential scanning calorimetry (DSC)**: Differential scanning calorimetry was performed on a calorimeter (DSC3+, Mettler-Toledo). Rate: 10 K min$^{-1}$. Cooling/heating profiles were recorded and analyzed using the Mettler-Toledo STARe software system.

**Dielectric spectroscopy**: Dielectric relaxation spectroscopy was performed ranging between 1 Hz and 1 MHz using an impedance/gain-phase analyzer (SI 1260, Solartron

Metrology) and a dielectric interface (SI 1296, Solartron Metrology). Prior to starting the measurement of the LC sample, the capacitance of the empty cell was determined as a reference.

***P–E* hysteresis measurement**: *P–E* hysteresis measurements were performed in the temperature range of the $N_F$ phase under a triangular-wave electric field (10 kV cm$^{-1}$, 200 Hz) using a ferroelectricity evaluation system (FCE 10, TOYO Corporation), which is composed of an arbitrary waveform generator (2411B), an IV/QV amplifier (model 6252) and a simultaneous A/D USB device (DT9832).

**SHG measurement**: The SHG investigation was carried out using a Q-switched DPSS Nd: YAG laser (FQS-400-1-Y-1064, Elforlight) at $\lambda$ = 1064 nm with a 5 ns pulse width (pulse energy: 400 μJ). The primary beam was incident on the LC cell followed by the detection of the SHG signal. The electric field was applied normally to the LC cell. The optical setup is shown in **Figure S1**.

**Wide- and small-angle X-ray scattering (WAXS, SAXS) analysis**: Two-dimensional WAXS and SAXS measurements were carried out at BL05B1 in the SPring-8 synchrotron radiation facility (Hyogo, Japan). The samples held in a glass capillary (1.5 mm in diameter) were measured under a magnetic field at a constant temperature using a temperature controller and a hot stage (mk2000, INSTEC) with high temperature-resistance samarium cobalt magnets (≈1 T, Shimonishi Seisakusho Co., Ltd.). The scattering vector $q$ ($q = 4\pi\sin\theta\,\lambda^{-1}$; $2\theta$ and $\lambda$ = scattering angle and wavelength of an incident X-ray beam [1.0 Å (for WAXS) and 1.0 Å (for SAXS)], respectively) and position of an incident X-ray beam on the detector were calibrated using several orders of layer diffractions from silver behenate ($d$ = 58.380 Å). We employed three types of measurement system for samples without any alignment treatment, with *M*-field and *E*-field (**Figure S2**). The sample-to-detector distances were 255.2203 mm (for WAXS) and 1886.416 mm (for SAXS), where acquired scattering 2D images were integrated along the Debye–Scherrer ring by using software (Igor Pro with Nika-plugin), affording the corresponding one-dimensional profiles.

**UV-vis spectra measurement**: The UV-vis spectra were recorded using a UV-Vis-NIR spectrophotometer (V-670, JASCO). For solution and LC sampes, we employed a 1 cm-thick quartz cuvette and quartz sandwich cell, respectively.

**Micro CD spectra measurement**: The microscopic CD spectra were recorded using a CD spectrophotometer based on a homemade inverted microscope equipped with a camera.

The optical setup is shown in **Figure S3**. A non-polarized light from a broadband EUV-VIS light source (EQ99, Energetiq/Hamamatsu) was incident to the sample cell on a hot stage from the bottom window. Due to the circular and linear dichroism, the transmitted light is partially and slightly elliptically polarized. This polarized transmitted light was collected and collimated by a 100x UV achromatic objective lens and then led into a photoelastic modulator (PEM100, HINDS instruments) and a prism polarizer. The intensity modulated signals detected by a photomultiplier were analyzed with a lock-in amplifier (7265, Signal Recovery/Ametek) at the 1$f$ detection mode and a dc voltmeter. The typical spatial resolution of the system was less than 5μm in diameter.

**Confocal laser scanning microscopy**: The confocal fluorescence microscopic image was taken using a confocal laser scanning microscope (A-1, Nikon) customized by Y.O. To observed images during heating, we used a homemade hot stage, and an objective lens (Plan Apo VC, Nikon, 100× magnify) equipped with a homemade jacket heater. Here, we set the upper-temperature limit of jacket heater (80 °C) to avoid fatal damage to the objective lens. We employed BBOT ($\lambda_{ex}$ ≈405 nm, $\lambda_{em}$ ≈450 nm) as a dichroic dye to observe the fluorescence images. The optical setup is shown in **Figure S4**.

**Information of used liquid crystalline (LC) Cellos**:

*Bare glas cell (homemade)*:
   - Sandwich-type film using two cover glasses
   - Experiments: Spectra studies (thickness: 10.0 μm)

*Antiparallel-rubbed cell (EHC)*:
   - PI-coated type
   - Alignment layer: LX-1400
   - Experiments: POM and micro-CD Studios (thickness: 5.0, 10.0 μm)

*Parallel-rubbed cell (EHC)*:
   - PI-Cate type
   - Alignment layer: LX-1400
   - Experiments: POM Stadies (thickness: 10.0 μm)

*ITO glass cell (EHC)*:
   - ITO-coated type, electrode area: 5 × 10 mm (D-type)
   - Experiments: POM (thickness: 10.0 μm) and DR (thickness: 9.0 μm) studies

*IPS cell (EHC)*:
   - PI-coated type, electrode distance: 500 μm, electrode length: 18 mm, thickness: 5.0 μm
   - Alignment layer: LX-1400
   - Rubbing condition: antiparallel
   - Experiments: *PE* hysteresis

*Antiparallel-rubbed cell (homemade)*:
   - PI-coated type (a cover glass + a slide glass)
   - Alignment layer: AL1254
   - Experiments: confocal laser scanning microscopy studies (thickness: 5.6, μm)

*Hybrid cell (homemade)*:
   - rubbed-PI + non-rubbed-PS
   - Alignment layer 1: AL1254
   - Alignment layer 2: PS (prepared according to Kumari's method [S2])
   - Experiments: confocal laser scanning microscopy studies (thickness: 5.6 μm)

## 2. Synthesis of nBOE-NO₂ and nDIOLT

### 2.1. Synthetic route

**nBOE-NO₂** (n = 3–5) and **nDIOLT** (n = 1–3) were synthesized by following pathway (**Scheme S1**). The blue and green colored pathway indicate solution chemical (SC) and mechanochemical (MC) synthesis, respectively. The detail of MC synthesis has been reported in our previous papers.[S3,S4]

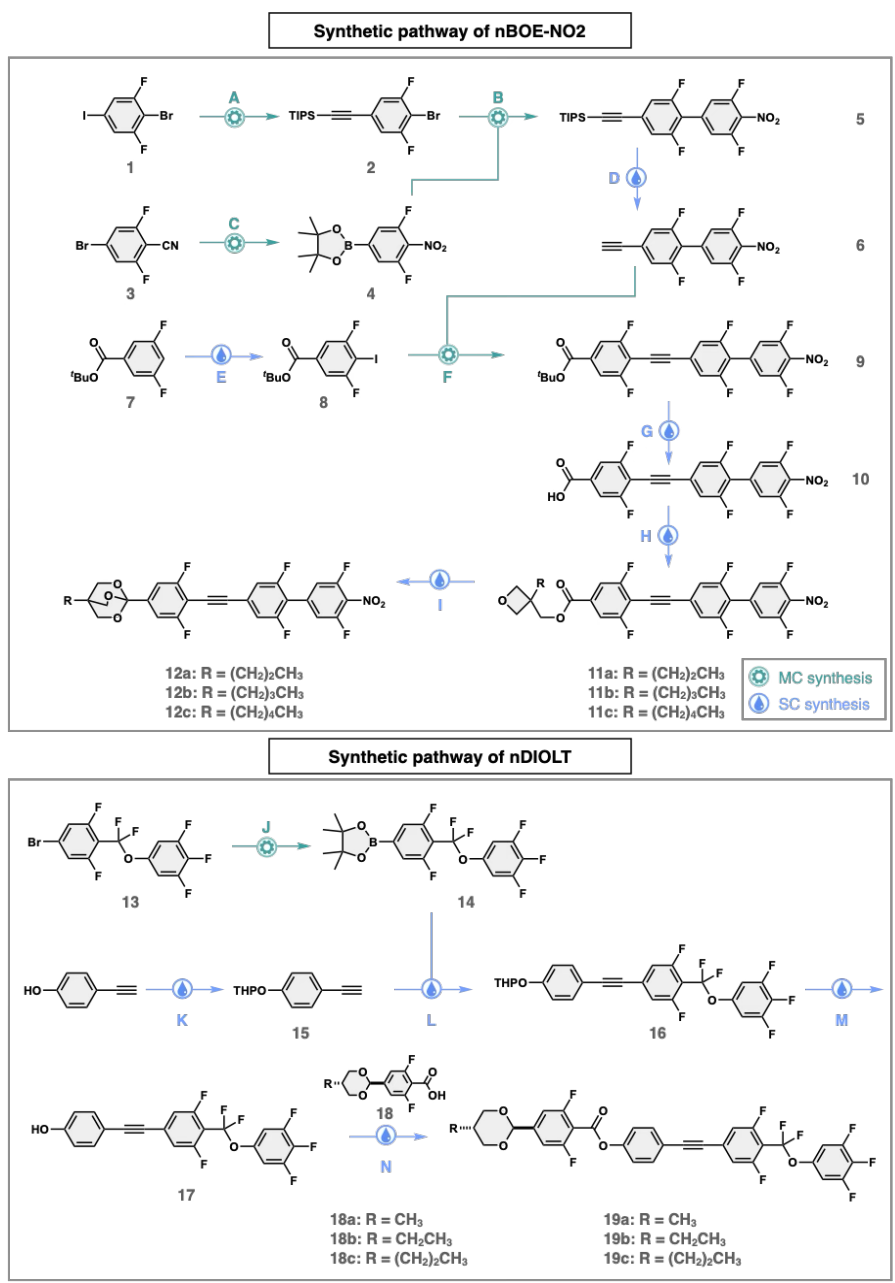

**Scheme S1** Synthetic pathway of **nBOE-NO₂** (n = 3–5) and **nDIOLT** (n = 1–3). Key reactions: Sonogashira-Hagihara coupling (A,L), Suzuki-Miyaura coupling (B,F), Miyaura-Ishiyama borylation (C,J), deprotection (D,G,M), Iodization (E), Esterification (H,N), orthoesterification (I), protection (J,K).

**2.2.1. *Synthesis of ((4-bromo-3,5-difluorophenyl)ethynyl)triisopropylsilane (2)***

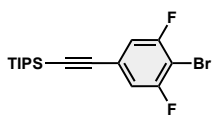

The synthetic procedure of **2** was reported in previous our works.[S3,4]

**2.2.2. *Synthesis of 2-(3,5-difluoro-4-nitrophenyl)-4,4,5,5-tetramethyl-1,3,2-dioxaborolane (4)***

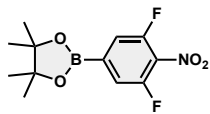

The synthetic procedure of **4** was reported in previous our works.[S4]

**2.2.3. *Synthesis of triisopropyl((2,3',5',6-tetrafluoro-4'-nitro-[1,1'-biphenyl]-4-yl)ethynyl)silane (5)***

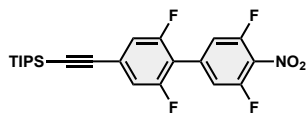

The synthetic procedure of **5** was reported in previous our works.[S4]

**2.2.4. *Synthesis of 4-ethynyl-2,3',5',6-tetrafluoro-4'-nitro-1,1'-biphenyl (6)***

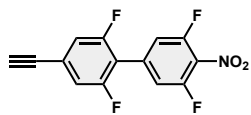

The synthetic procedure of **6** was reported in previous our works.[S4]

**2.2.5. *Synthesis of tert-butyl 3,5-difluoro-4-iodobenzoate (8)***

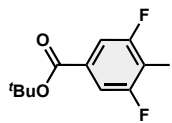

The synthetic procedure of **8** was reported in previous our works.[S4]

**2.2.6. *Synthesis of tert-butyl 3,5-difluoro-4-((2,3',5',6-tetrafluoro-4'-nitro-[1,1'-biphenyl]-4-yl)ethynyl)benzoate (9)***

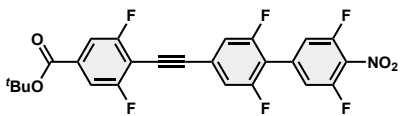

The synthetic procedure of **9** was reported in previous our works.[S4]

## 2.2.7. Synthesis of 3,5-difluoro-4-((2,3',5',6-tetrafluoro-4'-nitro-[1,1'-biphenyl]-4-yl)ethynyl)benzoic acid (10)

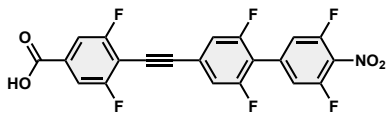

The synthetic procedure of **10** was reported in previous our works.[S4]

## 2.2.8. Synthesis of compounds (11a–c)

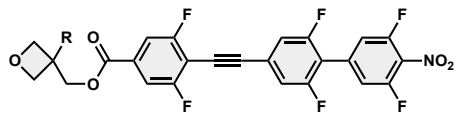

Compounds **11a** and **11c** were synthesized by following the procedure described in our previous paper.[S3] The synthetic procedure of **11b** was reported in previous our works.[S4]

### (3-propyloxetan-3-yl)methyl 3,5-difluoro-4-((2,3',5',6-tetrafluoro-4'-nitro-[1,1'-biphenyl]-4-yl)ethynyl)benzoate (11a)

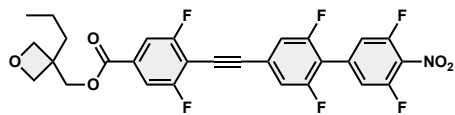

$^1$H{$^{19}$F}-NMR (500 MHz, CDCl$_3$): δ 7.64 (s, 2H), 7.30 (s, 2H), 7.28 (s, 2H), 4.57 (d, $J$ = 6.2 Hz, 2H), 4.52 (d, $J$ = 6.2 Hz, 2H), 4.50 (s, 2H), 1.80–1.77 (m, 2H), 1.41–1.34 (m, 2H), 0.99 (t, $J$ = 7.3 Hz, 3H)

$^{19}$F{$^1$H}-NMR (471 MHz, CDCl$_3$): δ −104.9, −112.9, −118.2

### (3-pentyloxetan-3-yl)methyl 3,5-difluoro-4-((2,3',5',6-tetrafluoro-4'-nitro-[1,1'-biphenyl]-4-yl)ethynyl)benzoate (11c)

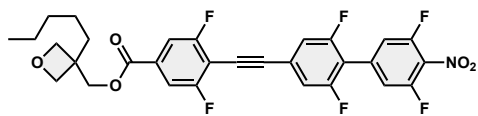

$^1$H{$^{19}$F}-NMR (500 MHz, CDCl$_3$): δ 7.65 (s, 2H), 7.29 (s, 2H), 7.28 (s, 2H), 4.57 (d, $J$ = 6.2 Hz, 2H), 4.51 (d, $J$ = 6.0 Hz, 2H), 4.49 (s, 2H), 1.81–1.78 (m, 2H), 1.37–1.28 (m, 6H), 0.92–0.89 (m, 3H)

$^{19}$F{$^1$H}-NMR (471 MHz, CDCl$_3$): δ −104.9, −112.9, −118.2

## 2.2.9. Synthesis of compounds (12a–c)

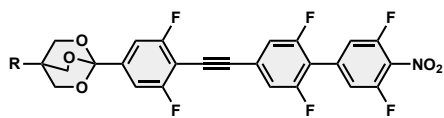

Compounds **12a** and **12c** were synthesized by following the procedure described in our previous paper.[S4] The synthetic procedure of **12b** was reported in previous our works.[S4]

*1-(3,5-difluoro-4-((2,3',5',6-tetrafluoro-4'-nitro-[1,1'-biphenyl]-4-yl)ethynyl)phenyl)-4-propyl-2,6,7-trioxabicyclo[2.2.2]octane (12a, 3BOE-NO2)*

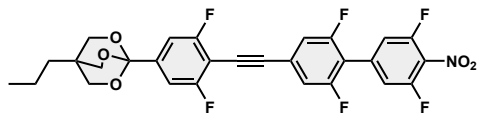

$^1H\{^{19}F\}$-NMR (500 MHz, CDCl$_3$): δ 7.27 (s, 2H), 7.25 (s, 2H), 7.23 (s, 2H), 4.11 (s, 6H), 1.31–1.22 (m, 4H), 0.94 (t, J = 6.5 Hz, 3H)

$^{19}F\{^1H\}$-NMR (471 MHz, CDCl$_3$): δ −106.3, −113.3, −118.3

$^{13}C\{^1H,^{19}F\}$-NMR (126 MHz, CDCl$_3$): δ 162.5, 159.2, 154.3, 141.2, 134.2, 129.1, 125.9, 115.5, 115.3, 115.0, 109.4, 106.3, 101.6, 96.0, 80.0, 72.1, 33.6, 31.9, 16.6, 14.7

APCI-HRMS (*m/z*, M⁻) calc for 563.1173; found, 563.1185.

*1-(3,5-difluoro-4-((2,3',5',6-tetrafluoro-4'-nitro-[1,1'-biphenyl]-4-yl)ethynyl)phenyl)-4-pentyl-2,6,7-trioxabicyclo[2.2.2]octane (12c, 5BOE-NO2)*

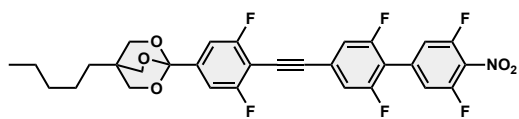

$^1H\{^{19}F\}$-NMR (500 MHz, CDCl$_3$): δ 7.27 (s, 2H), 7.26 (s, 2H), 7.24 (s, 2H), 4.11 (s, 6H), 1.35–1.21 (m, 8H), 0.90 (t, J = 7.1 Hz, 3H)

$^{19}F\{^1H\}$-NMR (471 MHz, CDCl$_3$): δ −106.3, −113.3, −118.3

$^{13}C\{^1H,^{19}F\}$-NMR (126 MHz, CDCl$_3$): δ 162.5, 159.2, 154.3, 141.2, 134.2, 129.1, 125.9, 115.5, 115.3, 115.0, 109.5, 106.3, 101.6, 96.0, 80.0, 72.1, 33.5, 32.3, 29.7, 22.8, 22.3, 13.9

APCI-HRMS (*m/z*, M⁻) calc for 591.1480; found, 591.1489.

### 2.2.10. Synthesis of 2-(4-(difluoro(3,4,5-trifluorophenoxy)methyl)-3,5-difluorophenyl)-4,4,5,5-tetramethyl-1,3,2-dioxaborolane (14)

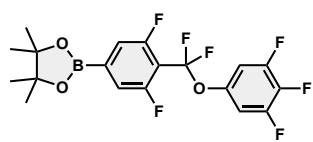

5-Bromo-2-(difluoro(3,4,5-trifluorophenoxy)methyl)-1,3-difluorobenzene (**13**) (3.89 g, 10.0 mmol), 4,4,4',4',5,5,5',5'-octamethyl-2,2'-bi(1,3,2-dioxaborolane) (2.79 g, 11.0 mmol), Pd(dppf)Cl$_2$-CH$_2$Cl$_2$ (245 mg, 0.3 mmol), KOAc (2.94 g, 30.0 mmol), and 1,4-dioxane (4.9 mL) were loaded in a grinding jar (30 mL) with one grinding ball. After closing the jar without purging with inert gas, be equipped with MM400 (10 min, *f* = 30 Hz) and with a heat gun ($T_{pre}$ = 150°C, $T_{int}$ = 110°C). After 10 min, heating was stopped, and the reaction jar was left to stand until it was cooled down to room temperature. The jar was opened, and the reaction mixture was transferred int a flask with DCM. The resulting solution was washed with H₂O, and the organic layer was dried over anhydrous Na₂SO₄ and evaporated. The residue was purified by flash column chromatography on silica gel (AcOEt/hexane, 5/95–80/20) to afford the target compound (**13**) in 98.5% yield (4.29 g) as an off-white solid.

$^1$H{$^{19}$F}-NMR (500 MHz, CDCl₃): δ 7.38 (s, 2H), 6.96 (d, $J$ = 5.7 Hz, 2H), 1.35 (s, 12H)

$^{19}$F{$^1$H}-NMR (471 MHz, CDCl₃): δ −61.7, −61.7, −61.8, −111.6, −111.6, −111.7, −132.4, −132.4, −163.0, −163.1, −163.1

### 2.2.11. *Synthesis of 2-(4-ethynylphenoxy)tetrahydro-2H-pyran* (15)

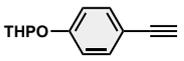

To a solution of 4-((tetrahydro-2H-pyran-2-yl)oxy)benzaldehyde (1.88 g, 9.13 mmol) and K₂CO₃ (2.52 g, 18.3 mmol) in MeOH (30 mL) was added dropwise dimethyl (1-diazo-2-oxopropyl)phosphonate (2.0 mL, 13.7 mmol) at 0 °C. The resulting mixture was stirred under Ar atmosphere for 0.5 hours at 0 °C and 3.5 hours at room temperature. The reaction mixture was quenched with H₂O and extracted with Et₂O. The combined organic layer was dried over anhydrous Na₂SO₄ and concentrated under reduced pressure. The residue was purified by flash column chromatography on silica gel (AcOEt/hexane, 0/100–10/90) to afford the target compound in 74.2 % yield (1.37 g, 6.77 mmol) as an oil.

$^1$H{$^{19}$F}-NMR (500 MHz, CDCl₃): δ 7.43–7.40 (m, 2H), 7.00–6.98 (m, 2H), 5.43 (t, $J$ = 3.2 Hz, 1H), 3.89−3.84 (m, 1H), 3.62−3.58 (m, 1H), 2.99 (s, 1H), 2.04−1.96 (m, 1H), 1.90−1.84 (m, 2H), 1.73−1.64 (m, 2H), 1.63−1.58 (m, 1H)

### 2.2.12. *Synthesis of 2-(4-((4-(difluoro(3,4,5-trifluorophenoxy)methyl)-3,5-difluorophenyl)ethynyl)phenoxy)tetrahydro-2H-pyran* (16)

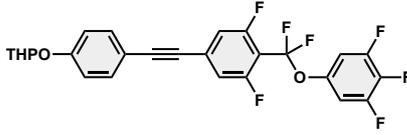

Compound (**15**) (1.37 g, 6.77 mmol), 5-bromo-2-(difluoro(3,4,5-trifluorophenoxy)methyl)-1,3-difluorobenzene (2.63 g, 6.77 mmol), Pd(PPh₃)₂Cl₂ (143 mg, 0.2 mmol), CuI (12.9 mg, 0.07 mmol), and triethylamine (1.89 mL, 13.5 mmol) were loaded in a grinding jar (30 mL) with one grinding ball. After closing the jar without purging with inert gas, be equipped with MM400 (30 min, $f$ = 30 Hz) and with a heat gun ($T_{pre}$ = 80°C, $T_{int}$ = 60°C). After 30 min, heating was stopped, and the reaction jar was left to stand until it was cooled down to room temperature. The jar was opened, and the reaction mixture was transferred int a flask with DCM. The resulting solution was washed with H₂O, and the organic layer was dried over anhydrous Na₂SO₄ and evaporated. The residue was purified by flash column chromatography on silica gel (DCM/hexane, 15/85–25/75) to afford the target compound in 53.5% yield (819 mg, 1.60 mmol) as a white solid.

$^1$H{$^{19}$F}-NMR (500 MHz, CDCl$_3$): δ 7.46 (dd, *J* = 11.3, 2.6 Hz, 2H), 7.10 (s, 2H), 7.05 (d, *J* = 8.7 Hz, 2H), 6.97 (d, *J* = 5.7 Hz, 2H), 5.47 (t, *J* = 3.2 Hz, 1H), 3.90–3.85 (m, 1H), 3.62 (dt, *J* = 11.4, 3.9 Hz, 1H), 2.04–1.97 (m, 1H), 1.89–1.86 (m, 2H), 1.74–1.60 (m, 3H)

$^{19}$F{$^1$H}-NMR (471 MHz, CDCl$_3$): δ −61.6 (t, *J* = 27.6 Hz), −110.9 (t, *J* = 27.6 Hz), −132.3 (d, *J* = 18.4 Hz), −162.9 (t, *J* = 20.3 Hz)

### 2.2.13. *Synthesis of 4-((4-(difluoro(3,4,5-trifluorophenoxy)methyl)-3,5-difluorophenyl)ethynyl)phenol* (17)

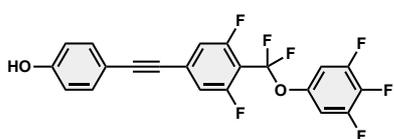

To a solution of (**16**) (819 mg, 1.60 mmol) in DCM/MeOH (5/5 mL) was added *p*-TsOH-H$_2$O (15.2 mg, 0.08 mmol) under Ar atmosphere, and the resulting mixture was stirred for 1 hour at 40 °C. The reaction mixture was diluted with H$_2$O and extracted with DCM. The organic layer was dried over anhydrous Na$_2$SO$_4$ and concentrated under reduced pressure. The residue was purified by flash column chromatography on silica gel (DCM/hexane, 65/30–100/0) to afford the target compound in 86.1 % yield (587 mg, 1.38 mmol) as a white solid.

$^1$H{$^{19}$F}-NMR (500 MHz, CDCl$_3$): 7.44 (dd, *J* = 11.2, 2.6 Hz, 2H), 7.10 (s, 2H), 6.97 (d, *J* = 5.7 Hz, 2H), 6.84 (dd, *J* = 11.3, 2.6 Hz, 2H), 5.12 (s, 1H)

$^{19}$F{$^1$H}-NMR (471 MHz, CDCl$_3$): δ −61.6 (t, *J* = 25.9 Hz), −110.8 (t, *J* = 27.6 Hz), −132.3 (d, *J* = 25.9 Hz), −162.9 (t, *J* = 20.0 Hz)

### 2.2.14. *Synthesis of compounds (18a–c)*

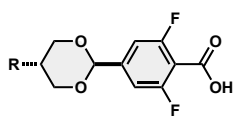

Compounds **18a** and **18b** were synthesized by following the procedure described in our previous paper.[S3] The synthetic procedure of **18c** was reported in previous our works.[S3]

*2,6-difluoro-4-((2r,5r)-5-methyl-1,3-dioxan-2-yl)benzoic acid (18a)*

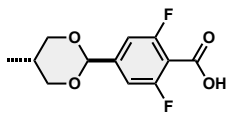

$^1$H{$^{19}$F}-NMR (500 MHz, acetone-d6): δ 7.17 (s, 2H), 5.51 (s, 1H), 4.16 (dd, *J* = 11.7, 4.7 Hz, 2H), 3.56 (t, *J* = 11.5 Hz, 2H), 2.18–2.08 (m, 1H), 0.77 (d, *J* = 6.7 Hz, 3H)

$^{19}$F{$^1$H}-NMR (471 MHz, CDCl$_3$): δ −112.3

*4-((2r,5r)-5-ethyl-1,3-dioxan-2-yl)-2,6-difluorobenzoic acid (18b)*

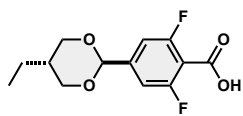

$^1$H{$^{19}$F}-NMR (500 MHz, CDCl$_3$): δ 7.14 (s, 2H), 5.37 (s, 1H), 4.26 (dd, *J* = 11.7, 4.6 Hz, 2H), 3.52 (t, *J* = 11.5 Hz, 2H), 2.08–1.99 (m, 1H), 1.20–1.14 (m, 2H), 0.94 (t, *J* = 7.5 Hz, 3H)

$^{19}$F{$^1$H}-NMR (471 MHz, CDCl$_3$): δ −107.8

**2.2.15. *Synthesis of compounds (19a–c)***

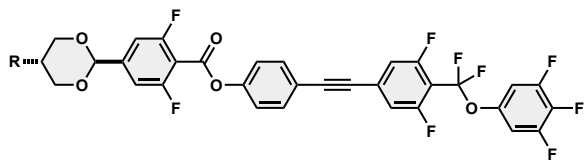

To a solution of carboxylic acid (**18**) (155 mg, 0.60 mmol) in DCM (5 mL) were added (**17**) (256 mg, 0.6 mmol), EDAC-HCl (138 mg, 0.72 mmol), and DMAP (7.3 mg, 0.06 mmol) at 0 °C under Ar. After 0.5 h stirring at the temperature, the solution was warmed up to r.t. and further stirred for 1.5 h. The reaction mixture was concentrated under reduced pressure, and the residue was purified by column chromatography on silica gel (CHCl$_3$/hexane, 30/70–80/20) to afford the target compound in 54.5 % yield (218 mg, 0.33 mmol) as a white solid. Another compound was also synthesized according to the above procedure.

*4-((4-(difluoro(3,4,5-trifluorophenoxy)methyl)-3,5-difluorophenyl)ethynyl)phenyl 2,6-difluoro-4-((2r,5r)-5-methyl-1,3-dioxan-2-yl)benzoate (19a, DIOLT1)*

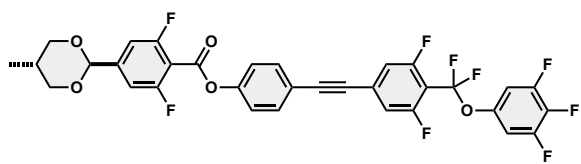

$^1$H{$^{19}$F}-NMR (500 MHz, CDCl$_3$): δ 7.60 (d, *J* = 8.2 Hz, 2H), 7.30 (d, *J* = 8.2 Hz, 2H), 7.19 (s, 2H), 7.15 (s, 2H), 6.98 (d, *J* = 5.0 Hz, 2H), 5.40 (s, 1H), 4.21 (dd, *J* = 11.6, 4.6 Hz, 2H), 3.52 (t, *J* = 11.2 Hz, 2H), 2.27–2.19 (m, 1H), 0.80 (d, *J* = 6.7 Hz, 3H)

$^{19}$F{$^1$H}-NMR (471 MHz, CDCl$_3$): δ −61.7 (t, *J* = 27.6 Hz), −112.1 (t, *J* = 25.7 Hz), −134.3 (d, *J* = 18.4 Hz), −165.0 (t, *J* = 20.3 Hz)

$^{13}$C{$^1$H,$^{19}$F}-NMR (126 MHz, CDCl$_3$): δ 160.9, 159.8 (Overlapped two carbons), 159.4, 151.0, 145.4, 144.5, 138.5, 133.2, 128.7, 122.0 (t, *J* = 261 Hz), 120.0, 119.8, 115.6, 110.3, 109.9 (t, *J* = 31.5 Hz), 109.7, 107.5, 98.6, 93.0, 86.2, 73.6, 29.2, 12.2

APCI-HRMS (*m/z*, [M+H]$^+$) calc for 667.1167; found, 667.1145.

*4-((4-(difluoro(3,4,5-trifluorophenoxy)methyl)-3,5-difluorophenyl)ethynyl)phenyl 4-((2r,5r)-5-ethyl-1,3-dioxan-2-yl)-2,6-difluorobenzoate (19b, DIOLT2)*

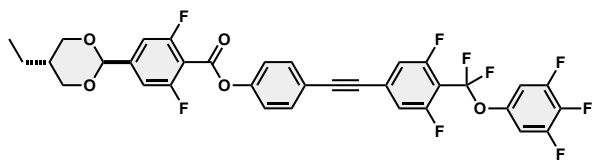

$^1$H{$^{19}$F}-NMR (500 MHz, CDCl$_3$): δ 7.62–7.59 (m, 2H), 7.31–7.29 (m, 2H), 7.19 (s, 2H), 7.15 (s, 2H), 6.98 (d, *J* = 5.6 Hz, 2H), 5.40 (s, 1H), 4.28 (dd, *J* = 11.7, 4.6 Hz, 2H), 3.54 (t, *J* = 11.5 Hz, 2H), 2.09–2.01 (m, 1H), 1.21–1.15 (m, 2H), 0.95 (t, *J* = 7.5 Hz, 3H)

$^{19}$F{$^1$H}-NMR (471 MHz, CDCl$_3$): δ −61.6 (t, J = 27.6 Hz), −108.6, −110.5 (t, *J* = 25.7 Hz), −132.3 (d, *J* = 18.4 Hz), −162.9 (t, *J* = 22.1 Hz)

$^{13}$C{$^1$H,$^{19}$F}-NMR (126 MHz, CDCl$_3$): δ 160.9, 159.8 (Overlapped two carbons), 159.4, 151.0, 145.4, 144.5, 138.5, 133.2, 128.7, 122.0, 120.0 (t, *J* = 262 Hz), 119.8, 115.6, 110.2, 109.9 (t, *J* = 32.0 Hz), 109.7, 107.5, 98.8, 93.0, 86.2, 72.4, 35.7, 21.1, 10.9

APCI-HRMS (*m/z*, [M+H]$^+$) calc for 680.1324; found, 681.1315.

*4-((4-(difluoro(3,4,5-trifluorophenoxy)methyl)-3,5-difluorophenyl)ethynyl)phenyl 2,6-difluoro-4-((2r,5r)-5-propyl-1,3-dioxan-2-yl)benzoate (19c, DIOLT3)*

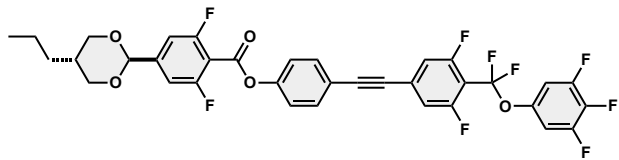

$^1$H{$^{19}$F}-NMR (500 MHz, CDCl$_3$): δ 7.60 (d, *J* = 8.7 Hz, 2H), 7.30 (d, *J* = 8.6 Hz, 2H), 7.19 (s, 2H), 7.15 (s, 2H), 6.98 (d, *J* = 5.6 Hz, 2H), 5.40 (s, 1H), 4.25 (dd, *J* = 11.7, 4.7 Hz, 2H), 3.54 (t, *J* = 11.5 Hz, 2H), 2.19–2.10 (m, 1H), 1.35 (dt, *J* = 22.8, 7.4 Hz, 2H), 1.10 (dd, *J* = 15.5, 7.3 Hz, 2H), 0.94 (t, *J* = 7.3 Hz, 3H)

$^{19}$F{$^1$H}-NMR (471 MHz, CDCl$_3$): δ −61.7 (t, *J* = 25.7 Hz), −108.5, −110.5 (t, *J* = 27.6 Hz), −132.3 (d, *J* = 22.1 Hz), −162.8 (t, *J* = 20.3 Hz)

$^{13}$C{$^1$H,$^{19}$F}-NMR (126 MHz, CDCl$_3$): δ 160.9, 159.8 (Overlapped two carbons), 159.4, 151.0, 145.4, 144.5, 138.5, 133.2, 128.7, 122.0, 120.0 (t, J = 263 Hz), 119.8, 115.6, 110.2, 109.9 (t, *J* = 32 Hz), 109.7, 107.5, 98.8, 93.0, 86.2, 72.6, 33.9, 30.2, 19.5, 14.2.

APCI-HRMS (*m/z*, [M+H]$^+$) calc for 695.1480; found, 695.1479.

**Supporting Notes (Notes S1, S2)**

**Note S1 | Characterization of intermediate phase between T-$N_F$ and T-$SmA_F$ states**

Let us examine the structure of the transient phase between the T-$N_F$ and T-$SmA_F$ states. In the T-$N_F$ structure, both **n** and **P** twist 180° from the top to the bottom of the substrates. As the temperature decreases, the correlation length within the plane of the T-$N_F$ structure increases, leading to the formation of a thin polar slab. These *pseudo*-smectic slabs are reflected in the T-$N_F$ structure, creating a twisted stacking arrangement (Figure S5a). Although twisting occurs at the center of rotation of the smectic slab, a defect arises within the coplanar due to the misalignment of the slabs (Figure S5a,b). Consequently, both twist and defect occur throughout the bulk, resulting in the appearance of a multiple triangular texture during the $N_F$–$SmA_F$ phase transition. This state is transient, and to minimize the energy cost of twist elasticity, each slab grows and fuses along the twist axis, ultimately forming smectic blocks. These blocks then undergo further twisting, leading to the establishment of the T-$SmA_F$ structure.

**Note S2 | Confocal laser scanning microscopy**

Figure S6a presents confocal fluorescence microscopy (CFM) images of the T-SmX$_F$ state (85 °C) for **2DIOLT** confined within an antiparallel rubbed cell (thickness: 5.6 μm). Due to the upper-temperature limit of the jacket heater, we could not observe the CFM images in the T-SmA$_F$ state. However, the structural characteristics of the T-SmX$_F$ state were clearly visible. Figure S6b shows changes in the CFM image captured at a specific confocal position from the cover glass to the slide glass side. Near the cover glass, a stripe texture aligned almost parallel to the rubbing direction was observed. The stripes disappeared at the mid-plane and reappeared near the slide glass, indicating that the stripe texture is likely a result of smectic layer undulation near the surface. The cross-section, indicated by the green line in Figure S6a, is displayed in Figure S6d–g. Figure S6d,e and S6f,g show detailed images of the regions (A) and (B) from Figure S6c. Notably, in region (A), areas with alternating fluorescence intensity were observed from the cover glass to the slide glass sides. Based on the data in Figure S6b, it is inferred that three smectic blocks existed in region (A). The intensity profile indicates that the two blocks (yellow and magenta) near the surface were aligned with the rubbing direction, while the middle block (blue) was oriented differently, suggesting that the smectic blocks were twisted (Figure S6d,e). Similarly, in region (B), two smectic blocks (with a thickness of ≈2.8 μm) were observed to be spontaneously twisted (Figure S6d,e). Observation of the T-SmA$_F$ state will be explored in future studies.

**Supporting Figures (Figures S1–S39)**

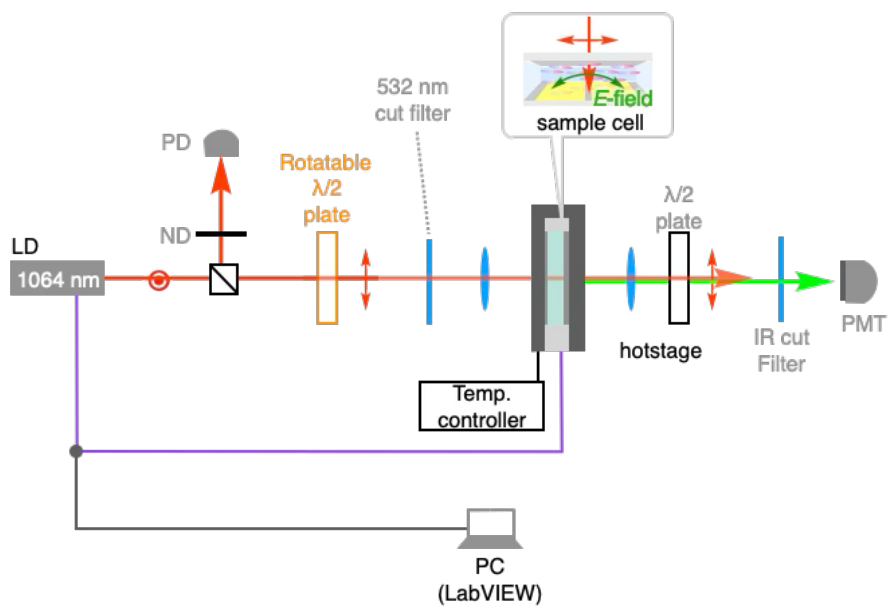

**Figure S1** Optical setup for SHG studies.

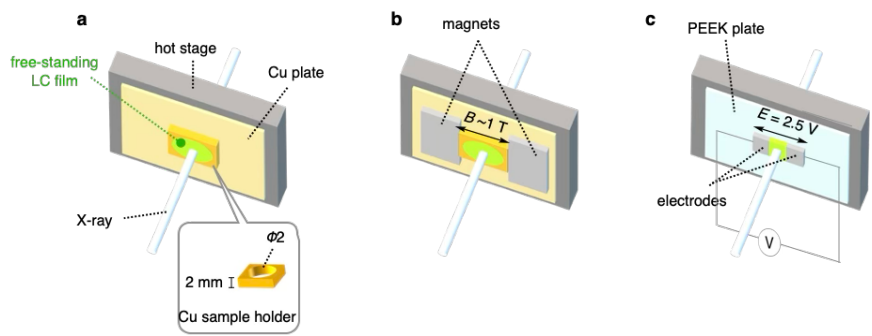

**Figure S2 Measurement system for XRD studies.** a) System without *M*/*E*-field, b) system with *M*-field and c) system with *E*-field.

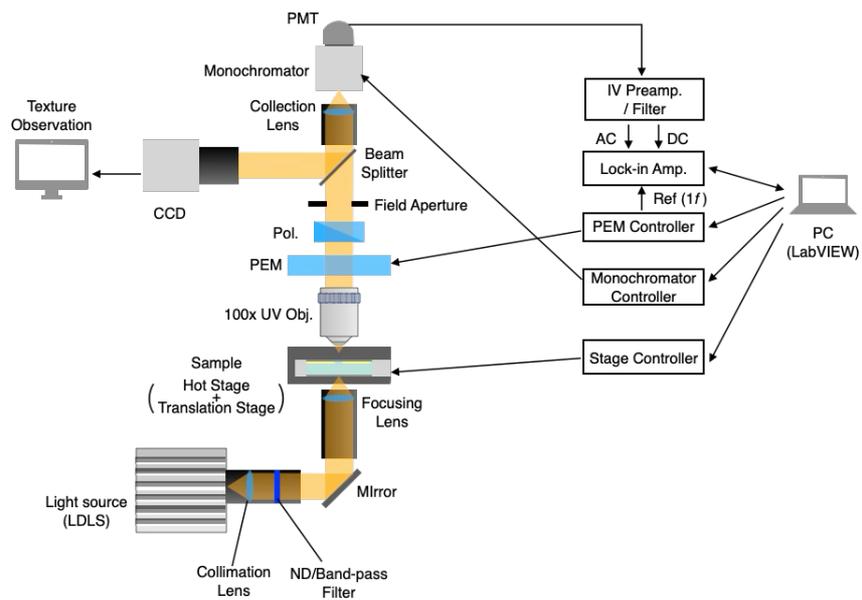

**Figure S3 Setup for micro-CD studies.**

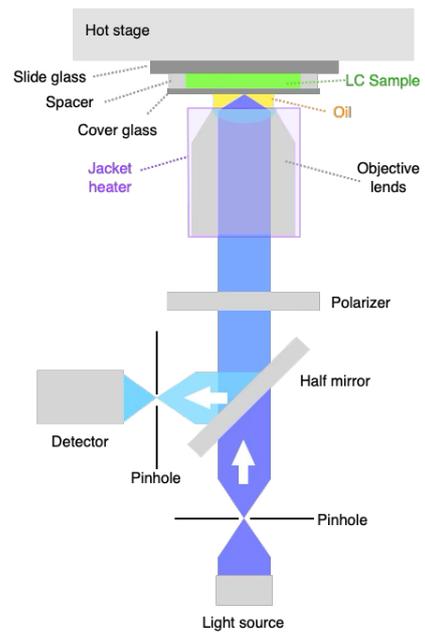

**Figure S4 Setup for confocal laser scanning microscopy.**

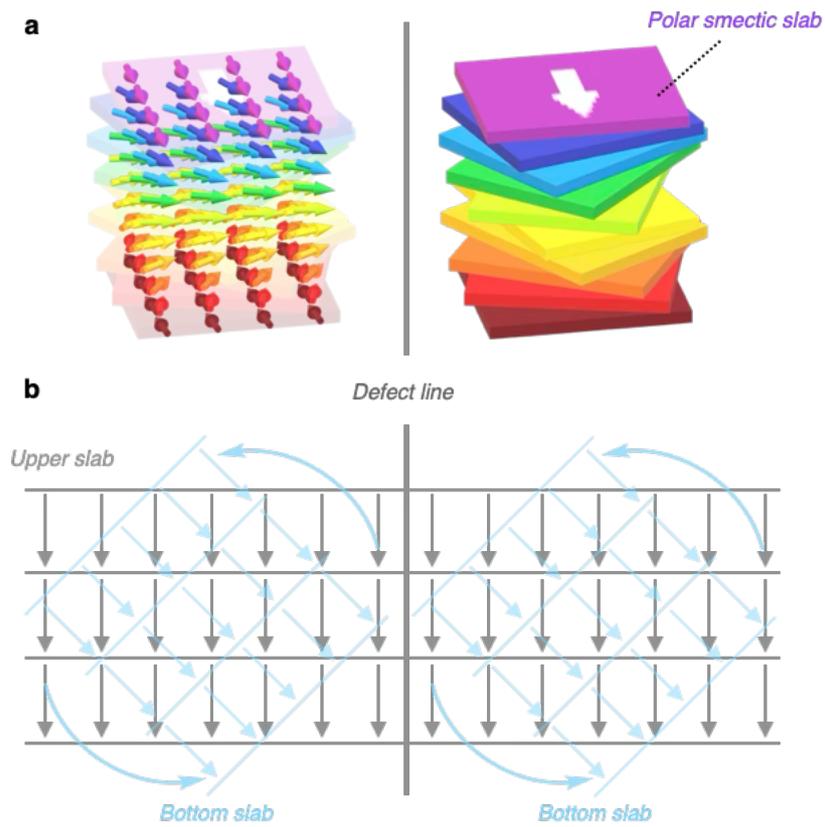

**Figure S5** Plausible model of the intermediate phase between T-N$_F$ and T-SmA$_F$ phases. a) a twisted structure piled up by *pseudo*-smectic slabs. b) Top view of the model indicated in the panel (a). Here, two types of slabs (gray and blue colored) are displayed to easily grab imagination.

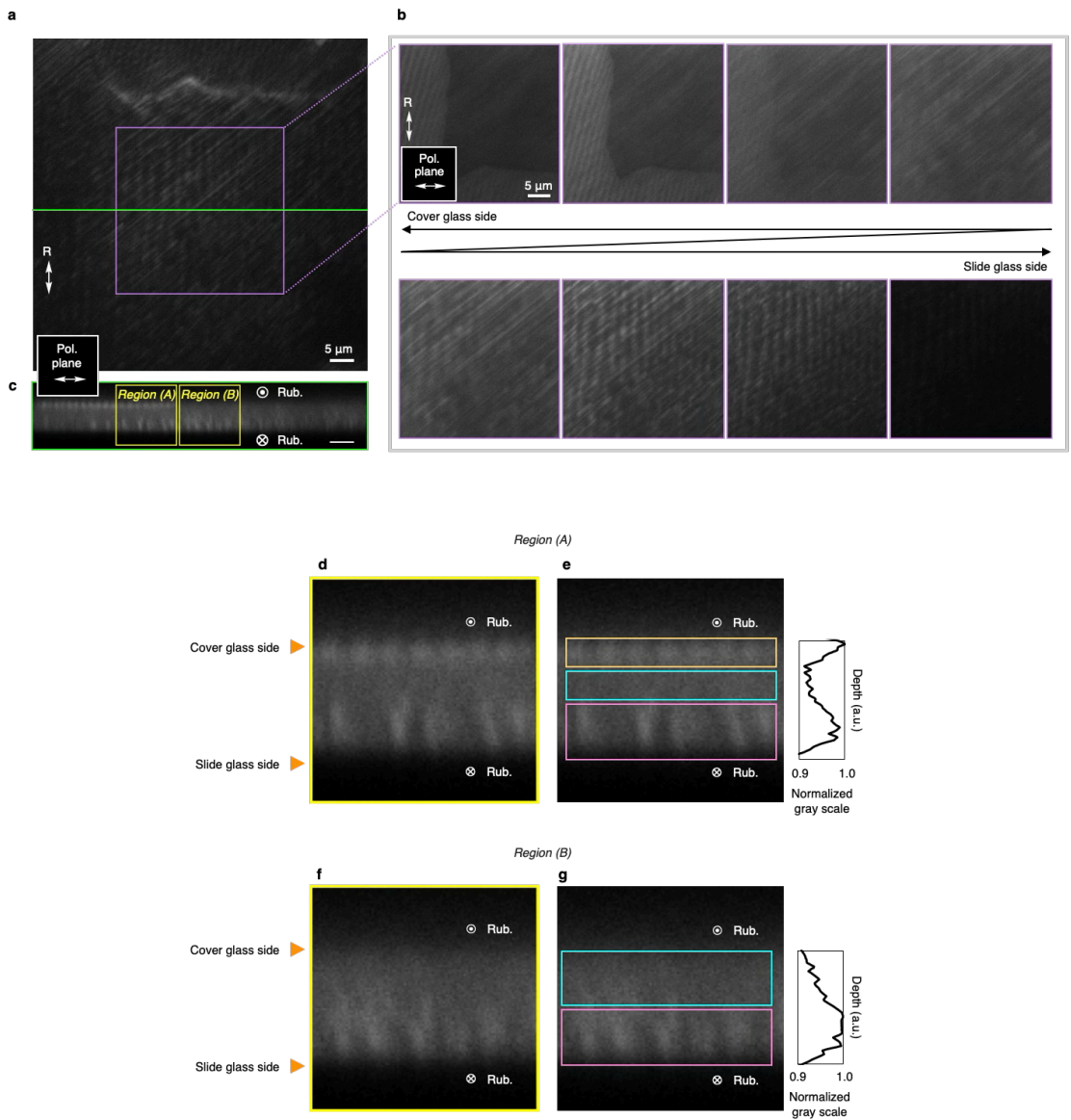

**Figure S6 CFM images of SmX$_F$ phase for 2DIOLT.** a) A snapshot in a certain confocal plane. b) Z-axis scan in the magenta-colored area in the panel (a). c) X-Z plane indicated by green line in the panel (a). d,e) X-Z plane in the region (A). f,g) X-Z plane in the region (B). Scale bar: 5 μm.

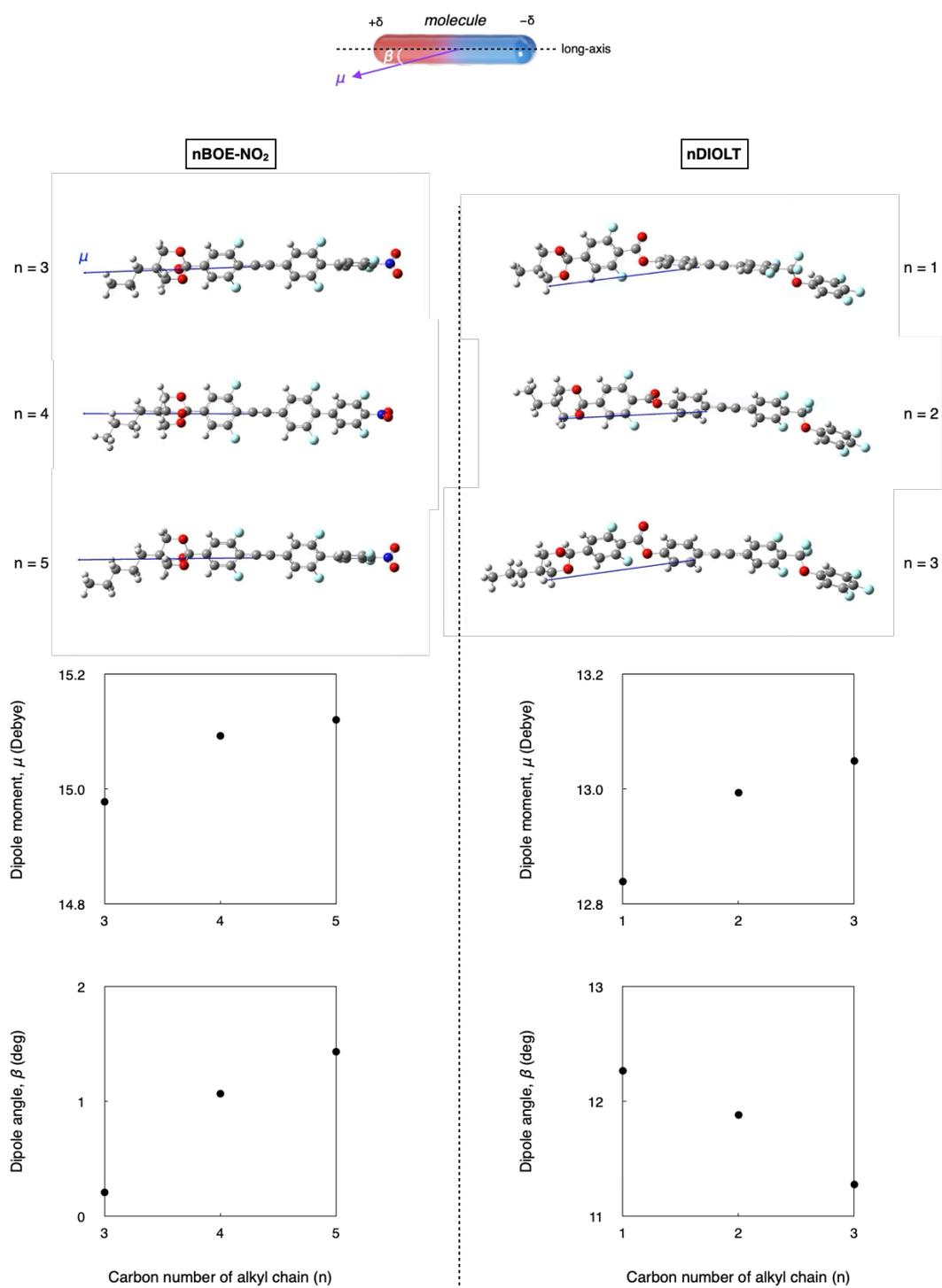

**Figure S7** Optimized structures, calculated dipole moment ($\mu$) and calculated dipole angle ($\beta$) for **nBOE-NO$_2$** (n = 3–6) and **nDIOLT** (n = 1–3).

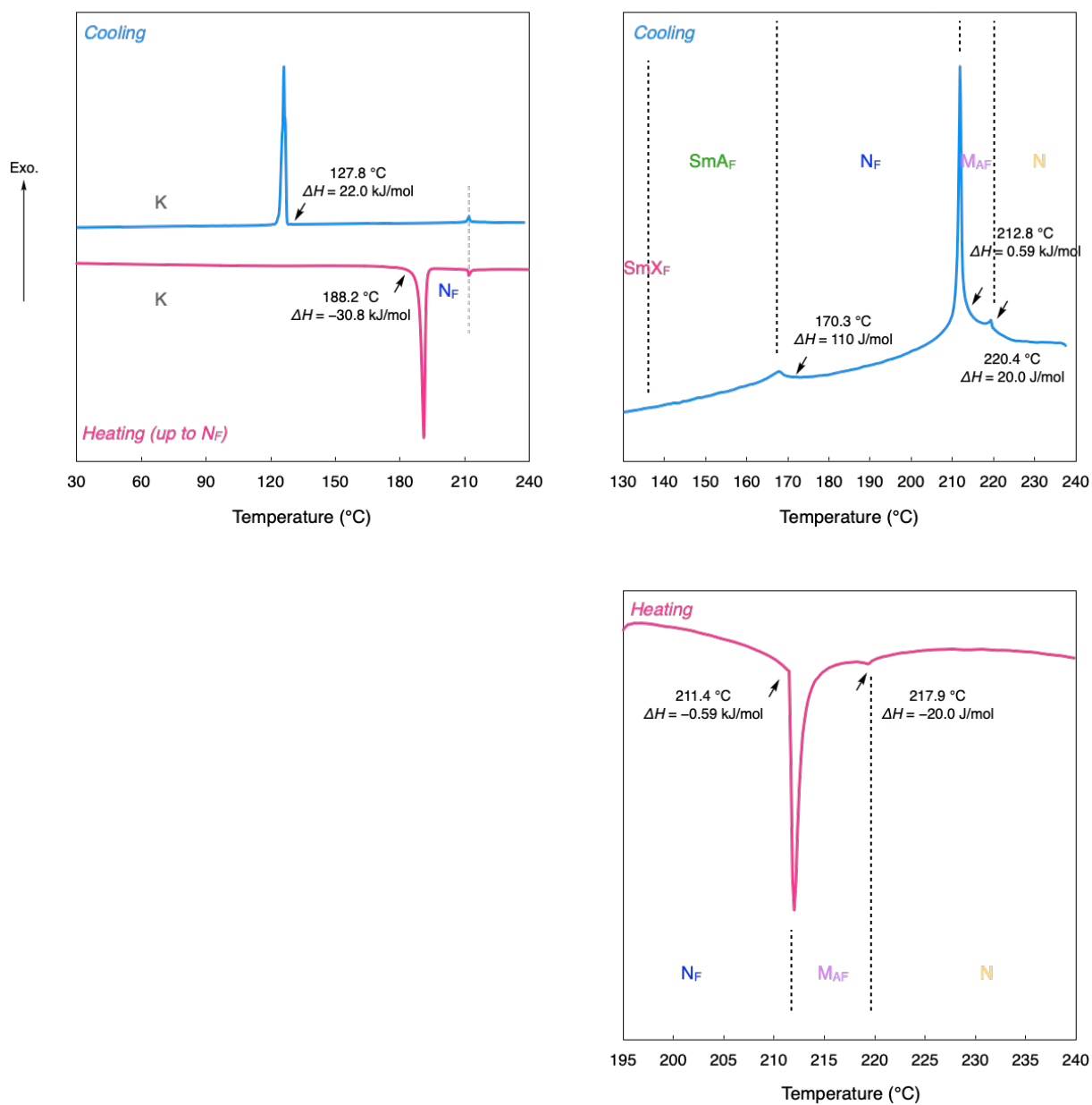

Figure S8 DSC curves for 5BOE-NO₂. Scan rate: 10 K min⁻¹.

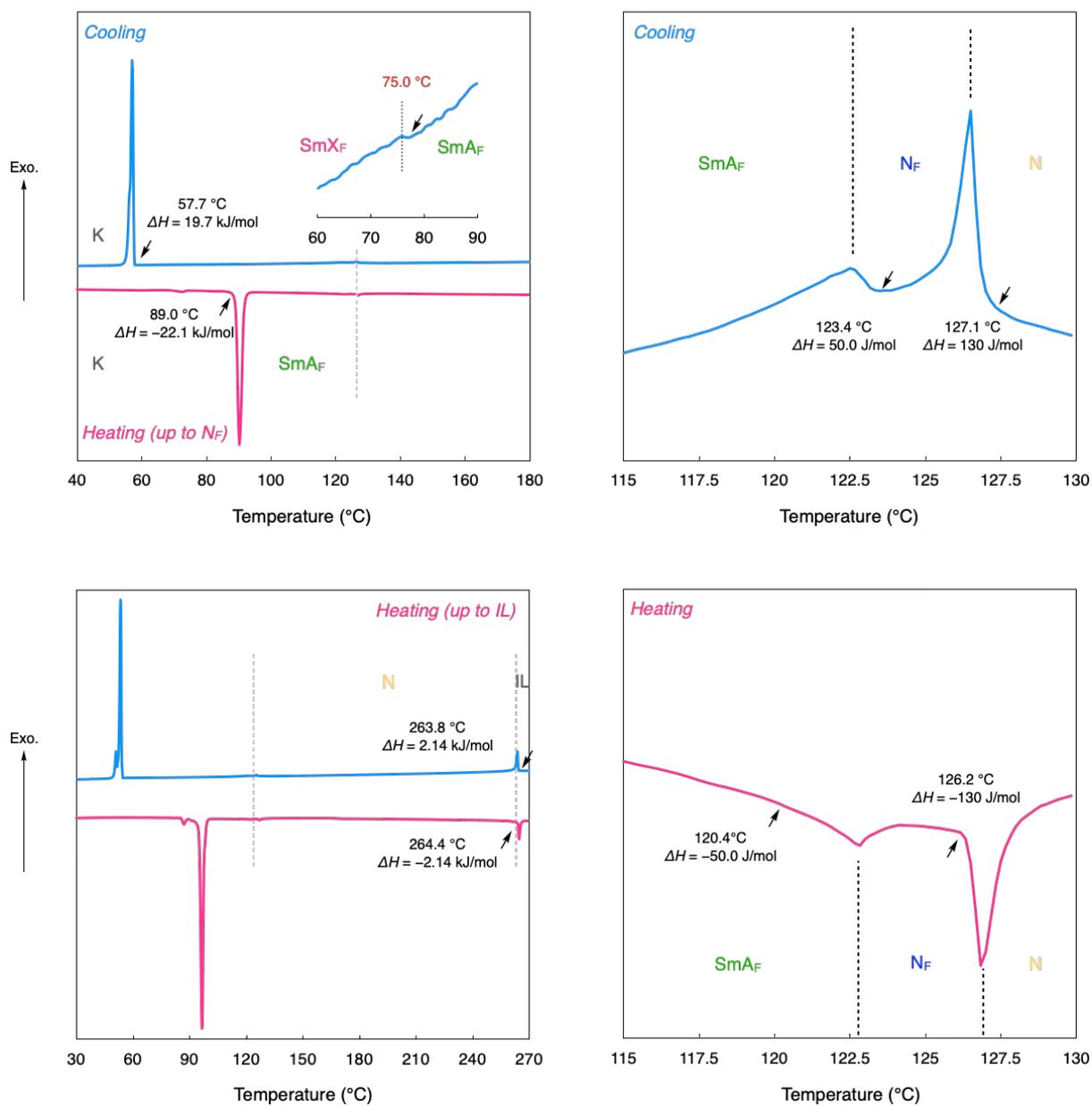

**Figure S9** DSC curves for **3DIOLT**. Scan rate: 10 K min$^{-1}$

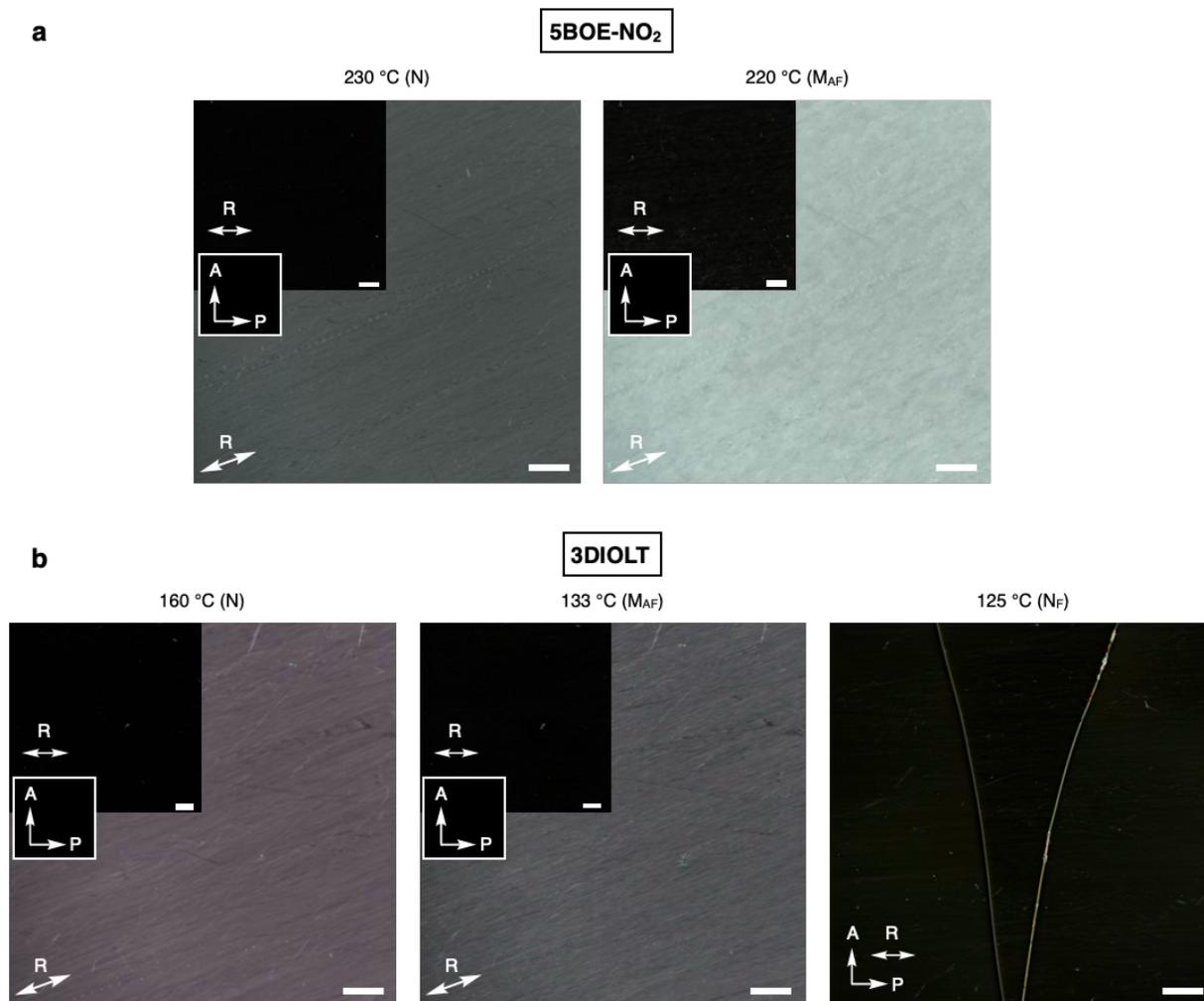

**Figure S10** Extra POM images for **5BOE-NO$_2$** (a) and **3DIOLT** (b) in the parallel-rubbed cell. Thickness: 10 μm. Scale bar: 100 μm.

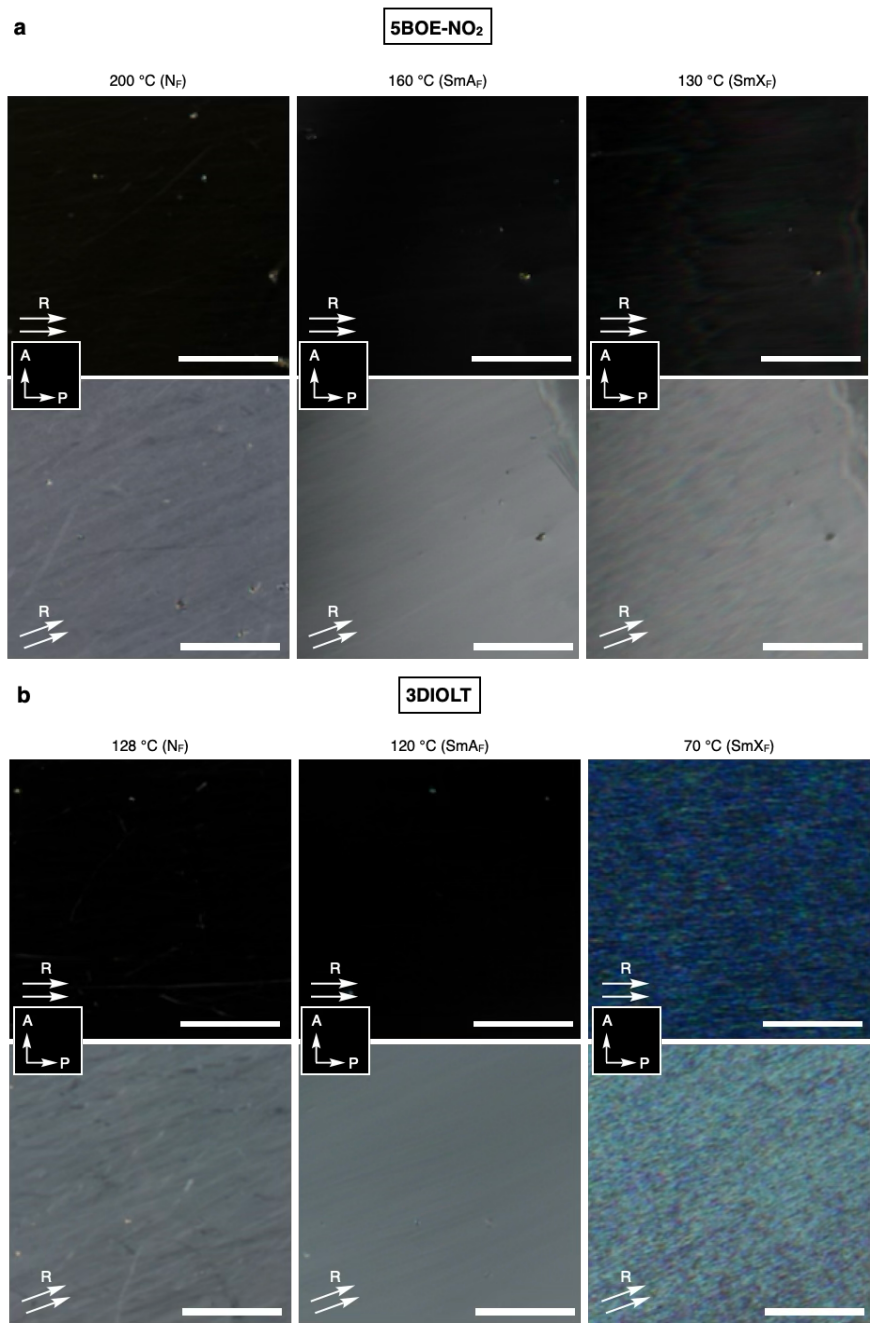

**Figure S11** POM images for **5BOE-NO₂** (a) and **3DIOLT** (b) in the parallel-rubbed cell. Thickness: 10 μm. Scale bar: 100 μm.

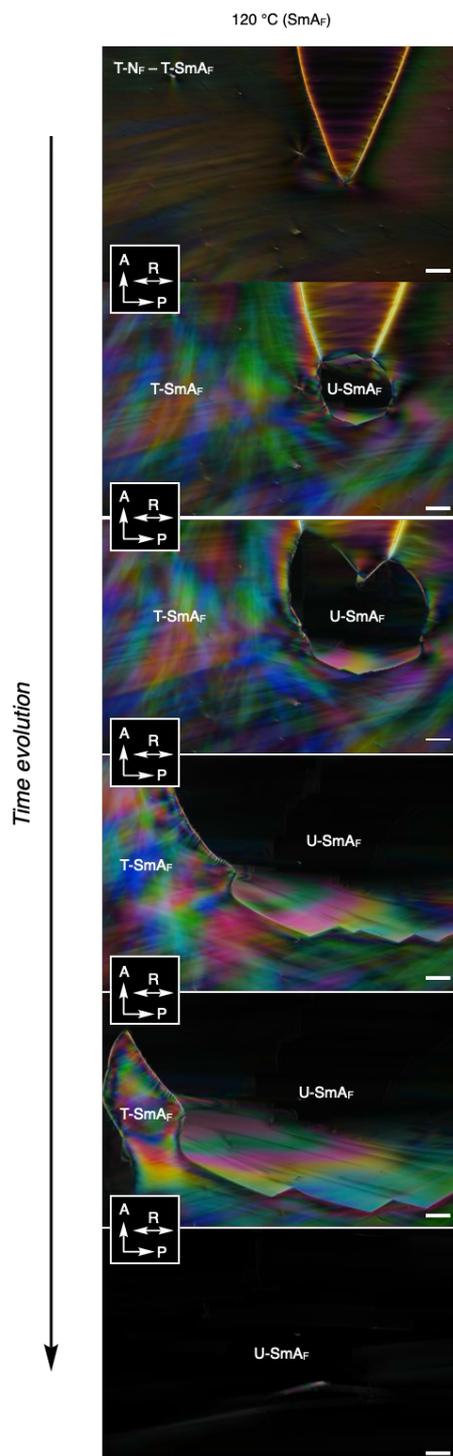

**Figure S12** POM texture changes in the antiparallel cell (thickness: 10 μm) during cooling for **3DIOLT**.

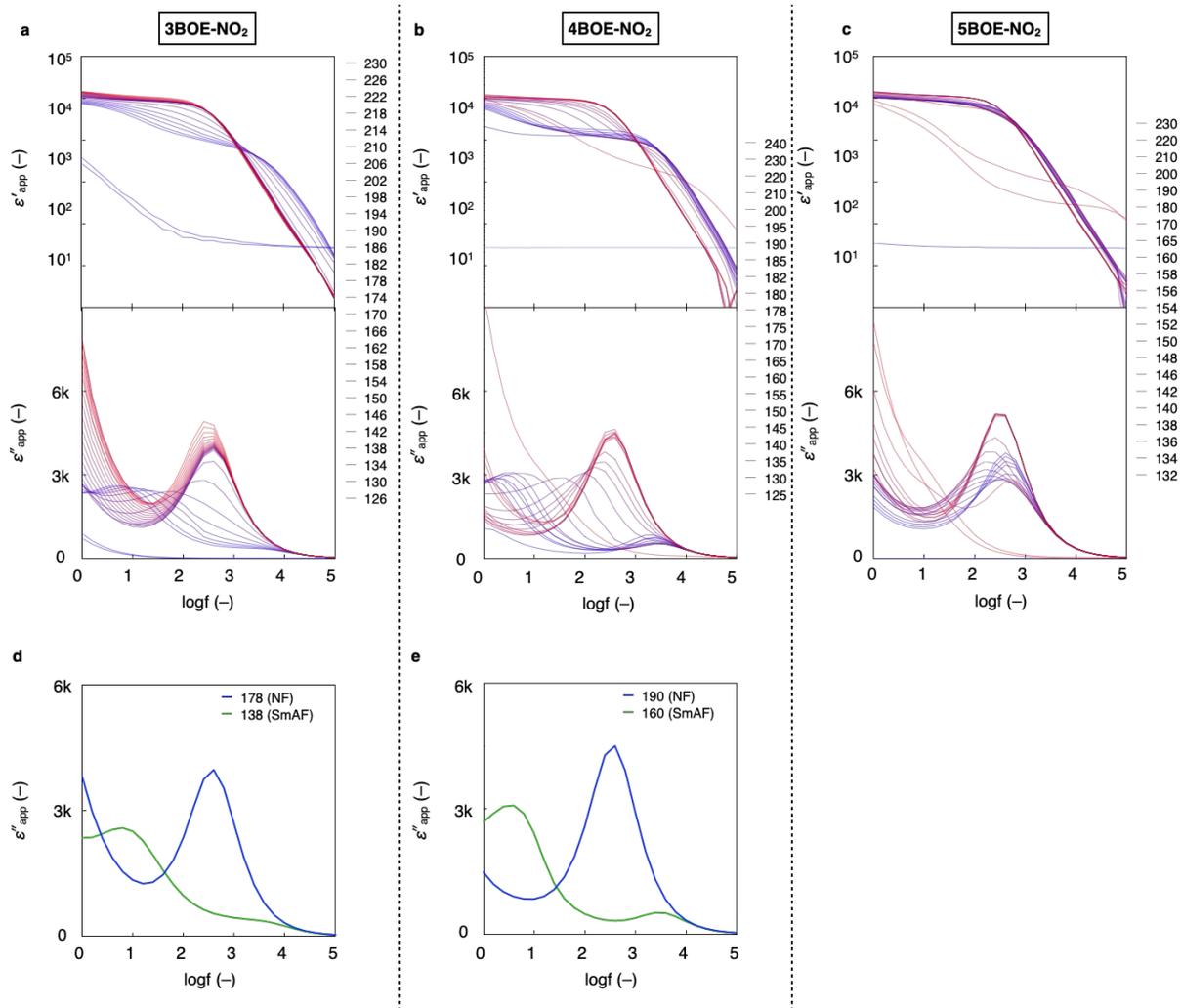

**Figure S13** Complete DR spectra for **nBOE-NO$_2$** (n = 3–5).

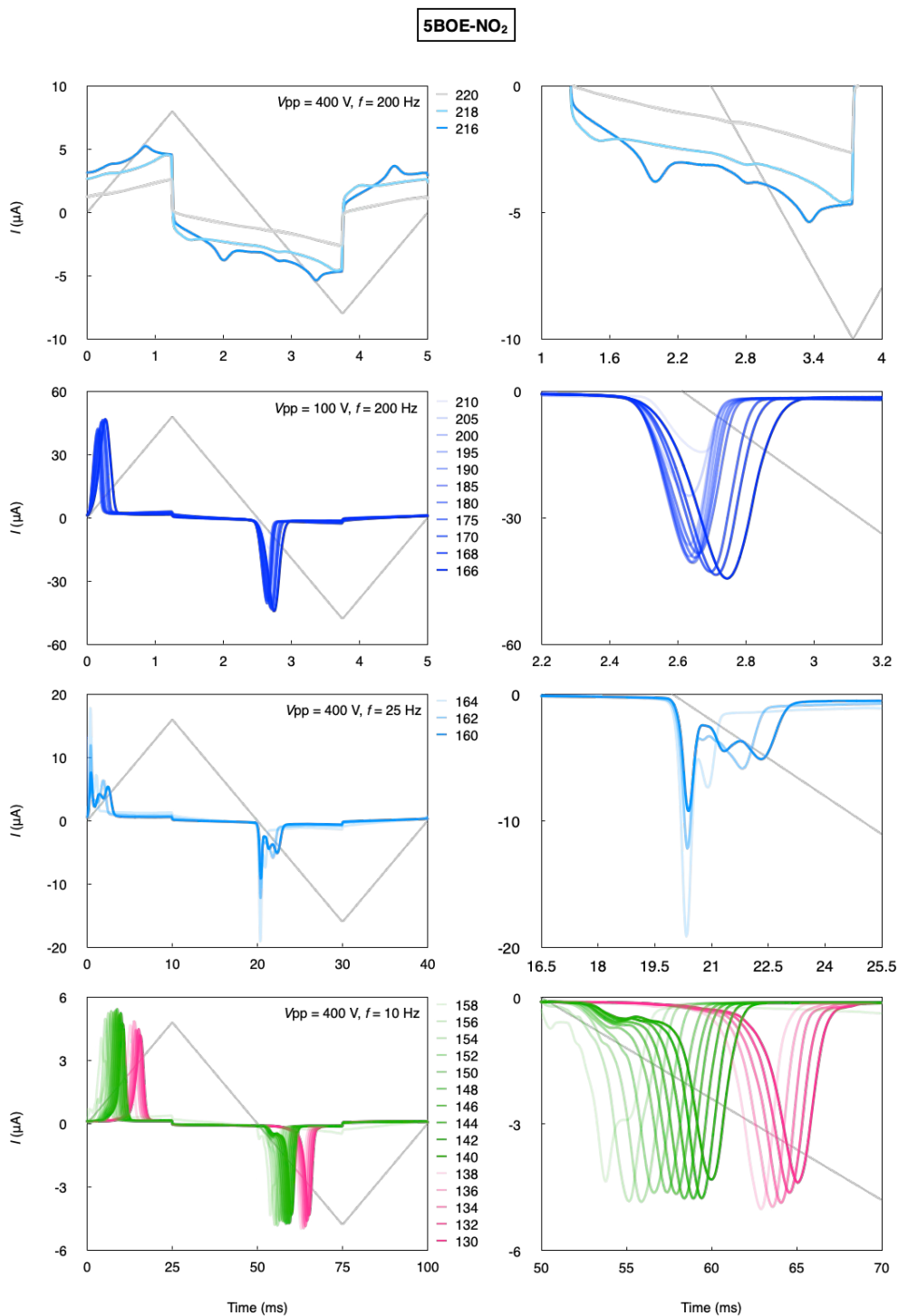

**Figure S14** Complete polarization reversal current data for **5BOE-NO₂** in various temperature.

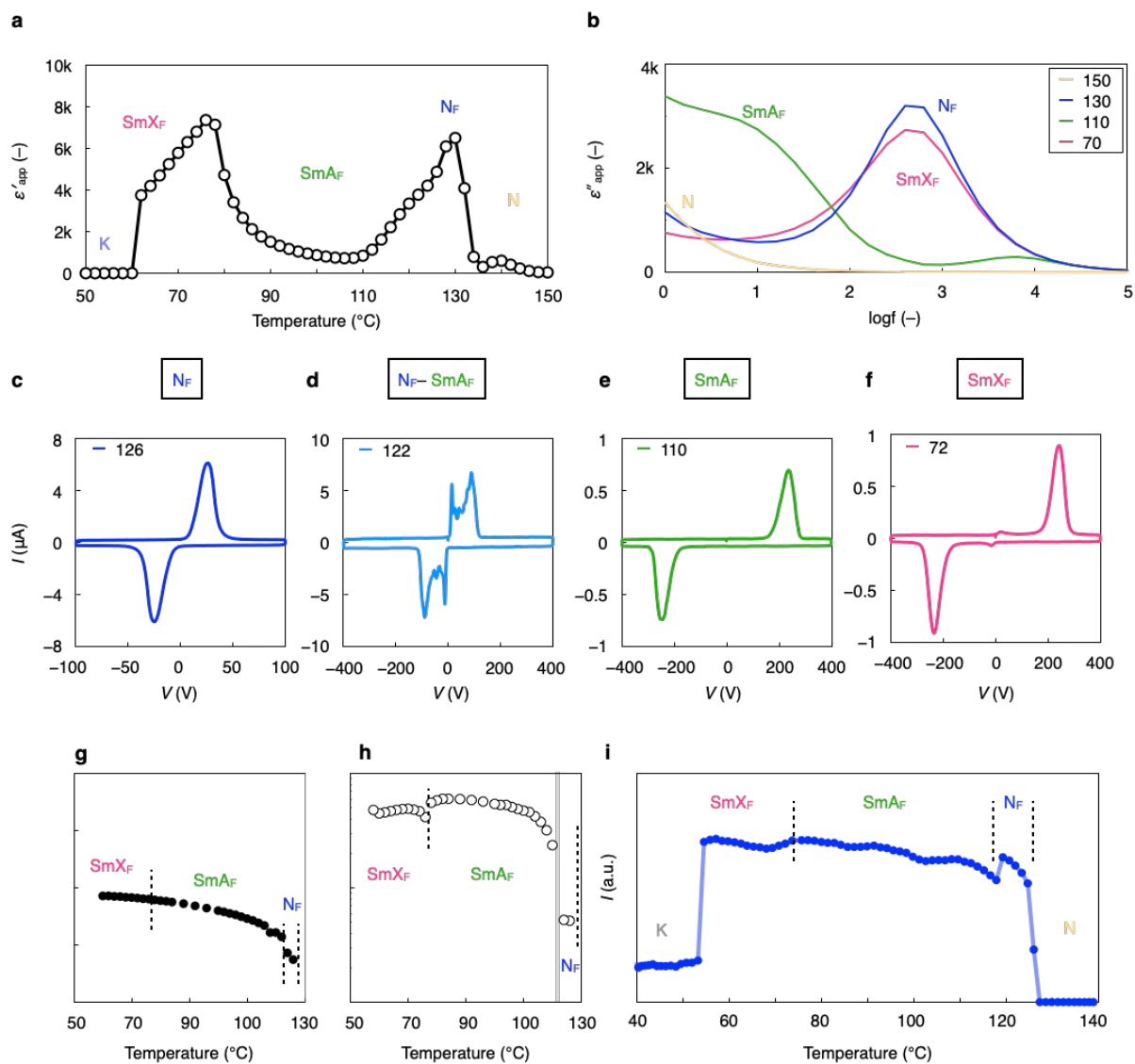

**Figure S15** Polarization behavior for **3DIOLT**. DR properties: a) apparent dielectric permittivity vs temperature, b) apparent dielectric loss vs frequency. c–f) current vs applied voltage, g) polarization density (*P*) vs temperature, h) coercive electric field ($E_c$) vs temperature, g) SHG properties: SH intensity vs Temperature. PRC and SHG studies were performed using an anti-parallel rubbed IPS cell (PRC: 5 μm thickness, SHG: 10 μm thickness).

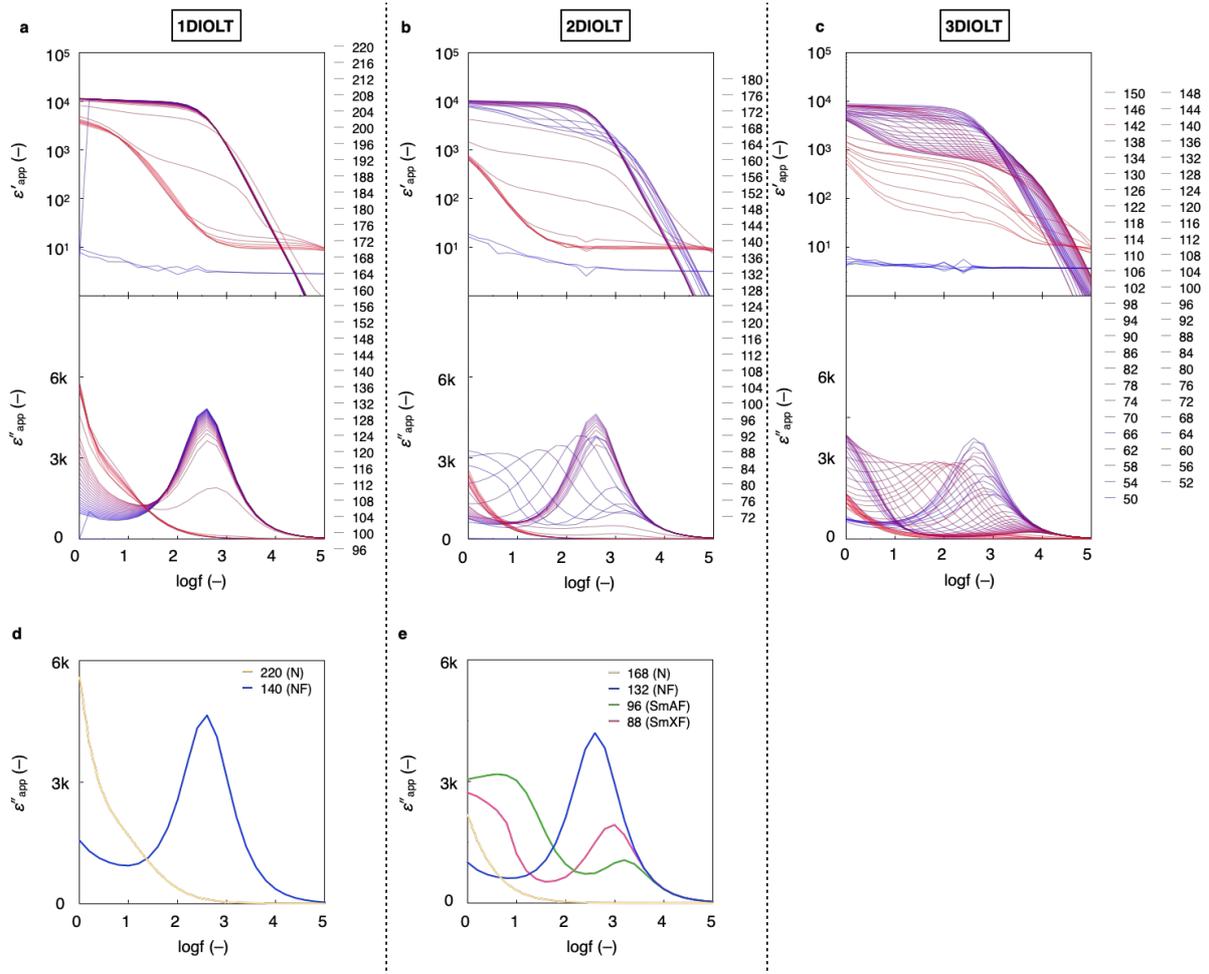

**Figure S16** Complete DR spectra for **nDIOLT** (n = 1–3).

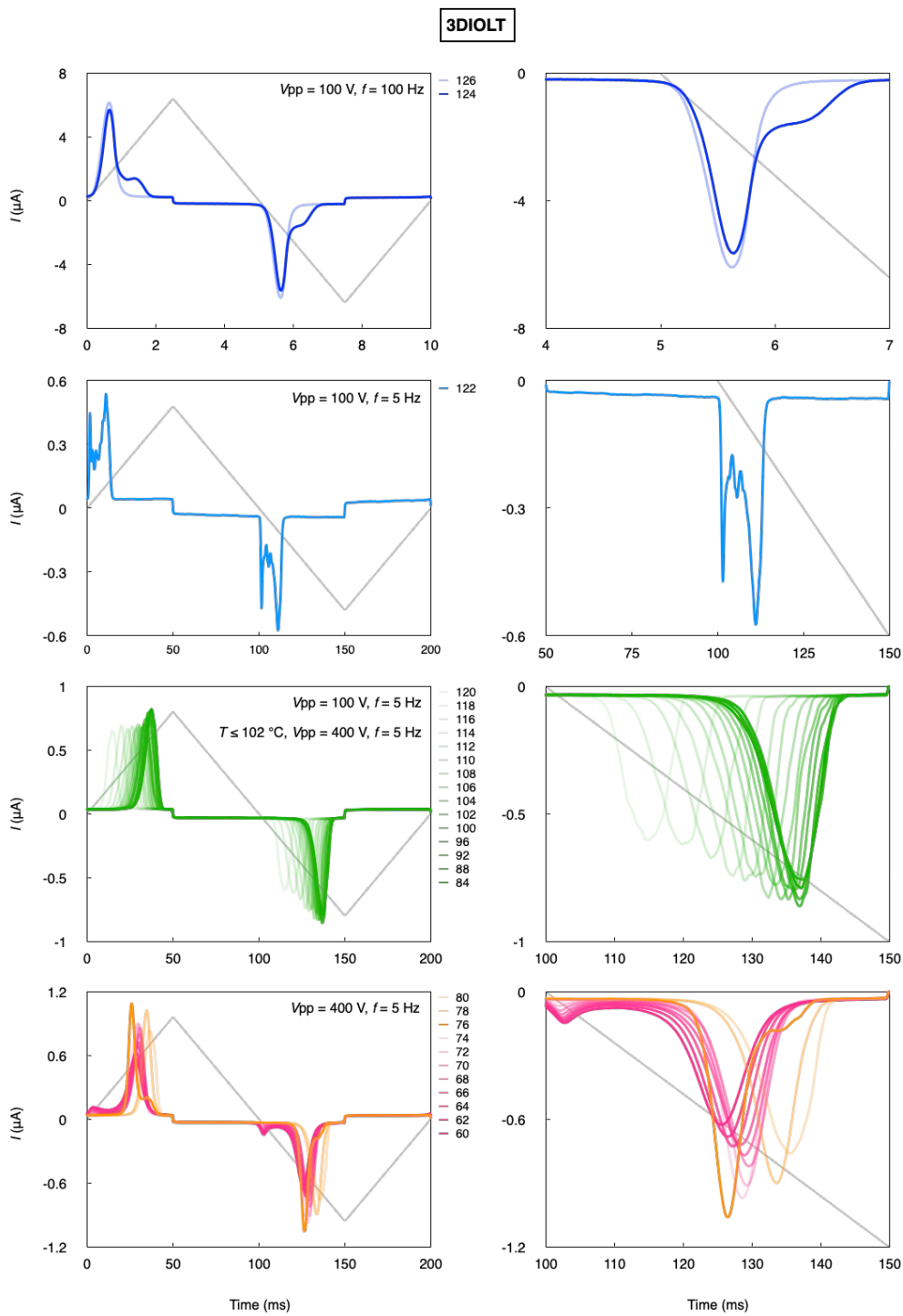

**Figure S17** Complete polarization reversal current data for **3DIOLT** in various temperature.

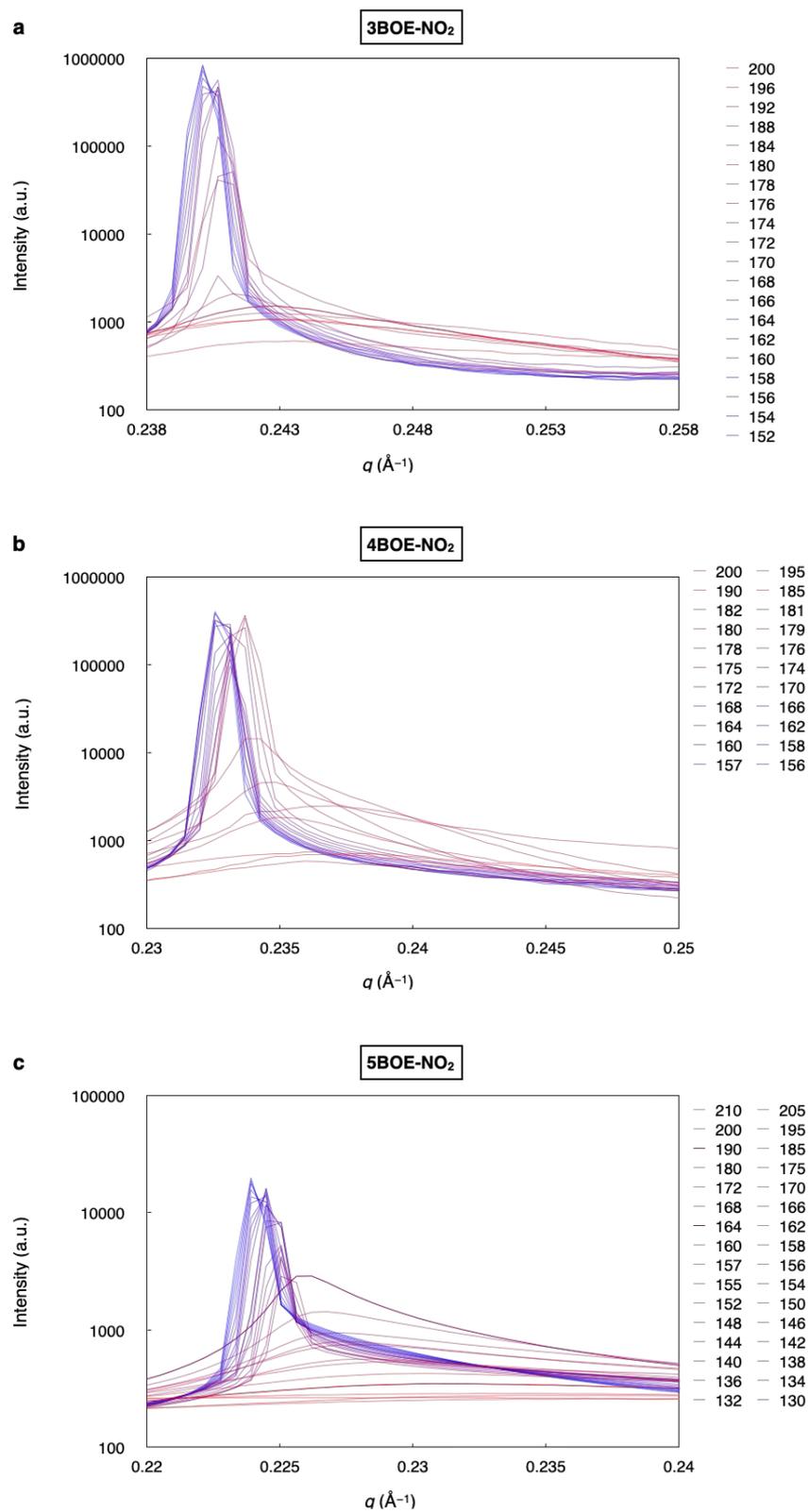

**Figure S18** Complete 1D XRD pattern for **nBOE-NO₂** (n = 3–5) in various temperature.

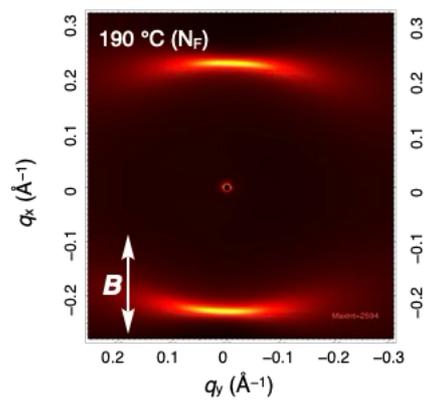

**Figure S19** 2D SAXS pattern with *M*-field (≈1 T) in the N$_F$ phase for **5BOE-NO$_2$**.

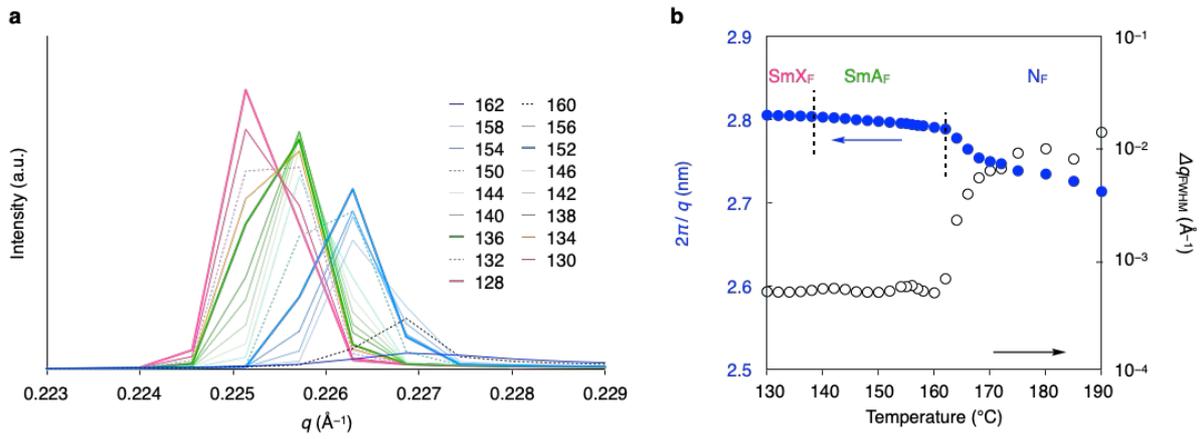

**Figure S20** a) Line scanned diffractogram of **5BOE-NO$_2$** (non-alignment sample) in various temperatures. b) $2\pi/q$ and $\Delta q_{FWHM}$ as a function of temperature.

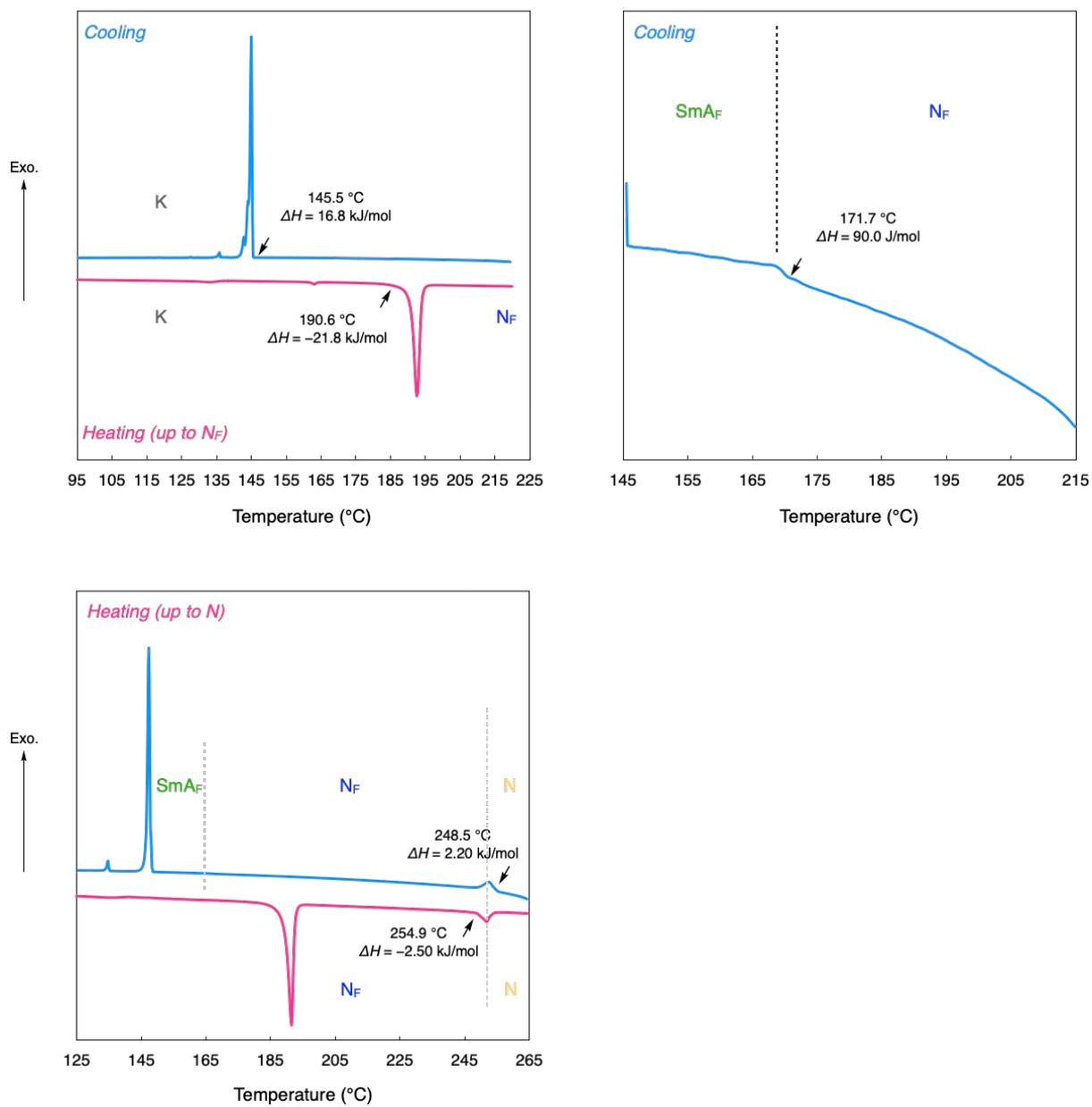

**Figure S21** DSC curves for **3BOE-NO₂**. Scan rate: 10 K min⁻¹.

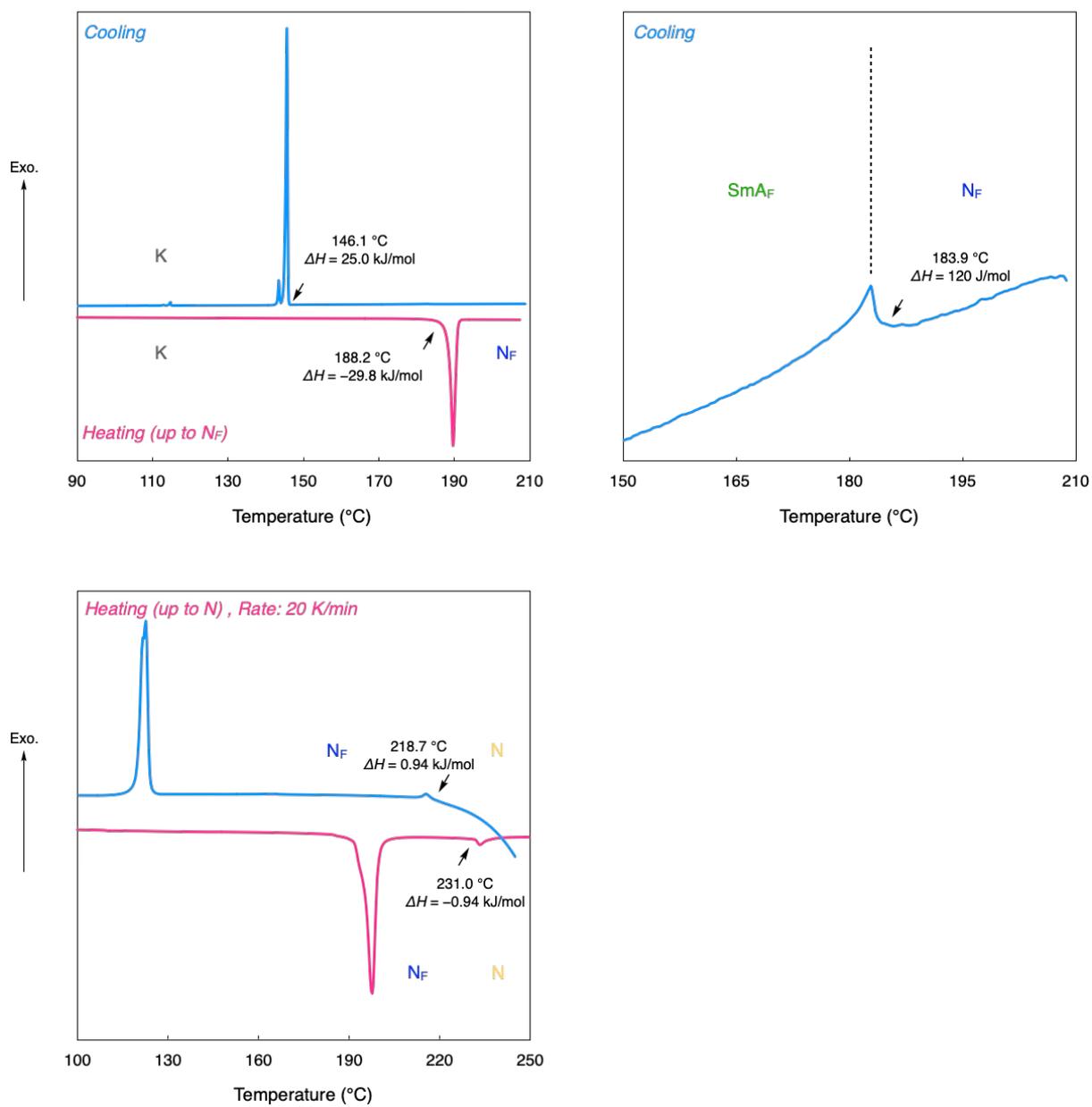

**Figure S22** DSC curves for **4BOE-NO$_2$**. Scan rate: 10 and 20 K min$^{-1}$.

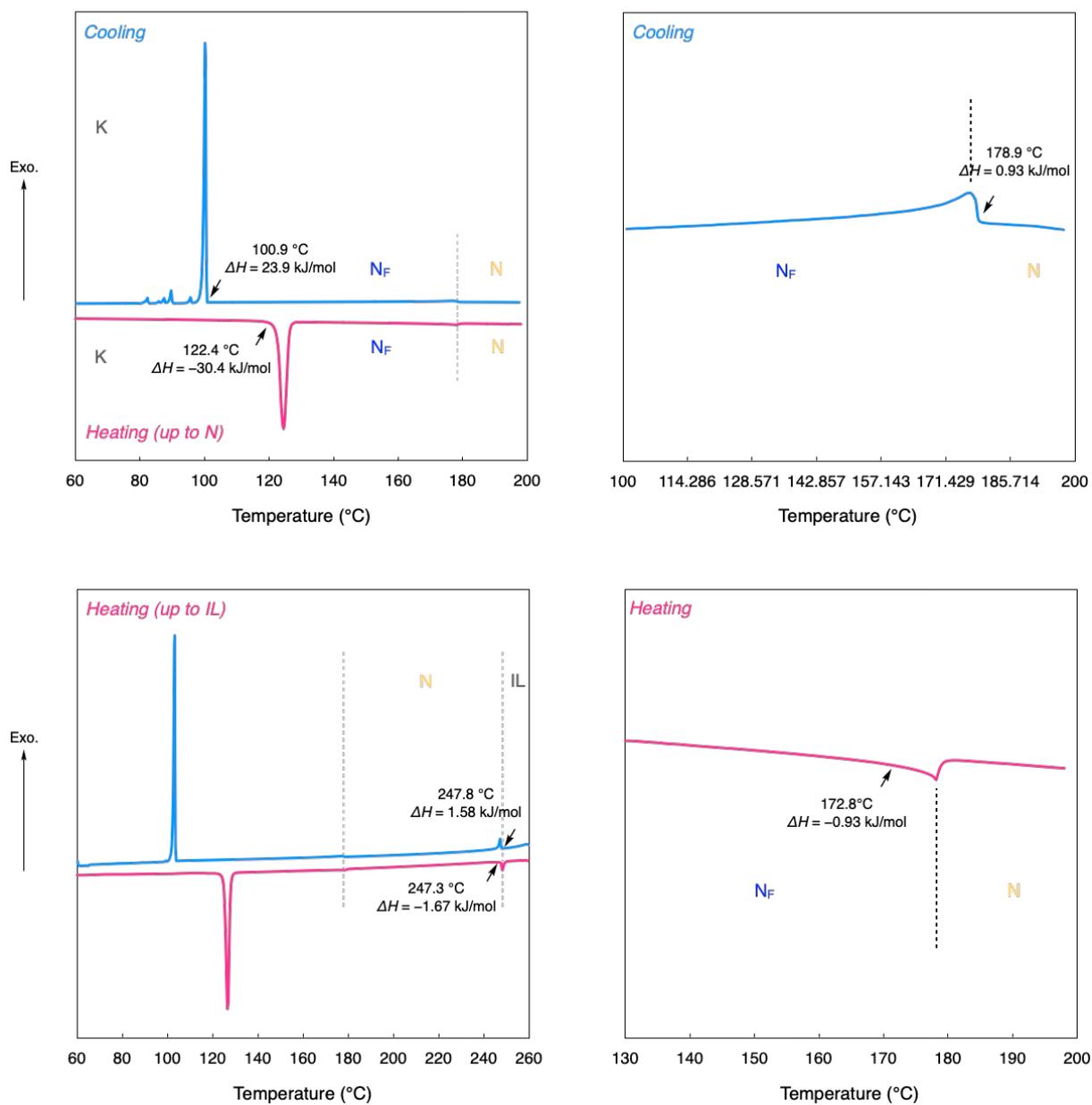

**Figure S23** DSC curves for **1DIOLT**. Scan rate: 10 K min$^{-1}$.

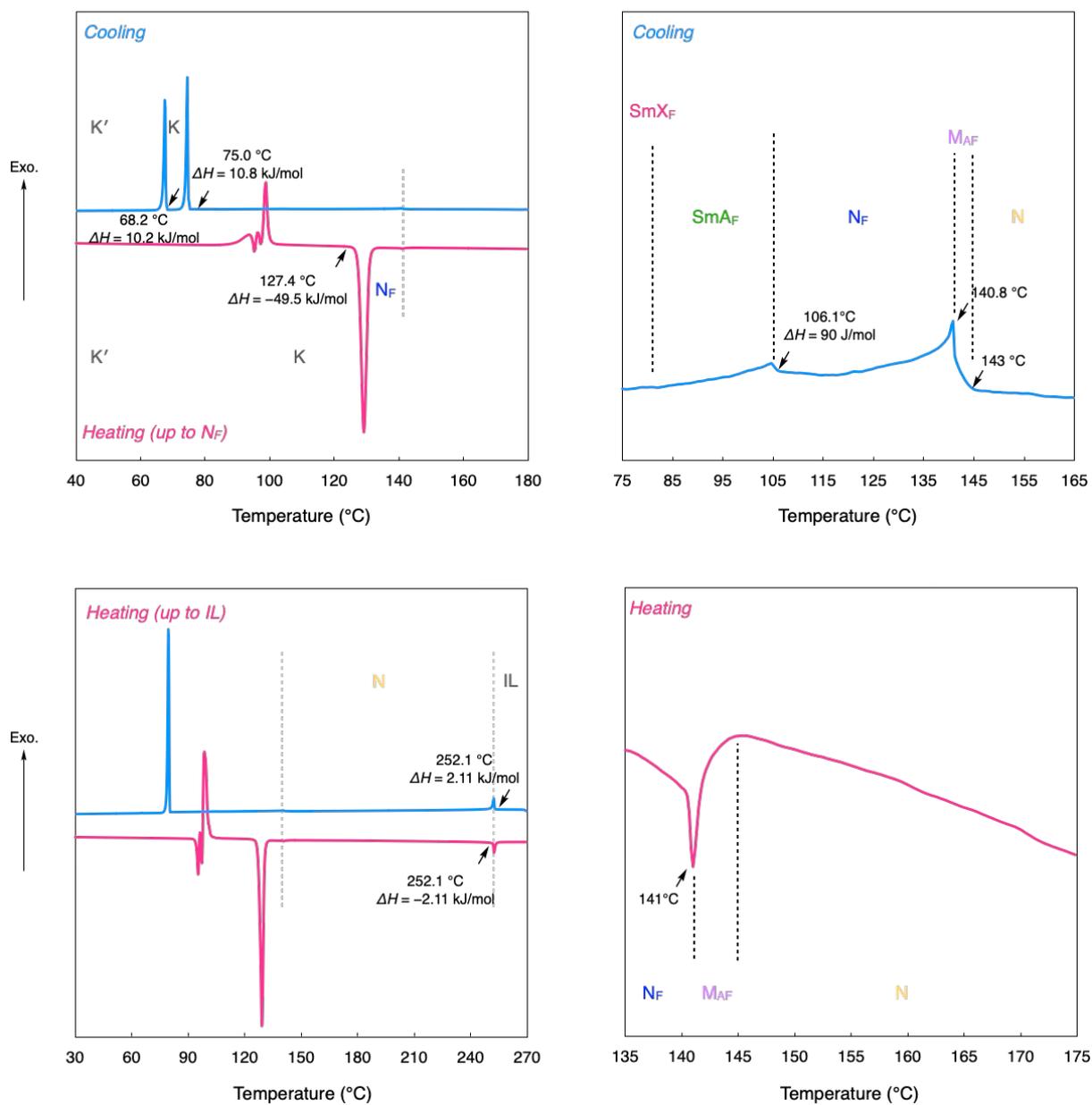

**Figure S24** DSC curves for **2DIOLT**. Scan rate: 10 K min$^{-1}$.

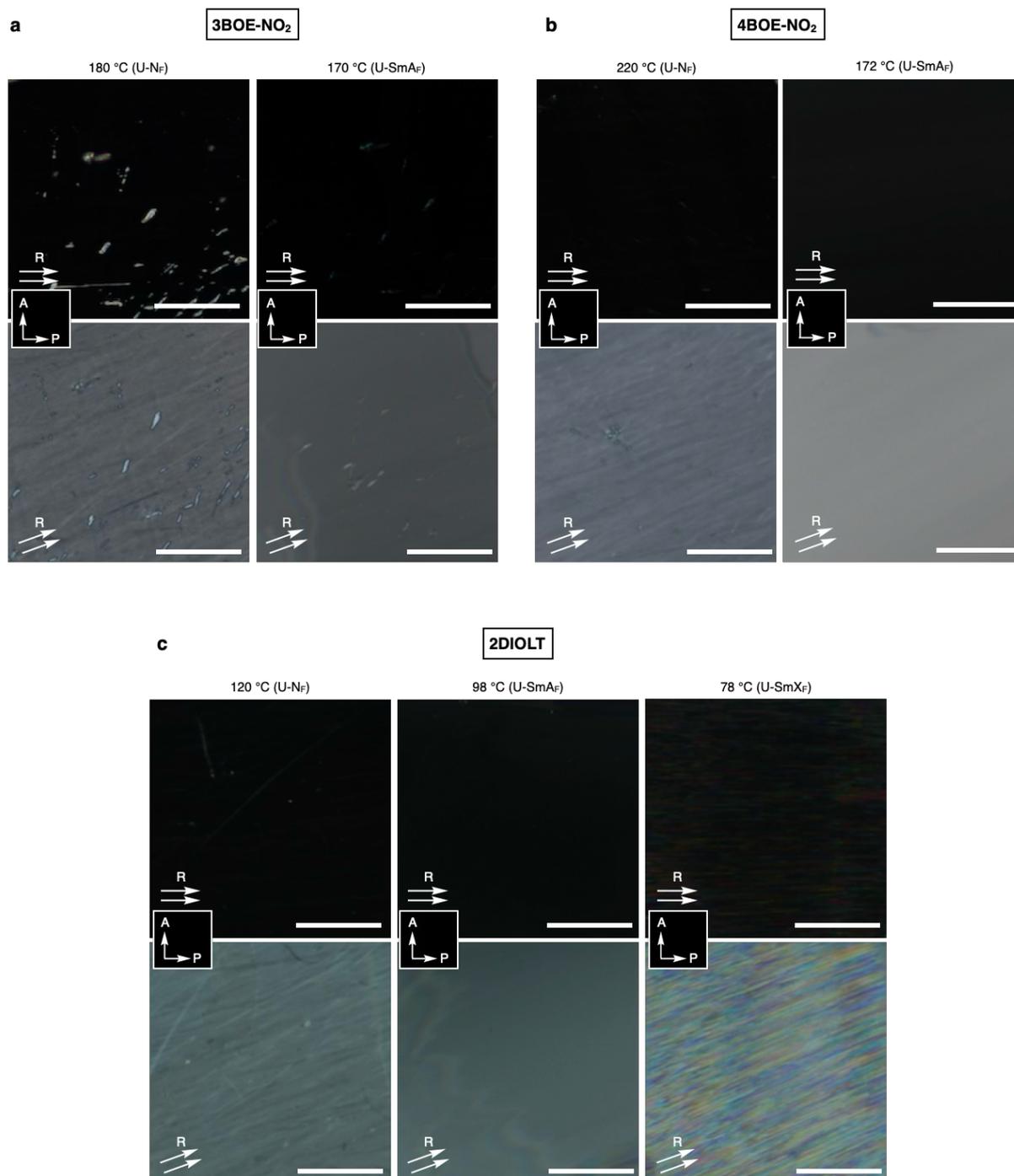

**Figure S25** POM images in the parallel rubbed cell for **3BOE-NO₂** (a), **4BOE-NO₂** (b) and **2DIOLT** (c). Thickness: 10 μm. Scale bar: 100 μm.

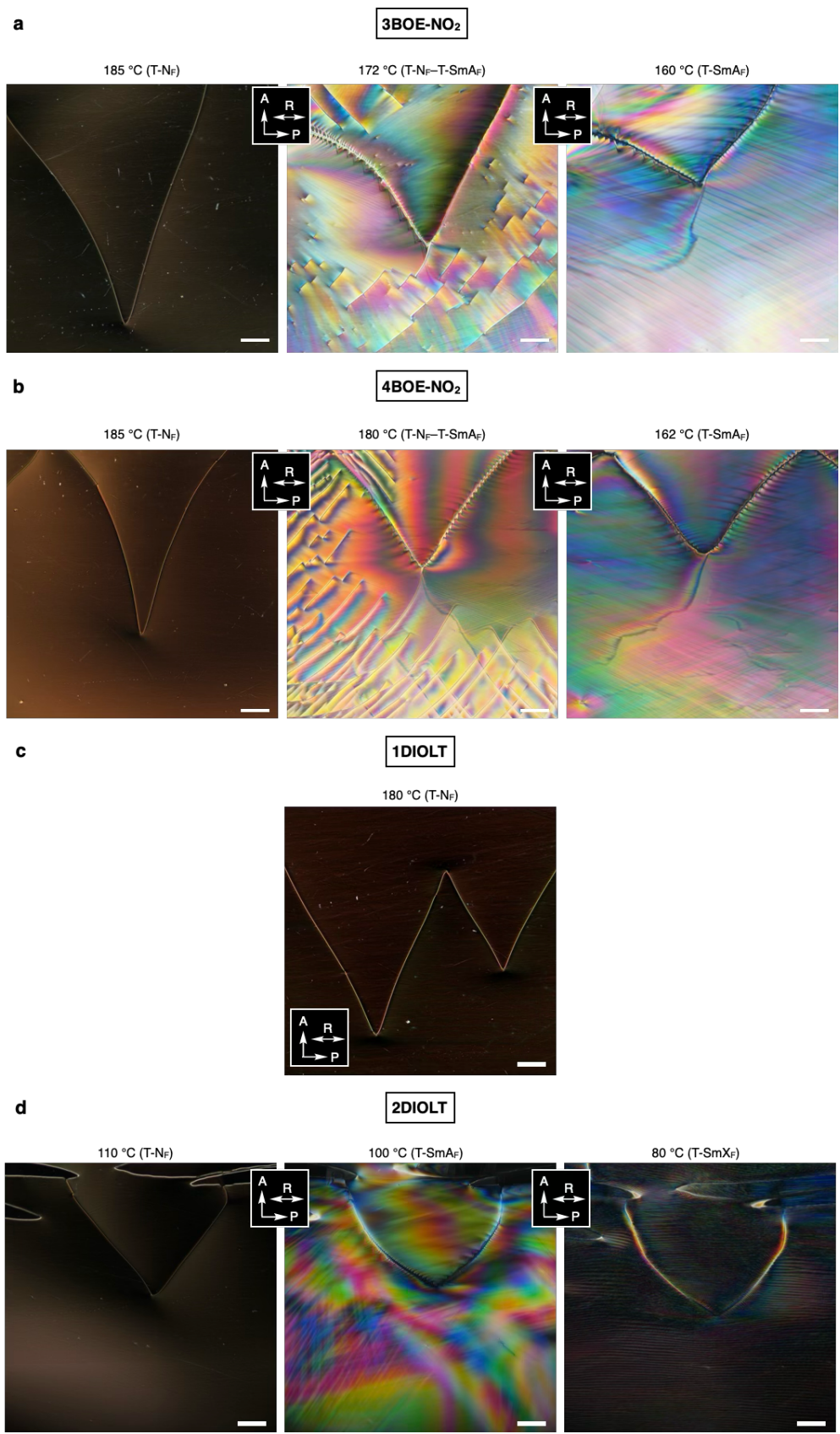

**Figure S26** POM images in the antiparallel rubbed cell for **3BOE-NO₂** (a), **4BOE-2NO₂** (b), **1DIOLT** (c) and **2DIOLT** (d). Thickness: 10 μm. Scale bar: 100 μm.

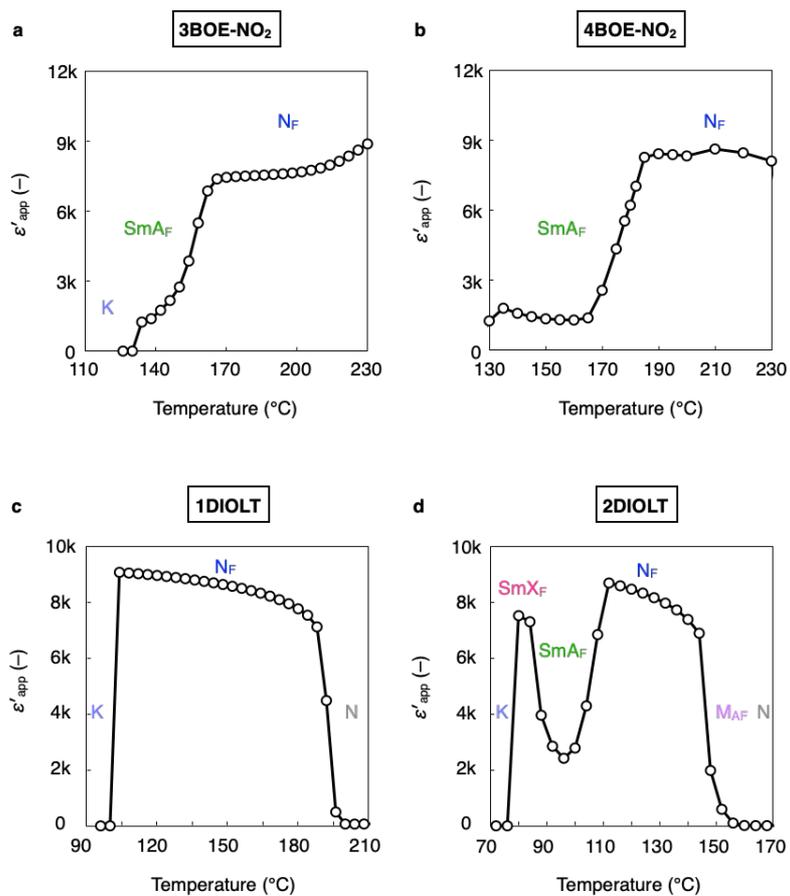

**Figure S27** Relative dielectric permittivity vs temperature for **nBOE-NO$_2$** (n = 3 and 4) and **nDIOLT** (n = 1 and 2).

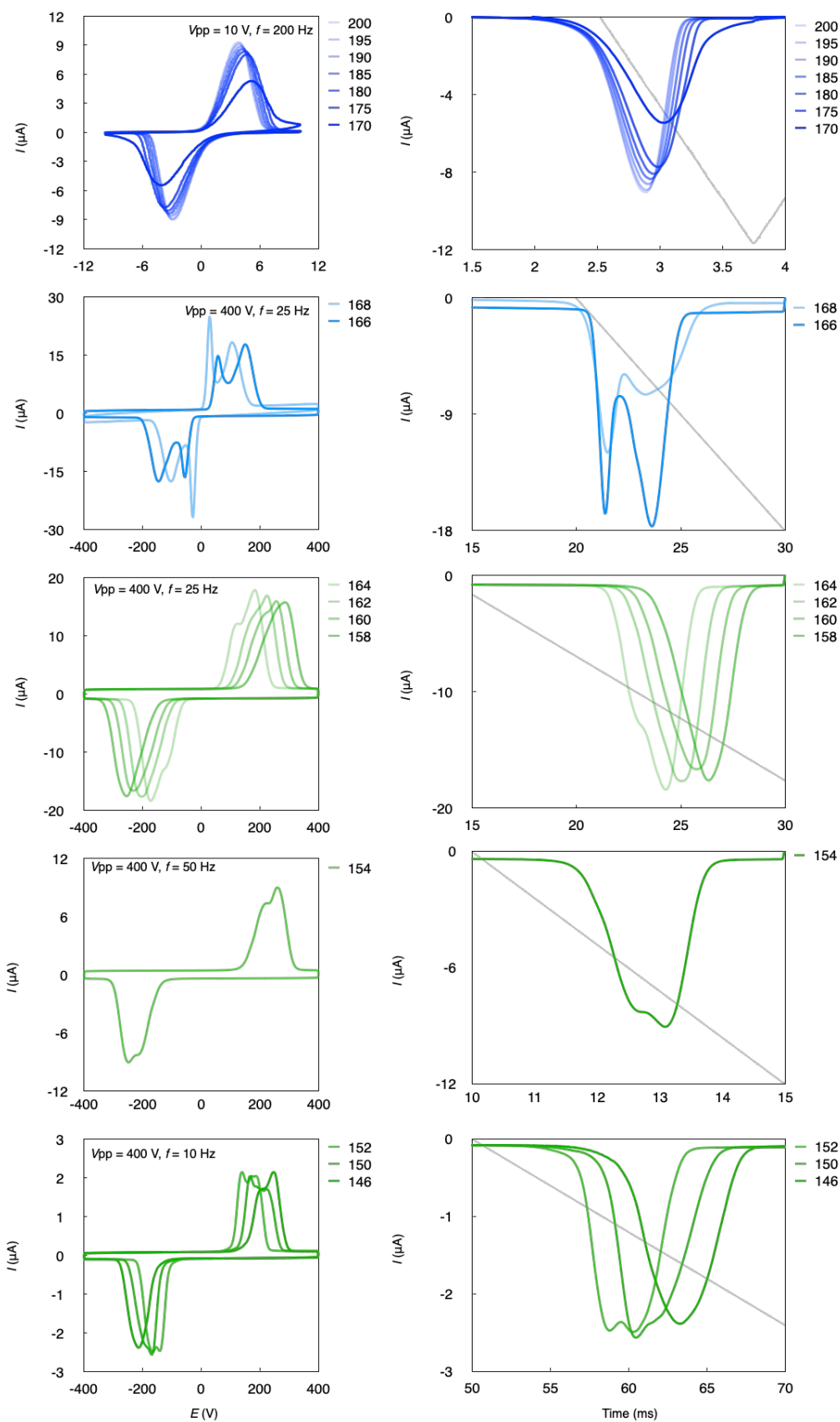

**Figure S28** Complete polarization reversal current data for **3BOE-NO₂** in various temperature.

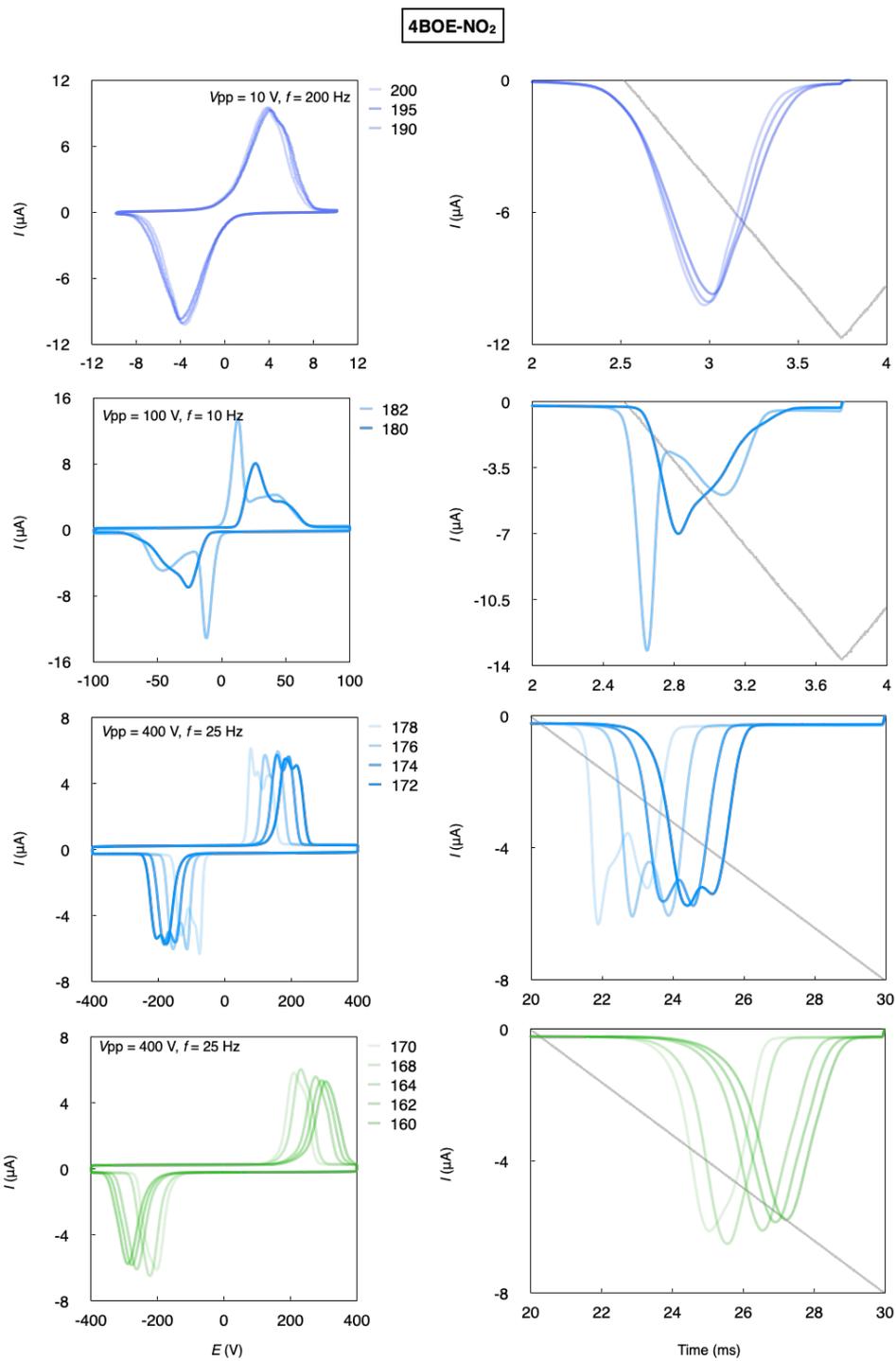

**Figure S29** Complete polarization reversal current data for **4BOE-NO₂** in various temperature.

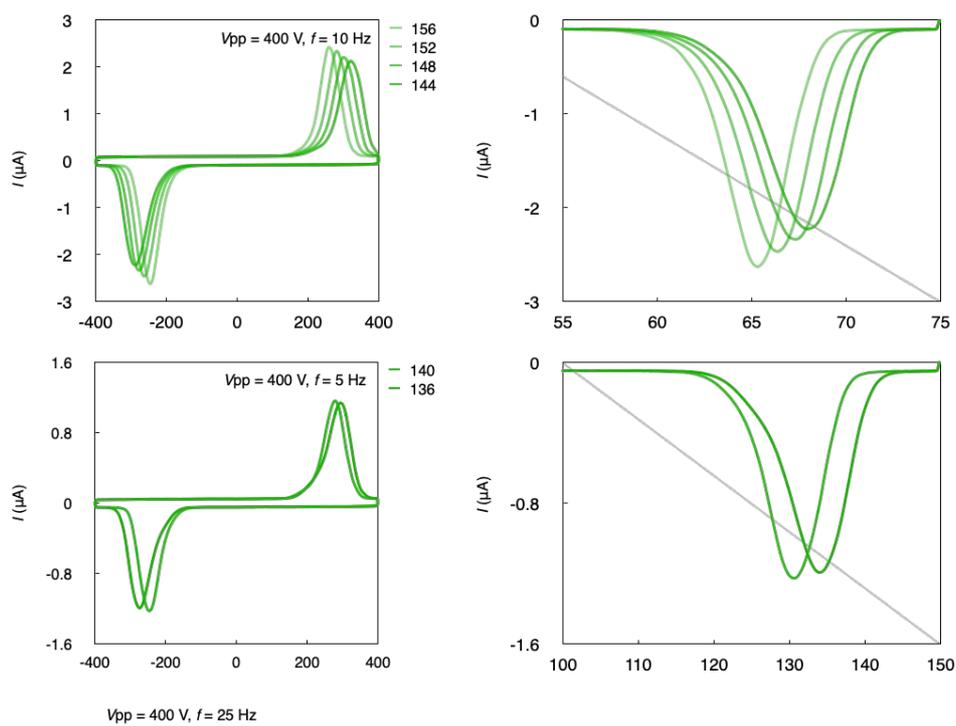

**Figure S30** (Continued) Complete polarization reversal current data for **4BOE-NO₂** in various temperature.

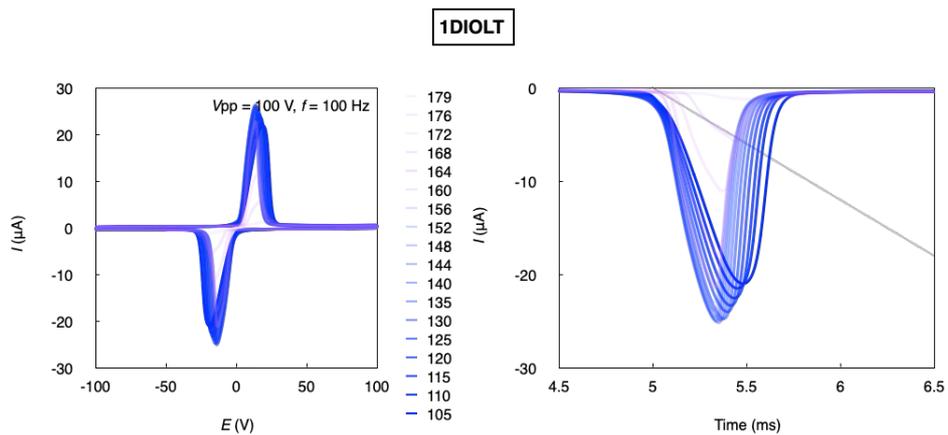

**Figure S31** Complete polarization reversal current data for **1DIOLT** in various temperature.

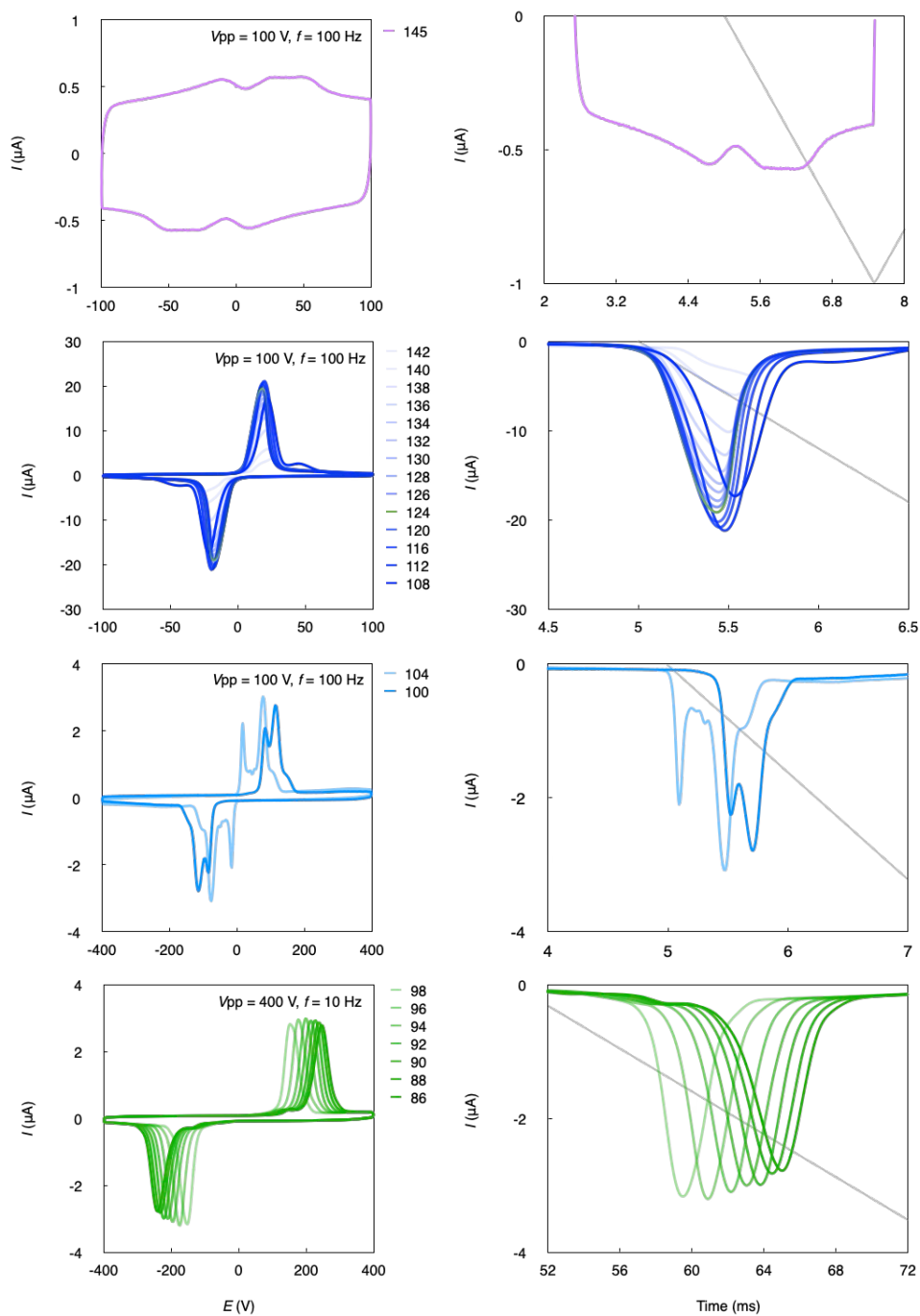

**Figure S32** Complete polarization reversal current data for **2DIOLT** in various temperature.

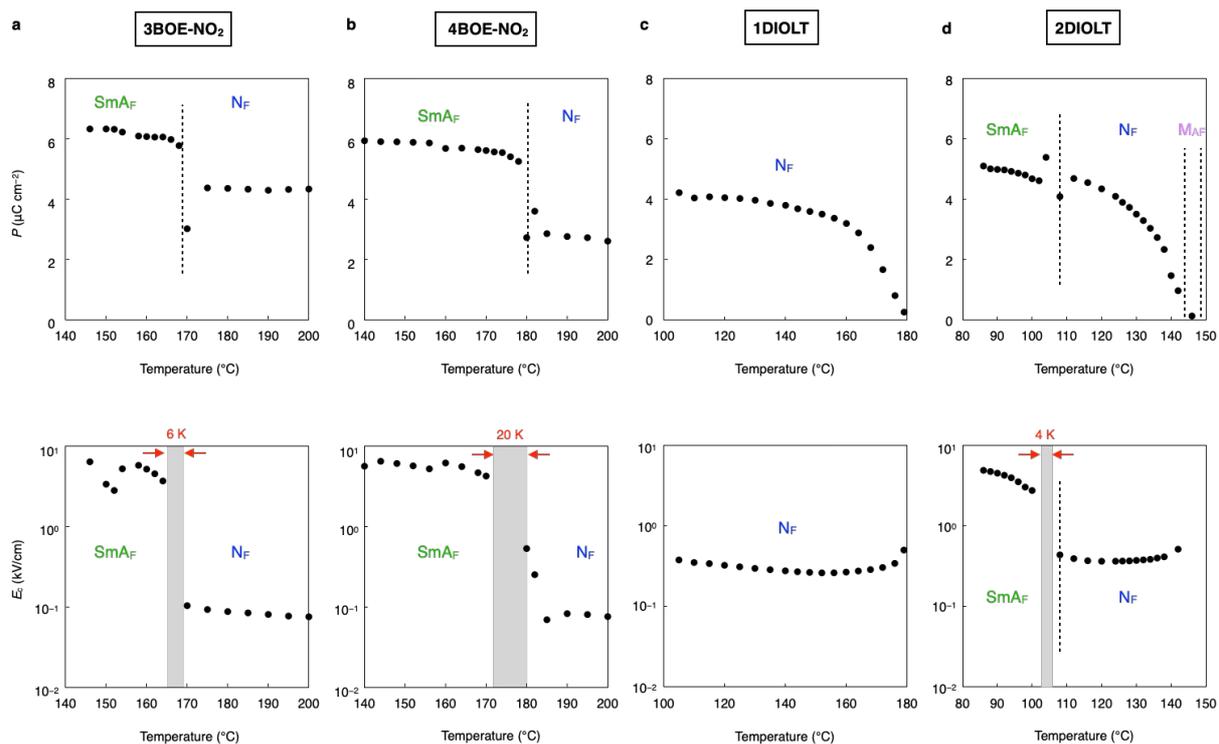

**Figure S33** $P_s$ and $E_c$ vs temperature for **3BOE-NO₂** (a), **4BOE-NO₂** (b), **1DIOLT** (c) and **2DIOLT** (d). Note: In the gray color region, $E_c$ could not be estimated because there were multiple peaks.

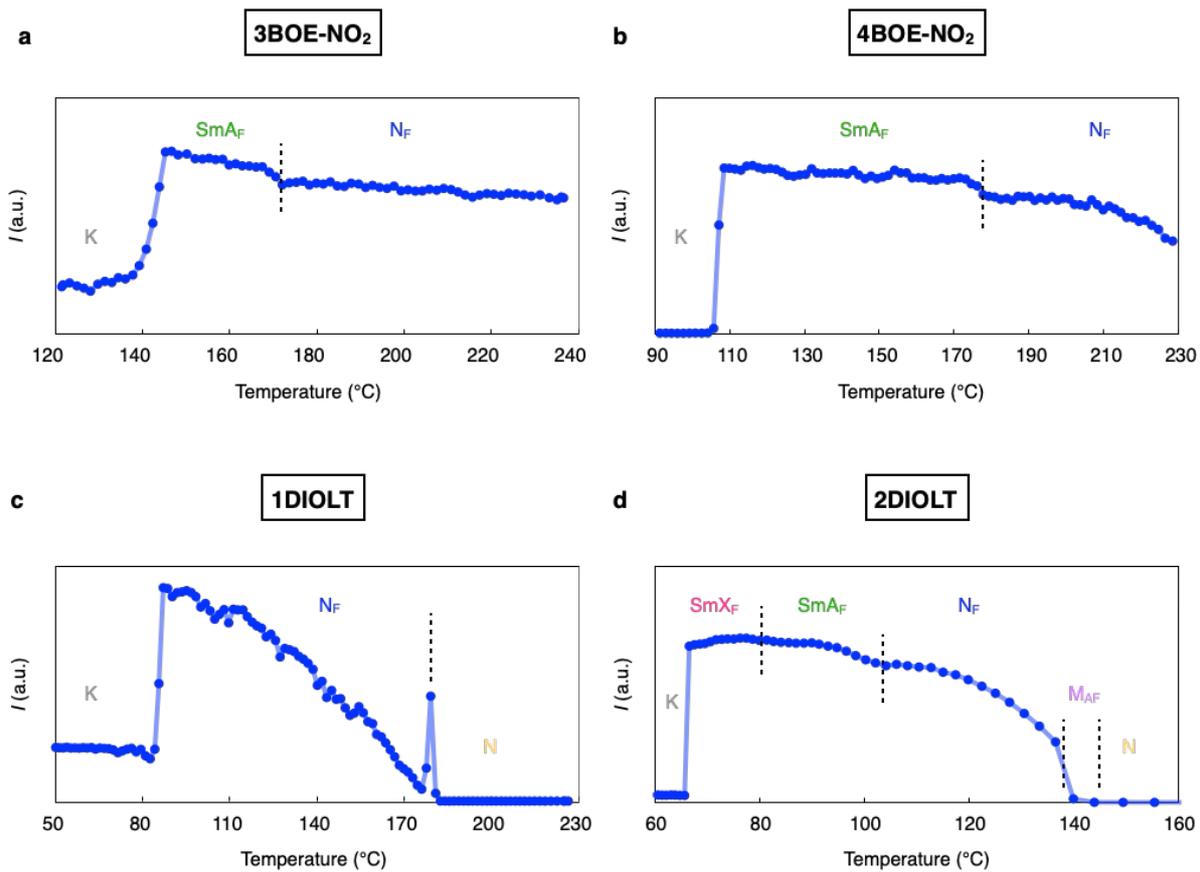

**Figure S34** SH intensity (*I*) vs temperature for **3BOE-NO$_2$** (a), **4BOE-NO$_2$** (b), **1DIOLT** (c) and **2DIOLT** (d).

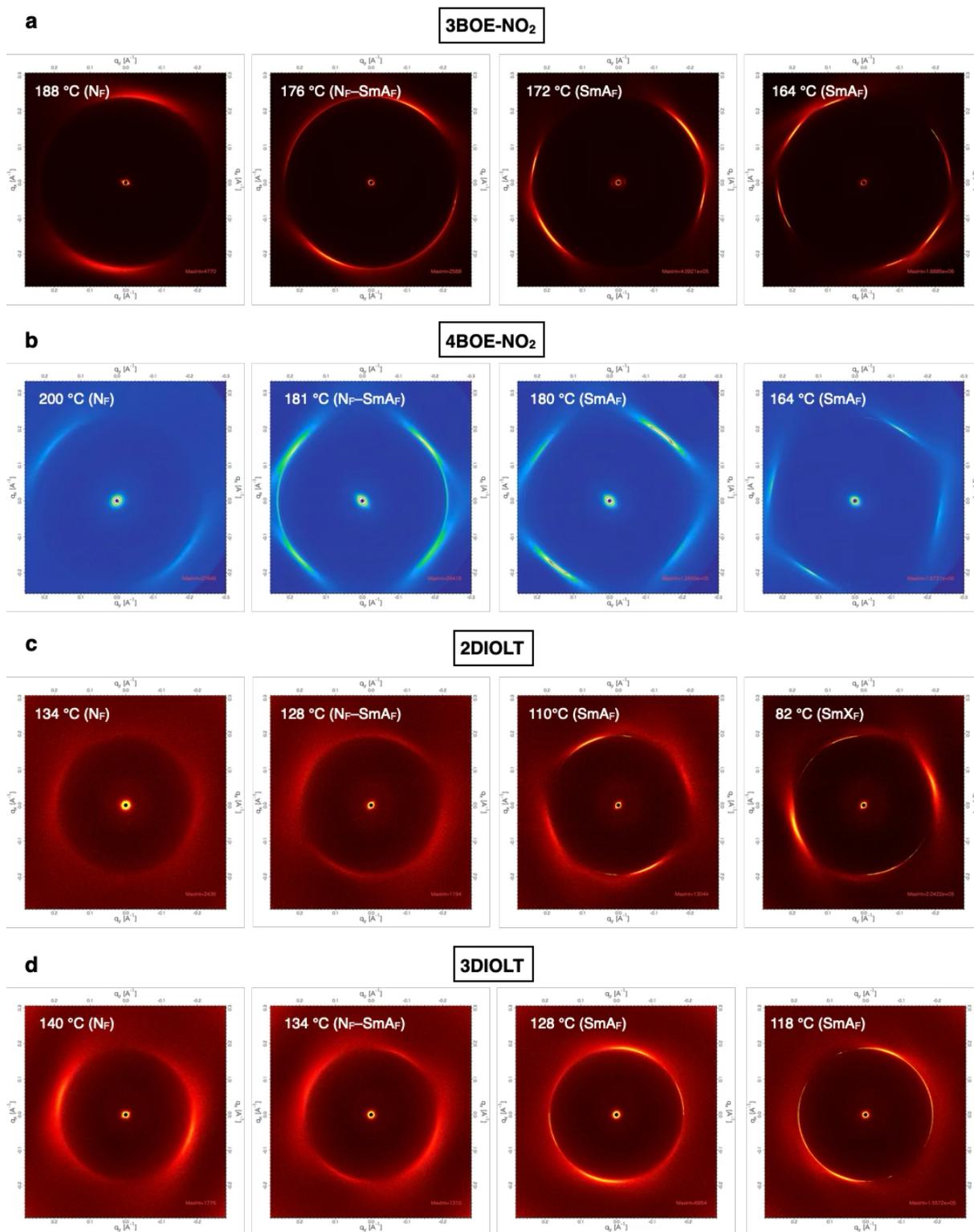

**Figure S35** 2D SAXS pattern without alignment for **3BOE-NO$_2$** (a), **4BOE-NO$_2$** (b), **2DIOLT** (c) and **3DIOLT** (d). Note: As to **3BOE-NO$_2$**, several pairs of diffraction peaks were observed, indicating that several smectic blocks are twisted. In this case, the difference in the intensity of each block is due to the different block thicknesses. For **4BOE-NO$_2$**, the intensities of the paired smectic blocks are different, suggesting this is due to the difference in the thickness of the blocks.

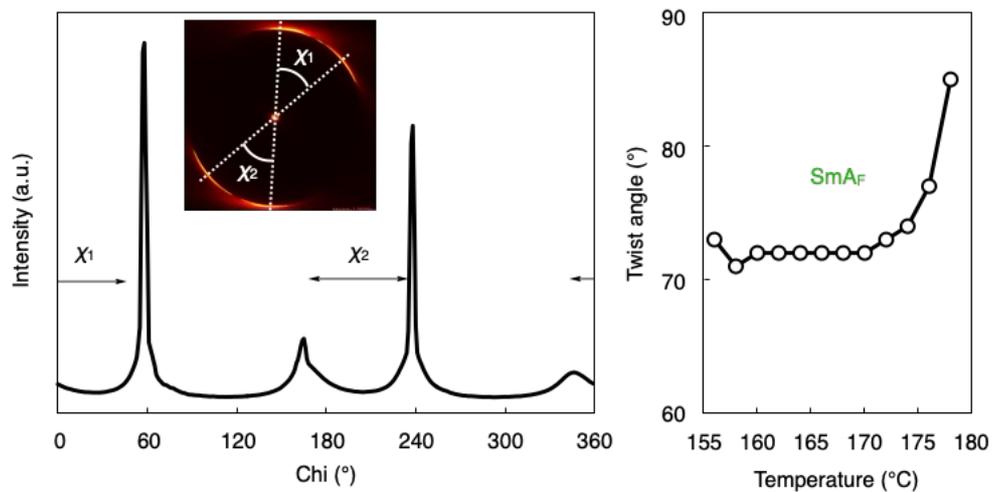

**Figure S36** Additional XRD data for **4BOE-NO$_2$**. a) Intensity vs chi angle. b) Twist angle between two pair of diffraction peaks as a function of temperature.

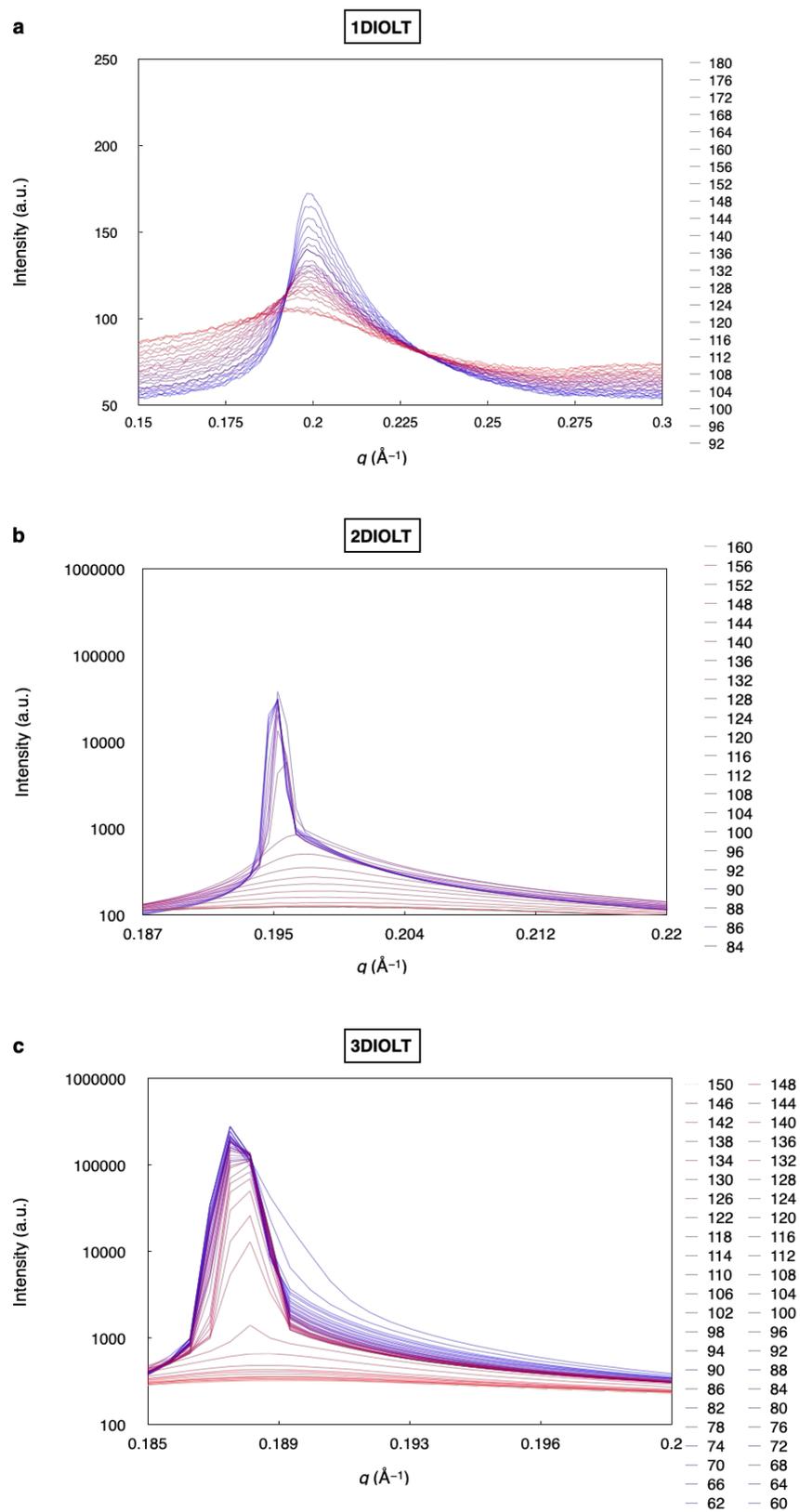

**Figure S37** Complete 1D XRD pattern for **nDIOLT** (n = 1–3) in various temperature.

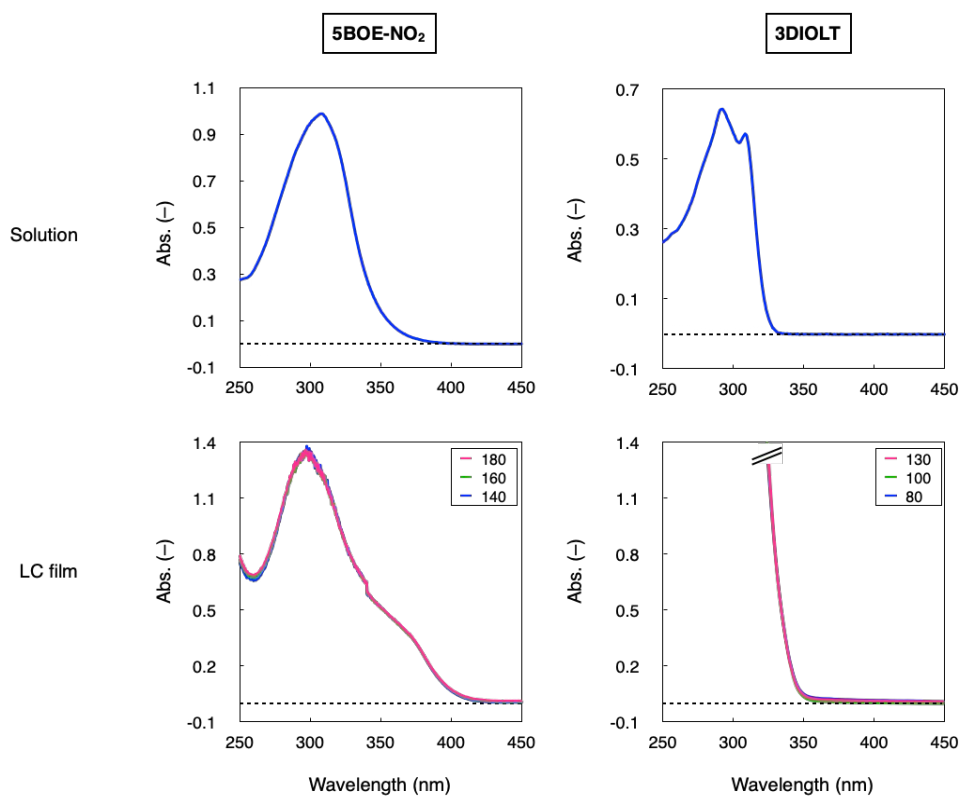

**Figure S38** UV-vis spectra for **5BOE-NO₂** and **3DIOLT** in chloroform (upper) and in LC film (bottom). Concentration in chloroform: 0.26 mM (**5BOE-NO₂**); 0.20 mM (**3DIOLT**). In principle, the CD spectra can be recorded for **3DIOLT**; however, the sample was damaged during spectra scan below 350 nm due to photoreaction of the tolan unit. To measure spectra for **5BOE-NO₂**, we scan the spectra over 370 nm through a long-pass filter (> 370 nm).

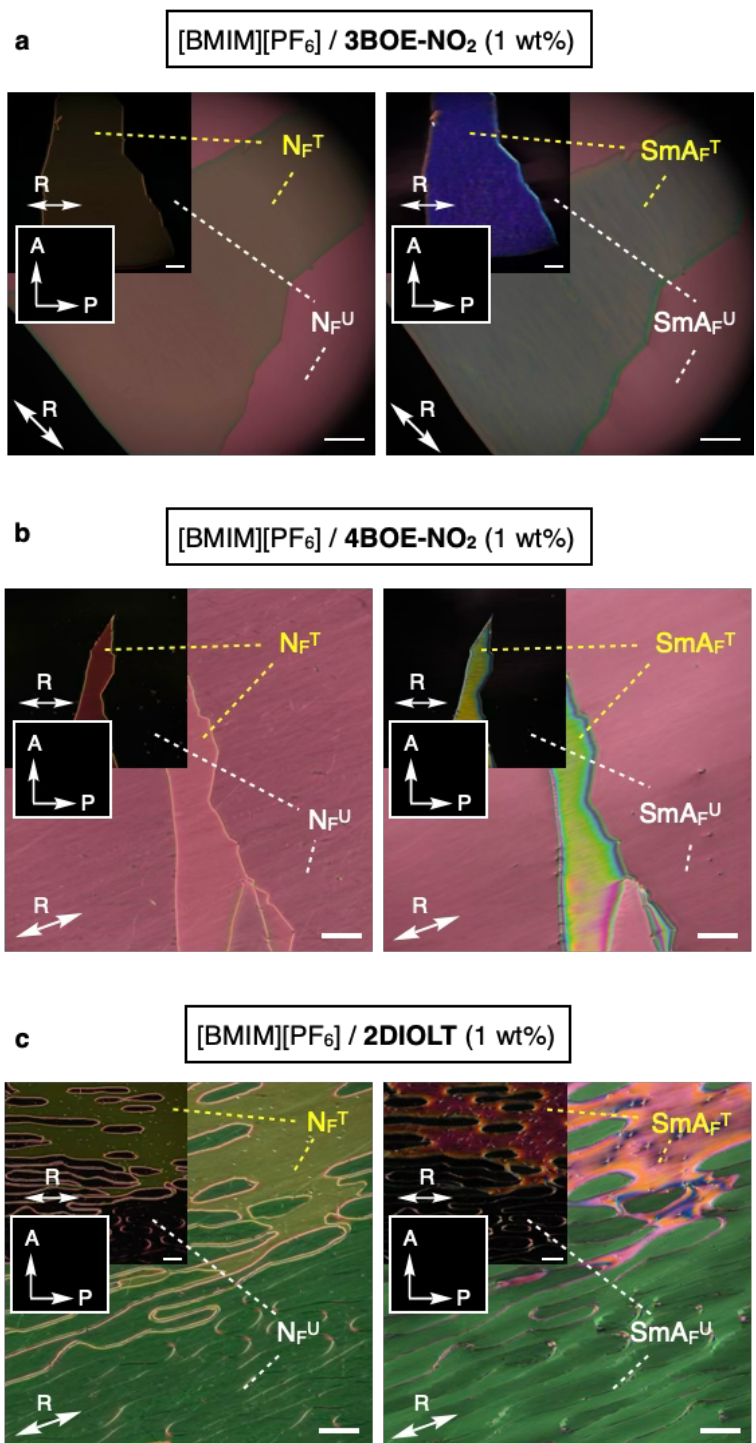

**Figure S39** POM texture in the antiparallel cell (thickness: 10 μm) for [BMIM][PF$_6$] / **3BOE-NO$_2$** (a), [BMIM][PF$_6$] / **4BOE-NO$_2$** (b) and [BMIM][PF$_6$] / **2DIOLT** (c). Conc.([BMIM][PF$_6$]) = 1 wt%. Scale bar: 100 μm.

**Supporting Tables (Tables S1–S3)**

**Table S1** The molecular parameters of energy-minimized conformations calculated by MM2/DFT for **nBOE-NO$_2$** (n = 3–5), **nDIOLT** (n = 1–3) and reference molecules.

| Entry | X | Y | Z | $\mu$ (D) | $\beta$ (deg)[a] |
|---|---|---|---|---|---|
| **3BOE-NO$_2$** | 14.978 | 0.036 | −0.041 | 14.978 | 0.207 |
| **4BOE-NO$_2$**[b] | 15.091 | 0.266 | 0.090 | 15.093 | 1.066 |
| **5BOE-NO$_2$** | 15.116 | 0.343 | 0.158 | 15.121 | 1.432 |
| **1DIOLT** | 12.546 | −1.097 | −2.497 | 12.839 | 12.26 |
| **2DIOLT** | 12.716 | −1.000 | −2.481 | 12.994 | 11.881 |
| **3DIOLT** | 12.798 | −0.776 | −2.431 | 13.049 | 11.276 |

a) an angle between the permanent dipole moment ($\mu$) and long molecular axis; b) reported in the previous work [S4]

**Table S2** Phase transition temperature (°C) and enthalpy changes (kJ mol$^{-1}$, in parenthesises) for **nBOE-NO$_2$** (n = 3–5).

| n | m.p. | process | SmX$_F$ | | SmA$_F$ | | N$_F$ | | M$_{AF}$ | | N | | IL |
|---|---|---|---|---|---|---|---|---|---|---|---|---|---|
| 3 | 202.0[a] (32.8) | 1C | | | • | 171.7 (0.09) | • | | | 248.5 (2.20)[c] | • | –[e] | • |
| 3 | 190.6[b] (21.8) | 2H | | | | | • | | | 254.9 (−2.50)[c] | • | –[e] | • |
| 4 | 188.5[a] (30.3) | 1C | | | • | 183.9 (0.12) | • | | | 218.7 (0.94)[c] | • | –[e] | • |
| 4 | 188.2[b] (29.8) | 2H | | | | | • | | | 231.0 (−1.47)[c] | • | –[e] | • |
| 5 | 190.1[a] (30.8) | 1C | • | 138[d] | • | 170.3 (0.11) | • | 212.8 (0.59) | • | 220.4 (0.02) | • | –[e] | • |
| 5 | 188.7[b] (30.8) | 2H | | | | | • | 211.4 (−0.60) | • | 217.9 (−0.02) | • | –[e] | • |

a) m.p. at 1st heating, b) m.p. at 2nd heating, c) heating up to N, d) identified by POM studies, e) not identified due to decomp. > ≈250 °C.

**Table S3** Phase transition temperature (°C) and enthalpy changes (kJ mol$^{-1}$, in parenthesises) for **nDIOLT** (n = 1–3).

| n | m.p. | Process | SmX$_F$ | | SmA$_F$ | | N$_F$ | | M$_{AF}$ | | N | | IL |
|---|---|---|---|---|---|---|---|---|---|---|---|---|---|
| 1 | 124.9[a] | 1C | | | | | 100.9 (23.9) | • | 178.9 (0.93) | • | 247.8 (1.58)[c] | • | |
| | 122.4[b] (30.4) | 2H | | | | | | • | 173.1 (−0.93) | • | 247.3 (−1.67)[c] | • | |
| 2 | 128.1[a] (50.3) | 1C | • | 88[d] | • | | 106.6 (0.1) | • | 140.8 (x)[f] | • | 143.0[d] (y)[f] | • | 252.1 (2.11)[c] | • |
| | 127.4[b] (49.5) | 2H | | | | | | • | 141.0 (x)[g] | • | 145.6[d] (y)[g] | • | 252.1 (−2.11)[c] | • |
| 3 | 95.2 (30.6)(a | 1C | • | 75[d] | • | | 123.4 (0.05) | • | 127.1 (0.13) | | | • | 263.8 (2.14)[c] | • |
| | 89.0[b] (22.1) | 2H | | | | • | 120.4 (−0.05) | • | 126.2 (−0.13) | | | • | 264.4 (−2.14)[c] | • |

a) m.p. at 1st heating, b) m.p. at 2nd heating, c) heating up to IL, d) identified by POM studies, e) not identified due to decomp, f) x + y = 0.22 kJ mol$^{-1}$, g) x + y = −0.22 kJ mol$^{-1}$.

**Supporting Movie**

**Movie S1. POM image changes between T-N$_F$ and T-SmA$_F$ states for 5BOE-NO$_2$.**

**Supporting References**